\documentclass[11pt,aps,showpacs,nofootinbib,prd,aps,epsf,floats,
               amsmath,amssymb,amsfonts]{revtex4}
\usepackage{amsmath, amssymb}
\newcommand{\mathsym}[1]{{}}

\usepackage{graphicx}
\usepackage{amsmath}
\usepackage{amssymb}
\usepackage{bm}
\newcommand{\be}{\begin{equation}}
\newcommand{\ee}{\end{equation}}
\newcommand{\bea}{\begin{eqnarray}}
\newcommand{\eea}{\end{eqnarray}}

\newcommand{\rem}[1]{}
\newsavebox{\PSLASH}
 \sbox{\PSLASH}{$p$\hspace{-1.8mm}/}
 
\renewcommand{\theequation}{\thesection.\arabic{equation}}
\newcounter{saveeqn}
\newcommand{\add}{\addtocounter{equation}{1}}
\newcommand{\alpheqn}{\setcounter{saveeqn}{\value{equation}}%
\setcounter{equation}{0}%
\renewcommand{\theequation}{\mbox{\thesection.\arabic{saveeqn}{\alph{equation}}}}}
\newcommand{\reseteqn}{\setcounter{equation}{\value{saveeqn}}%
\renewcommand{\theequation}{\thesection.\arabic{equation}}}

 \newsavebox{\notrightarrow}
 \sbox{\notrightarrow}{$\to$\hspace{-4mm}/}
 
 \newsavebox{\PARTIALSLASH}
 \sbox{\PARTIALSLASH}{$\partial$\hspace{-1.6mm}/}
 
 \newsavebox{\ASLASH}
 \sbox{\ASLASH}{$A$\hspace{-2.1mm}/}
 
 \newsavebox{\KSLASH}
 \sbox{\KSLASH}{$k$\hspace{-1.8mm}/}
 
 \newsavebox{\LSLASH}
 \sbox{\LSLASH}{$\ell$\hspace{-1.8mm}/}
 
 \newsavebox{\QSLASH}
 \sbox{\QSLASH}{$q$\hspace{-1.8mm}/}
 
 \newsavebox{\DSLASH}
 \sbox{\DSLASH}{$D$\hspace{-2.2mm}/}
 
 \newsavebox{\DbfSLASH}
 \sbox{\DbfSLASH}{${\mathbf D}$\hspace{-2.8mm}/}
 
 \newsavebox{\DELVECRIGHT}
 \sbox{\DELVECRIGHT}{$\stackrel{\rightarrow}{\partial}$}
 
 \newcommand{\blue}{\IfColor{\textCadetBlue}{}}
\newcommand{\black}{\IfColor{\textBlack}{}}
\newcommand{\red}{\IfColor{\textRed}{}}
\newcommand{\green}{\IfColor{\textOliveGreen}{}}
\newcommand{\lila}{\IfColor{\textRedViolet}{}}

\begin{document}

\title{Color neutral 2SC phase of cold and dense quark matter in the
presence of constant magnetic fields}
\author{Sh. Fayazbakhsh}\email{fayyazbaksh@physics.sharif.ir}
\author{N. Sadooghi}\email{sadooghi@physics.sharif.ir}
\affiliation{Department of Physics, Sharif University of Technology,
P.O. Box 11155-9161, Tehran-Iran}
\begin{abstract}
\noindent The color neutral two-flavor superconducting (2SC) phase
of cold and dense quark matter is studied in the presence of
constant magnetic fields and at moderate baryon densities. In the
first part of the paper, a two-flavor effective Nambu--Jona-Lasinio
(NJL) model consisting of a chiral symmetry breaking ($\chi$SB) mass
gap $\sigma_{B}$, a color superconducting (CSC) mass gap
$\Delta_{B}$ and a color chemical potential $\mu_{8}$ is introduced
in the presence of a rotated $U(1)$ magnetic field
$\tilde{\mathbf{B}}$. To study the phenomenon of magnetic catalysis
in the presence of strong magnetic fields, the gap equations
corresponding to $\sigma_{B}$ and $\Delta_{B}$, as well as $\mu_{8}$
are solved in the lowest Landau level (LLL) approximation. In the
second part of the paper, a detailed numerical analysis is performed
to explore the effect of any arbitrary magnetic field on the above
mass gaps and the color chemical potential. The structure of the
$\chi$SB and CSC phases is also presented in the
$\mu_{c}-\tilde{e}B$ plane, and the effect of $\mu_{8}$ on the phase
structure of the model is explored. As it turns out, whereas the
transition from the $\chi$SB to CSC phase is of first order,
nonvanishing $\mu_{8}$ affects essentially the second order phase
transition from CSC to the normal phase.
\end{abstract}
 \maketitle
\section{Introduction}\label{introduction}
\par\noindent
Dense baryonic matter at low temperature and asymptotically large
chemical potential is known to be a color superconductor
\cite{old-CSC}. This can be shown in the framework of perturbative
Quantum Chromodynamics (pQCD). To explore the color superconducting
phase at moderate chemical potential, however, it is necessary to
use effective models, such as the well-known NJL model with
four-fermion interaction \cite{NJL}. Using an appropriate NJL type
model, one can show that at baryon densities $\mu_{c}\simeq 350$
MeV, i.e. only several times larger than the density of nuclear
matter, the two-flavor color superconducting (2SC) phase might be
present \cite{shuryak,huang2004} (see \cite{rajagopal2007} for
recent reviews on color superconductivity in dense quark matter).
Different astrophysical processes might therefore be influenced by
the color superconductivity that is supposed to exist inside the
compact stars. In \cite{ebert2005}, the competition between the
chiral symmetry breaking and the color symmetry breaking condensates
is investigated in the framework of a two-flavor color neutral NJL
type model, including meson and diquark condensates, $\sigma_{0}$
and $\Delta_{0}$. Imposing the color neutrality condition, it is
found that in the 2SC phase at $\mu>\mu_{c}=342$ MeV, the color
chemical potential $\mu_{8}$ acquires rather small values of about
$10$ MeV.\footnote{The underlying physics of color charge neutrality
is discussed in  \cite{neutrality}.} Here, $\mu_{c}$ is the critical
chemical potential. The diquark mass gap is numerically computed to
be $\Delta_{0}\simeq 100$ MeV. It is also shown that the appearance
of a coexistence regime (mixed phase) depends directly on the
relative strength of the meson and diquark coupling constants
$G_{S}$ and $G_{D}$. This is also indicated in \cite{berges1998,
mixed, huang2002}, where it is stated that neglecting the quark
masses and choosing $G_{D}<G_{S}$, no mixed phase appears at
$\mu>\mu_{c}$. The $\chi$SB and CSC phases can therefore be studied
separately under these conditions.
\par
In the present paper, we study the mesons and diquarks in the color
neutral 2SC phase of cold and dense quark matter in the presence of
constant magnetic fields. The aim is to study the effect of the
magnetic field on the formation of chiral as well as diquark
condensates, the dependence of mass gaps on the chemical potential
$\mu$ and the external magnetic field, the phase diagram $\mu$ vs.
$B$, and the effect of nonvanishing color chemical potential on the
type of phase transitions for different $\mu$ and $B$ at zero
temperature $T$.\footnote{The effect of finite temperature will be
presented elsewhere \cite{sadooghi2010}.}
\par
The study of quark matter in the presence of constant magnetic field
is relevant for the astrophysics of compact stars: Strong magnetic
fields exist on the surface of compact stars. For neutron stars the
magnetic fields $B\lesssim 10^{12}$ Gau\ss, whereas for magnetars,
they can be as large as $B\simeq 10^{16}$ Gau\ss~
\cite{thompson1996}. In the interiors of compact stars, the magnetic
field can be even several orders of magnitude larger
\cite{incera2010}. On the other hand, it is believed that the
superdense interior of compact stars may be composed of electric and
color neutral quark matter in the color superconducting phase. To
test the predictions of astrophysical signatures of color
superconductivity, a better understanding of the r\^{o}le of
magnetic fields on the CSC phase is important. The study of
superconducting phase in the presence of external magnetic fields is
also relevant for the physics of heavy ion collisions: According to
\cite{warringa,kharzeev51-STAR}, in off-central collisions, heavy
ions possess a very large angular momentum and very strong magnetic
fields can be created. In \cite{skokov}, it is shown that the
magnetic field presently created at RHIC is at most $eB\simeq 1.3\
m_{\pi}^{2}\sim 4.3 \times 10^{18}$ Gau\ss, and the estimated value
of the magnetic field strength for the LHC energy amounts to $15\
m_{\pi}^{2}\sim 5\times 10^{19}$ Gau\ss.\footnote{Here, the pion
mass, $m_{\pi}=140$ MeV.} Recently, the question of accessibility of
the 2SC phase in the future heavy ion collision experiments is
investigated in \cite{blaschke2010}. Here, the authors do not
consider the effect of the before mentioned magnetic fields. It
would be therefore important to study the effect of external
magnetic fields on the formation of 2SC diquark condensates, as well
as the corresponding phase structure in the presence of external
magnetic fields. As for the results presented in this paper, they
may be relevant only for the physics of the heavy ion collisions,
because in contrary to the electric and color neutrality requirement
of the superdense core of the compact stars, only the color
neutrality condition is considered in this paper.
\par
The effect of constant magnetic field on the formation of diquark
condensates has been investigated by several authors. In
\cite{rajagopal1999,gorbar2000}, it is shown that there is a linear
combination of photon and a gluon that remains massless. The
resulting ``rotated'' external magnetic field can therefore
penetrate the color superconducting region without being affected by
the Meissner effect. This has consequences for the structure of
compact star cores. In \cite{ferrer2006, ferrer-CFL}, the formation
of magnetic color-flavor locked (MCFL) phase, as well as the
transition to the paramagnetic-CFL (PCFL) phase are studied. In
\cite{warringa2007, shovkovy2007}, it is shown that for small
magnetic fields, the CFL mass gap as well as the corresponding
magnetization exhibit small oscillations, the van Alfven--de Haas
(vAdH) oscillations. This effect, which is well-known from condensed
matter physics, is predicted by Landau and observed experimentally
by van Alfven and de Haas (see \cite{alfven} for an investigation of
this effect in cold dense quark matter in a homogeneous magnetic
field). The transport properties of 2SC phase is investigated
recently in \cite{sedrakian2010}.
\par
Recently,  in \cite{mandal2009}, the formation of chiral and diquark
condensates as well as their competition in the 2SC phase at zero
temperature and moderate densities are studied using the same NJL
type model as in the present paper. It is shown that for vanishing
magnetic field, a mixed broken phase can be found where both chiral
and superconducting gaps are non-zero. For $\tilde{e}B=0.05$ GeV$^2$
(corresponding to $B\simeq 8.5\times 10^{18}$ Gau\ss) and moderate
diquark-to-chiral coupling ratios $G_{D}/G_{S}$, the chiral and
superconducting transitions become weaker. For large $G_{D}/G_{S}$,
strong magnetic fields disrupt the mixed broken phase region and a
first order phase transition is found between the $\chi$SB and the
CSC phase for $\tilde{e}B=0.05$ GeV$^2$. In contrast to
\cite{mandal2009}, our results include a detailed analytical and
numerical survey on the effect of external magnetic field and color
chemical potential on cold and dense as well as color neutral quark
matter in the presence of external magnetic fields.
\par
The organization of this paper is as follows: In Sec. II, starting
from an extended Lagrangian density of a gauged NJL model containing
two flavors, and following the method presented in
\cite{rajagopal1999, gorbar2000}, we introduce the rotated magnetic
field $\tilde{e}B$ and determine the Lagrangian density containing
the $\chi$SB and CSC mass gaps, $\sigma_{B}$ and $\Delta_{B}$,
respectively. In Sec. III, the one-loop effective action and
thermodynamic potential of the model are determined at zero
temperature and finite quark chemical potential. In Sec. IV,
assuming very strong magnetic fields, we solve analytically the gap
equations corresponding to $\sigma_{B}$ and $\Delta_{B}$, as well as
the color chemical potential $\mu_{8}$ in an appropriate LLL
approximation. The $\chi$SB and the CSC phases are studied, in IV.A
and IV.B, separately. This is possible because of our specific
choice of free parameters, the quark mass $m_{0}$ and the meson and
diquark couplings $G_{S}$ and $G_{D}$. In the $\chi$SB phase,
characterized by $\sigma_{B}\neq 0$ and $\Delta_{B}=\mu_{8}=0$, the
magnetic field enhances the bound state formation. This is because
of the phenomenon of magnetic catalysis \cite{miransky1995,
catalysis} studied intensively in the past few years.\footnote{See
\cite{cosmology} for the application of magnetic catalysis in
cosmology, \cite{condensed} for its application in condensed matter
physics, and \cite{particle, inagaki2003, sato1998, providenca2008}
for its applications in particle physics.} In the CSC phase,
characterized by $\sigma_{B}=0$, $\Delta_{B}\neq 0$ and $\mu_{8}\neq
0$, we determine analytically the $\mu$ and $\tilde{e}B$ dependence
of $\Delta_{B}$ and $\mu_{8}$ in the regime of LLL dominance. In
Sec. V, a numerical analysis is performed to study the $\tilde{e}B$
dependence of the $\chi$SB and CSC mass gaps at $\mu=250$ MeV (in
the $\chi$SB regime) and $\mu=460$ MeV (in the CSC regime). For
small values of $\tilde{e}B$, we observe vAdH oscillations in the
mass gaps as well as the corresponding magnetizations, as expected.
These are also observed in \cite{warringa2007, shovkovy2007} for
three-flavor CFL phase at $\mu=500$ MeV.  At $\tilde{e}B\simeq
0.45-0.5$ GeV$^{2}$, the oscillations end up in a ``linear regime''.
Comparing eventually our numerical results for $\tilde{e}B\gtrsim
0.45$ GeV$^{2}$ with the analytical results arising in Sec. IV for
strong magnetic fields in the LLL approximation, we conclude that
this approximation is only reliable in the above linear regime. The
$\mu$-dependence of the mass gaps and the color chemical potential
is also discussed for various $\tilde{e}B$. Our results for
vanishing $\tilde{e}B$ coincide with the results in
\cite{ebert2005}. We also present the phase structure of $\chi$SB
and CSC phases in a $\mu_{c}-\tilde{e}B$ plane. In particular, we
are interested on the effect of the color chemical potential
$\mu_{8}$ on the phase structure of the model. As it turns out, for
$\mu_{8}=0$, a first order phase transition exists between the
$\chi$SB and the CSC phase in the regime $\mu_{c}\simeq 350-450$ MeV
and $\tilde{e}B\in[0,0,7]$ GeV$^{2}$, whereas the transition from
the CSC to the normal phase is of second order and occurs at
$\mu_{c}\simeq 750-800$ MeV. For $\mu_{8}\neq 0$, however, whereas
the phase transition between the $\chi$SB and the CSC phase is still
of first order, the second order phase transition between the CSC
and the normal phase goes over into a first order phase transition
between the CSC and the normal phase at $\mu_{c}\simeq 755$ MeV and
$\tilde{e}B\simeq 0.13$ GeV$^{2}$. Note that the first order nature
of the transition between the $\chi$SB and CSC phases was expected
from \cite{mandal2009}, where the type of phase transition between
these two phases is studied for a fixed $\tilde{e}B=0.05$ GeV$^{2}$
and various $G_{D}/G_{S}$ ratios. Our results confirm the findings
in \cite{mandal2009} for a wide range of $\tilde{e}B\in\{0,0.7\}$
GeV$^{2}$ and fixed value of $G_{D}/G_{S}=0.75$. Section VI is
devoted to a summary of our results and concluding remarks.
\section{Two flavor 2SC model at $T=0$, and $\mu,B\neq 0$}
\noindent Let us start with the fermionic part of the extended
Lagrangian density of a gauged NJL model\footnote{The gauge kinetic
term will be added to this Lagrangian in the last step.}
\begin{eqnarray}\label{F1}
\lefteqn{{\cal{L}}_{f}=\overline{\psi}(x)[i\gamma^{\mu}(\partial_{\mu}-ieQ
A_{\mu}-igT^{8}G^{8}_{\mu})-m_{0}+\hat{\mu}\gamma^{0}]\psi(x)}\nonumber\\
&&+G_{S}[(\overline{\psi}(x)\psi(x))^2+(\overline{\psi}(x)i\gamma_{5}\vec{\tau}\psi(x))^2]+G_{D}[(i\overline{\psi}^C(x)\epsilon_{f}\epsilon_{c}\gamma_{5}\psi(x))(i\overline{\psi}(x)
\varepsilon_{f}\epsilon_{c}^{3}\gamma_{5}\psi^C(x))].
\end{eqnarray}
Here, $\psi^C=C\overline{\psi}^T$ and $\overline{\psi}^C=\psi^TC$
are charge-conjugate spinors, and $C=i\gamma^2\gamma^0$ is
charge-conjugation matrix, $\vec{\tau}=(\tau_{1},\tau_{2},\tau_{3})$
are Pauli matrices. Moreover,
$(\epsilon^{3}_c)^{ab}\equiv(\epsilon_c)^{ab3}$ and
$(\varepsilon_{f})_{ij}$ are antisymmetric matrices in color and
flavor spaces, respectively. For a theory with two quark flavors,
$i,j=(1,2)=(u,d)$, and three color degrees of freedom
$a,b=(1,2,3)=(r,g,b)$. We assume that both quarks have the same
(bare) mass $m_u=m_d\equiv m_0$.\footnote{In Sec. IV and V, the bare
mass, $m_{0}$, will be chosen to be zero.} Further, $\hat{\mu}$ is
defined by $\hat{\mu}\equiv\mu+\sqrt3\mu_{8}\lambda_{8}$, where
$\mu$ is the quark chemical potential and is responsible for the
nonzero baryonic density of quark matter, and ${\mu}_8$ is inserted
by hand to impose the color neutrality after the process of
dynamical color symmetry breaking. Here,
$T^{8}=\frac{\lambda_{8}}{2}$ with
$\lambda_{8}=\frac{1}{\sqrt3}\mbox{diag}(1,1,-2)$ the
$8^{\mbox{\small{th}}}$ Gell-Mann $\lambda$-matrix. The scalar and
diquark couplings are denoted by $G_S$ and $G_D$, respectively.
Furthermore, $Q=Q_{f}\otimes {\mathbf{1}}_{c}$ with
$Q_{f}\equiv\mbox{diag}\left(2/3,-1/3\right)$ is the fermionic
charge matrix coupled to $U(1)$ gauge field $A_{\mu}$. The same
setup without the coupling to $A_{\mu}$ and $G_{\mu}^{8}$ is also
used in \cite{ebert2005}. To determine the effective action of the
above model, we introduce first the bosonized Lagrangian density
\begin{eqnarray}\label{F2}
{\cal{L}}_{f}&=&\overline{\psi}(x)[i\gamma^{\mu}(\partial_{\mu}-ieQA_{\mu}-igT^{8}G^{8}_{\mu})
+\hat{\mu}\gamma^{0}]\psi(x)-
\overline{\psi}(x)(m+i\gamma^5\vec{\tau}\cdot\vec{\pi})\psi(x)\nonumber\\
&&-\frac{1}{2}\Delta^{*3}(i\overline{\psi}^C(x)\varepsilon_{f}\epsilon_{c}^{3}\gamma_{5}\psi(x))-\frac{1}{2}\Delta^{3}(i\overline{\psi}(x)\varepsilon_{f}\epsilon_{c}^{3}\gamma_{5}\psi^C(x))-
\frac{\sigma^{2}+\vec{\pi}^{2}}{4G_{S}}-\frac{\Delta^{3}\Delta^{*3}}{4G_{D}},
\end{eqnarray}
with $m\equiv m_{0}+\sigma$, that includes the auxiliary mesonic
fields
\begin{eqnarray}\label{F3}
\sigma=-2 G_{S}(\overline{\psi}\psi), \qquad\qquad\vec{\pi}=-2
G_{S}(\overline{\psi}i\gamma^5\vec{\tau}\psi),
\end{eqnarray}
and diquarks
\begin{eqnarray}\label{F4}
\Delta^{3}=-2
G_{D}(i\overline{\psi}^C\varepsilon_{f}\epsilon_{c}^{3}\gamma_{5}\psi),\qquad\qquad\Delta^{*3}=-2
G_{D}(i\overline{\psi}\varepsilon_{f}\epsilon_{c}^{3}\gamma_{5}\psi^C).
\end{eqnarray}
From now on, we will skip the supperscript ``3'' for $\Delta$ and
$\Delta^{*}$. Using an appropriate mean field approximation, the
effective potential of this model can be determined as a function of
the condensates $\langle \sigma\rangle$, $\langle
\vec{\pi}\rangle$,$\langle \Delta\rangle$ and $\langle
\Delta^{*}\rangle$. For simplicity we set $\langle
\vec{\pi}\rangle=\vec{0}$. It is the purpose of this paper to study
the effect of a constant background $U(1)$ magnetic field on the
formation of these condensates. To do this, we have, principally, to
replace $A_{\mu}$ by a classical $A_{\mu}^{ext}$ and a dynamical
part $a_{\mu}$ and then integrate out the dynamical gauge field
$a_{\mu}$ and $G_{\mu}^{8}$. However, it turns out that for
non-vanishing $(\Delta,\Delta^{*})$, both gauge fields $A_{\mu}$ and
$G_{\mu}^{8}$ are massive and underlie the Meissner
effect.\footnote{As it turns out $\sigma$ is invariant under
$U_{V}(1)$ and $SU_{V}(3)$ groups. Thus
$Q\langle\sigma\rangle=T^{8}\langle\sigma\rangle=0$, whereas
$Q\langle\Delta\rangle\neq 0$ as well as
$T^{8}\langle\Delta\rangle\neq 0$.} They are therefore unappropriate
to be taken as external fields. But, as it is shown in
\cite{rajagopal1999, gorbar2000}, there is indeed a linear
combination of $A_{\mu}$ and $G_{\mu}^{8}$, that leads to a massless
``rotated'' $U(1)$ field,
$\tilde{A}_{\mu}=A_{\mu}\cos\theta-G_{\mu}^{8}\sin\theta$, and a
massive ``rotated'' $SU(3)$ field,
$\tilde{G}_{\mu}^{8}=A_{\mu}\sin\theta+G_{\mu}^{8}\cos\theta$.
According to \cite{gorbar2000}, the angle $\theta$ can be determined
from
\begin{eqnarray}\label{F11}
\cos\theta\equiv -\frac{\sqrt3g}{\sqrt{3g^2+e^2}},
\qquad\mbox{and}\qquad \sin\theta\equiv -\frac{e}{\sqrt{3g^2+e^2}}.
\end{eqnarray}
To rotate the fields, one uses the identity
\begin{eqnarray}\label{F7}
eQA_{\mu}+gT^{8}G^{8}_{\mu}\equiv
\tilde{e}\tilde{Q}\tilde{A}_{\mu}+\tilde{g}\tilde{T}\tilde{G}^{8}_{\mu},
\end{eqnarray}
and insert the combination ${\cal{O}}{\cal{O}}^{T}=1$ on the right
hand side (r.h.s.) of this identity. Here, ${\cal{O}}$ is an
appropriate rotation matrix including sine and cosine of $\theta$.
The identity (\ref{F7}) not only determines the new rotated fields
as a linear combination of the original non-rotated ones, it also
fixes the relation between the rotated and non-rotated couplings as
$\tilde{e}\tilde{Q}=eQ\cos\theta-gT^{8}\sin\theta$, as well as
$\tilde{g}\tilde{T}=eQ \sin\theta+gT^{8}\cos\theta$. In the rotated
system, one chooses $\tilde{Q}$ so that
$\tilde{Q}\langle\Delta\rangle=0$. This leads to
\begin{eqnarray}\label{F9}
\tilde{Q}=Q_{f}\otimes
{\mathbf{1}}_{c}-{\mathbf{1}}_{f}\otimes(\frac{T^{8}}{\sqrt3})_{c}.
\end{eqnarray}
The above relations between the rotated and non-rotated generators,
$(Q,T^{8})$ and $(\tilde{Q},\tilde{T})$, lead then to
$\tilde{T}\langle\Delta\rangle\neq 0$, which then yields a
non-vanishing mass for $\tilde{G}_{\mu}^{8}$. Hence, as long as the
diquark condensate $\Delta$ is non-vanishing, the rotated
$\tilde{G}_{\mu}^{8}$ is massive because of
$\tilde{T}\langle\Delta\rangle\neq 0$. In this case, the rotated
system is the true physical system. Once $\Delta=0$ and $\sigma\neq
0$, the rotated and non-rotated systems are equivalent, because the
identity
$\tilde{Q}\langle\sigma\rangle=\tilde{T}\langle\sigma\rangle=0$
holds automatically [see footnote 8]. Using (\ref{F9}) and the above
relation $\tilde{e}\tilde{Q}=eQ\cos\theta-gT^{8}\sin\theta$ between
the rotated $\tilde{e}\tilde{Q}$ and the non-rotated $eQ$, it turns
out that $\tilde{e}\equiv e\cos\theta$, as in the electroweak
Standard Model.\footnote{In a system including mesons and diquarks,
only diquarks play the role of electroweak Higgs field.} In the six
dimensional flavor-color representation,
$(u_r,u_g,u_b,d_r,d_g,d_b)$, the rotated $\tilde{Q}$ charges of
different quarks, in units of $\tilde{e}$, are presented in Table I.
\begin{table}[h] \centering
\begin{tabular}{c|c c c c c c}
    \hline\hline
    quarks & $u_{r}$ & $u_{g}$ & $u_{b}$ & $d_{r}$ & $d_{g}$ & $d_{b}$ \\
    \hline
    $\tilde{q}$ & $+\frac{1}{2}$ & $+\frac{1}{2}$ & 1 & $-\frac{1}{2}$ & $-\frac{1}{2}$ & 0 \\
    \hline\hline
   \end{tabular}
\caption{$\tilde{Q}$ charges of quarks in 2SC model in the presence
of rotated magnetic field $\tilde{\mathbf{B}}$ in units of
$\tilde{e}$.}\label{table1}
\end{table}
\par\noindent
Plugging (\ref{F7}) in (\ref{F2}), the resulting transformed
Lagrangian density is then given by (\ref{F2}) with
$eQA_{\mu}+gT^{8}G^{8}_{\mu}$ replaced by
$\tilde{e}\tilde{Q}\tilde{A}_{\mu}+\tilde{g}\tilde{T}\tilde{G}^{8}_{\mu}$,
and $\vec{\pi}=\vec{0}$ [see (\ref{F7})], and reads
\begin{eqnarray}\label{F14}
{\cal{L}}_{f}&=&\overline{\psi}(x)[i\gamma^{\mu}(\partial_{\mu}-i\tilde{e}\tilde{Q}\tilde{A}_{\mu}-i\tilde{g}\tilde{T}\tilde{G}_{\mu}^{8})+
\hat{\mu}\gamma^{0}]\psi(x)- m\overline{\psi}(x)\psi(x)\nonumber
\\&&-\frac{1}{2}\Delta^{*}(i\overline{\psi}^C(x)\varepsilon_{f}\epsilon_{c}^{3}\gamma_{5}\psi(x))-\frac{1}{2}\Delta(i\overline{\psi}(x)\varepsilon_{f}\epsilon_{c}^{3}\gamma_{5}\psi^C(x))
-\frac{\sigma^{2}}{4G_{S}}-\frac{|\Delta|^2}{4G_{D}},
\end{eqnarray}
with $|\Delta|^{2}\equiv \Delta\Delta^*$. To introduce the external
rotated $U(1)$ magnetic field in the third direction, we replace
$\tilde{A}_{\mu}\to \tilde{A}_{\mu}^{ext}+\tilde{a}_{\mu}$, with the
external rotated electromagnetic field $\tilde{A}_{\mu}^{ext}$ in
the Landau gauge $\tilde{A}_{\mu}^{ext}=(0,0,Bx,0)$, and integrate
out the remaining dynamical rotated fields  $\tilde{a}_{\mu}$ and
$\tilde{G}_{\mu}^{8}$. We arrive therefore at the full modified
bosonized Lagrangian $\tilde{\cal{L}}=
\tilde{\cal{L}}_{k}+\tilde{\cal{L}}_{f}$, with\footnote{Comparing to
(\ref{F14}), in (\ref{F15}), we have added the kinetic term of the
rotated $U(1)$ external gauge field
$-\frac{1}{4}(F_{\mu\nu})^{2}|_{\tilde{A}_{\mbox{\tiny{ext}}}}=-\frac{B^{2}}{2}$.}
\begin{eqnarray}\label{F15}
\tilde{\cal{L}}_{k}\equiv-\left(
\frac{\sigma^{2}}{4G_{S}}+\frac{|\Delta|^2}{4G_{D}}+\frac{B^{2}}{2}\right),
\end{eqnarray}
and
\begin{eqnarray}\label{F16}
\tilde{\cal{L}}_{f}&=&
\overline{\psi}(x)[i\gamma^{\mu}(\partial_{\mu}-i\tilde{e}\tilde{Q}\tilde{A}_{\mu}^{ext})+\hat{\mu}\gamma^{0}]\psi(x)-
m\overline{\psi}(x)\psi(x)\nonumber\\
&&-\frac{1}{2}\Delta^{*}(i\overline{\psi}^C(x)\varepsilon_f\epsilon^{3}_c\gamma_{5}\psi(x))-\frac{1}{2}\Delta(i\overline{\psi}(x)
\varepsilon_f\epsilon^{3}_c\gamma_{5}\psi^C(x)),
\end{eqnarray}
in a constant (rotated) background $U(1)$ magnetic field
$\tilde{\mathbf{B}}=B{\mathbf{e}}_{3}$. In what follows, we will
simplify (\ref{F16}) using the method presented in \cite{ferrer2006}
and arrive at an equivalent Lagrangian, which will then be used in
Sec. III to determine the effective potential of the above model in
the presence of a rotated background $U(1)$ magnetic field
$\tilde{\mathbf{B}}$. To do this, we introduce the rotated-charge
projectors $\Omega_{\tilde{q}}$, that satisfy the eigenvalue
equation $\tilde{Q}\Omega_{\tilde{q}}=\tilde{q}\Omega_{\tilde{q}}$.
They are given by
\begin{eqnarray}\label{F17}
\begin{array}{lcrclcr}
\Omega_{0}&=&\mbox{diag}(0,0,0,0,0,1),&\qquad&\Omega_{+\frac{1}{2}}&=&\mbox{diag}(1,1,0,0,0,0)\nonumber\\
\Omega_{1}&=&\mbox{diag}(0,0,1,0,0,0),&\qquad&\Omega_{-\frac{1}{2}}&=&\mbox{diag}(0,0,0,1,1,0),\nonumber\\
\end{array}
\end{eqnarray}
and satisfy
\begin{eqnarray}\label{F18} 
\sum_{\tilde{q}\in\{0,1,\pm
\frac{1}{2}\}}\Omega_{\tilde{q}}=1,\qquad\mbox{and}\qquad
\Omega_{\tilde{q}}\Omega_{{\tilde{q}}^{\prime}}=\delta_{\tilde{q}{\tilde{q}}^{\prime}}.
\end{eqnarray}
Using the definition $\psi_{\tilde{q}}(x)\equiv
\Omega_{\tilde{q}}\psi(x)$, the fermion field in the six dimensional
color-flavor representation can now given by
\begin{eqnarray}\label{F19}
\psi=\sum_{\tilde{q}\in\{0,1,\pm\frac{1}{2}\}}\psi_{\tilde{q}}.
\end{eqnarray}
Introducing, at this stage, the Nambu-Gorkov bispinor wave function
$$\Psi_{\tilde{q}}=\left(\begin{array}{c}
\psi_{\tilde{q}}\\
\psi^{C}_{-\tilde{q}}
\end{array}\right),$$
the part of the Lagrangian which is bilinear in $\psi$, i.e.
$\tilde{\cal{L}}_{f}$ from (\ref{F16}), can be brought in the
following form:
\begin{eqnarray}\label{F20} 
\tilde{\cal{L}}_{f}=\frac{1}{2}\sum_{\tilde
q\in\{0,1,\pm\frac{1}{2}\}}\overline{\Psi}_{\tilde
q}(x){\cal{S}}_{\tilde q}{\Psi}_{\tilde q}(x),
\end{eqnarray}
where ${\cal{S}}_{\tilde{q}}$ for $\tilde{q}\in\{0,1\}$ is given by
\begin{eqnarray}\label{F21}
{\cal{S}}_{\tilde{q}\in\{0,1\}}\equiv\left(
\begin{array}{cc}
[G^{+}_{(\tilde{q})}]^{-1}   & 0\;\;\;\;\;\; \\ 0\;\;\;\;\;\;\;\; & [G^{-}_{(\tilde{q})}]^{-1} \\
\end{array}
\right),
\end{eqnarray}
and for $\tilde{q}\in\{-\frac{1}{2},+\frac{1}{2}\}$ by
\begin{eqnarray}\label{F22}
{\cal{S}}_{\tilde{q}\in\{-\frac{1}{2},+\frac{1}{2}\}}\equiv\left(
\begin{array}{cc}
[G^{+}_{(\tilde{q})}]^{-1} & -\kappa\Omega_{-\tilde{q}}\;\;  \\ -\kappa^{\prime}\Omega_{\tilde{q}} & [G^{-}_{(\tilde{q})}]^{-1} \\
\end{array}
\right).
\end{eqnarray}
Here, $ [G^{\pm}_{(\tilde{q})}]^{-1}\equiv
[\gamma^{\mu}(i\partial_{\mu}+\tilde{e}\tilde{q}\tilde{A}_{\mu}\pm\hat{\mu}\delta_{\mu0})-m],
$ and $ \kappa^{ij,ab}_{\alpha\beta}\equiv
i\Delta\tau_{2}^{ij}\lambda_{2}^{ab}\gamma^{5}_{\alpha\beta}$ as
well as
$\kappa^{\prime}\equiv\gamma_{0}\kappa^{\dagger}\gamma_{0}=i\Delta^{*}\tau_{2}\lambda_{2}\gamma^{5}$.
They can be read from (\ref{F16}) and the relations (\ref{F18}) as
well as the definition of $\psi_{\tilde{q}}=\Omega_{\tilde{q}}\psi$.
Note that (\ref{F20}) can be equivalently expressed as
\begin{eqnarray}\label{F25} 
\tilde{\cal{L}}_{f}&=&\frac{1}{2}\sum_{\tilde
{q}\in\{0,1,\pm\frac{1}{2}\}}\bigg\{\overline{\psi}^{C}_{\tilde{q}}(x)[G^{-}_{(-\tilde{q})}]^{-1}\psi^{C}_{\tilde{q}}(x)+\overline{\psi}_{\tilde{q}}(x)[G^{+}_{(\tilde{q})}]^{-1}
\psi_{\tilde{q}}(x)
-\overline{\psi}_{\tilde{q}}^C(x)\tilde{\kappa}^{\prime}_{\tilde{q}}\psi_{\tilde{q}}(x)-\overline{\psi}_{\tilde{q}}(x)
\tilde{\kappa}_{\tilde{q}}\psi_{\tilde{q}}^C(x)\bigg\},
\nonumber\\
\end{eqnarray}
where,  $\tilde{\kappa}_{\tilde{q}}$ is defined by
$\tilde{\kappa}_{\tilde{q}}\equiv\Omega_{\tilde{q}'}\kappa\Omega_{\tilde{q}}$.
In (\ref{F25}), $\tilde{\kappa}_{\tilde{q}}$ is non-vanishing only
for $\tilde{q}'+\tilde{q}=0$ with $\tilde{q}'\neq \tilde{q}$. For
$\tilde{q}\in\{-\frac{1}{2}, +\frac{1}{2}\}$ we have therefore
\begin{eqnarray}\label{F26}
(\tilde{\kappa}_{\tilde{q}=-\frac{1}{2}})_{\rho\sigma}=
(\Omega_{\frac{1}{2}}\kappa\Omega_{-\frac{1}{2}})_{\rho\sigma}&=&(\kappa\Omega_{-\frac{1}{2}})_{\rho\sigma}=\left
\lbrace
  \begin{array}{ccrcl}
    +i\Delta\gamma^{5} & \mbox{if}&(\rho,\sigma)&=&(2,4),\\
    -i\Delta\gamma^{5} & \mbox{if}&(\rho,\sigma)&=&(1,5),
    \\ 0 &\mbox{otherwise},&&&
  \end{array} \right.
  \\
  (\tilde{\kappa}_{\tilde{q}=+\frac{1}{2}})_{\rho\sigma}=(\Omega_{-\frac{1}{2}}\kappa\Omega_{\frac{1}{2}})_{\rho\sigma}&=&(\kappa\Omega_{+\frac{1}{2}})_{\rho\sigma}=\left \lbrace
  \begin{array}{ccrcl}
    +i\Delta\gamma^{5} & \mbox{if}&(\rho,\sigma)&=&(4,2),\\
    -i\Delta\gamma^{5} & \mbox{if}&(\rho,\sigma)&=&(5,1),
   \\ 0 & \mbox{otherwise},
  \end{array} \right.
\end{eqnarray}
whereas for $\tilde{q}\in\{0,1\}$, we have
\begin{eqnarray}
\tilde{\kappa}_{ij}=(\Omega_{i}\kappa\Omega_{j})_{\rho\sigma}&=&0,\qquad\mbox{for}\qquad
i,j=0,1\qquad\mbox{or}\qquad i=j=\pm\frac{1}{2}.
\end{eqnarray}
This is in contrast to the case of three-flavor color-flavor locked
(CFL) phase, studied in \cite{ferrer2006}. In that case, there
exists a charge $\tilde{q}=-1$ and the combination of $(\Omega_{\mp
1}\kappa\Omega_{\pm 1})$ leads also to nonzero result.
\section{One-loop effective action and thermodynamic potential}
\setcounter{equation}{0}\noindent In what follows, the one-loop
effective action of the theory, $\Gamma$, will be determined in the
mean field approximation in terms of $\sigma\equiv
\langle\sigma(x)\rangle, \Delta\equiv \langle\Delta(x)\rangle$, and
$\Delta^*\equiv \langle\Delta^{*}(x)\rangle$. Using the following
path integral over the quark fields
\begin{eqnarray}\label{A1}
e^{i\Gamma[\sigma, \Delta, \Delta^{*}]}=\int
{\cal{D}}\psi{\cal{D}}\bar{\psi}\exp\left(i\int d^{4}x~
\tilde{\cal{L}}\right),
\end{eqnarray}
where, $\tilde{\cal{L}}\equiv
\tilde{\cal{L}}_{k}+\tilde{\cal{L}}_{f}$, with $\tilde{\cal{L}}_{k}$
and $\tilde{\cal{L}}_f$ from (\ref{F15}) and (\ref{F20}), the
effective action up to one-loop quantum corrections is given by
\begin{eqnarray}\label{A2}
\Gamma[\sigma, \Delta,
\Delta^{*}]=-\left(\frac{\sigma^{2}}{4G_{S}}+\frac{|\Delta|^2}{4G_D}+\frac{B^2}{2}\right){\cal{V}}+\Gamma_{\mbox{\tiny{eff}}}^{(1)}[\sigma,
\Delta, \Delta^{*}].
\end{eqnarray}
Here, ${\cal{V}}$ is the 4-dimensional space-time volume, and
$\Gamma_{\mbox{\tiny{eff}}}^{(1)}$  is the one-loop contribution to
the effective potential. It arises by integrating out the fermion
fields and reads
\begin{eqnarray}\label{A3}
\Gamma^{(1)}_{\mbox{\tiny{eff}}}=-\frac{i}{2}\sum_{\tilde
q}{\mbox{Tr}}_{\{NGcfsx\}}~\ln[{{\cal{S}}_{\tilde q}}^{-1}],
\end{eqnarray}
where ${\cal{S}}_{\tilde{q}}$ is defined in (\ref{F21}) and
(\ref{F22}). Here, the trace ``Tr'' operation in (\ref{A3}) includes
apart from a two-dimensional trace in the Nambu-Gorkov (NG) space, a
trace over the whole phase space. It is therefore defined by a trace
over the color ($c$), flavor ($f$), and spinor ($s$) degrees of
freedom, as well as over a four-dimensional space-time coordinate
($x$) \cite{ebert2005}. To compute (\ref{A3}), we have to notice
that, according to Table \ref{table1}, the blue quarks $(u_b, d_b)$
have $\tilde{q}=0,1$, whereas the green and red quarks $(u_r, u_g,
d_r,d_g)$ have $\tilde{q}=\pm \frac{1}{2}$. Thus relation (\ref{A3})
reduces to
\begin{eqnarray}\label{A4}
\Gamma^{(1)}_{\mbox{\tiny{eff}}}=\sum\limits_{\kappa\in\{r,g,b\}}\Gamma^{(1)/\kappa}_{\mbox{\tiny{eff}}},
\end{eqnarray}
where the one-loop effective action of the blue (b) and red/green
(r/g) are given by
\begin{eqnarray}\label{A5}
\Gamma^{(1)/\mbox{\tiny{b}}}_{\mbox{\tiny{eff}}}&=&-
\frac{i}{2}\mbox{Tr}_{\{NGcfsx\}}\ln[{{\cal{S}}_{0}}^{-1}]
-\frac{i}{2}\mbox{Tr}_{\{NGcfsx\}}\ln[{{\cal{S}}_{+1}}^{-1}],\nonumber\\
\sum\limits_{\kappa\in\{r,g\}}\Gamma^{(1)/\kappa}_{\mbox{\tiny{eff}}}&=&-
\frac{i}{2}\mbox{Tr}_{\{NGcfsx\}}\ln[{{\cal{S}}_{+\frac{1}{2}}}^{-1}]
-
\frac{i}{2}\mbox{Tr}_{\{NGcfsx\}}\ln[{{\cal{S}}_{-\frac{1}{2}}}^{-1}].
\end{eqnarray}
To perform the trace operation in the NG space, we use
\begin{eqnarray}\label{A6}
\det\left( \begin{array}{rr}
   A  & B \\ C & D \\ \end{array}
   \right)=\det\left(-BC+BDB^{-1}A\right)=\det\left(-CB+CAC^{-1}D\right).
\end{eqnarray}
Using further $\mbox{tr}\ln A=\ln \det A$, we arrive at
\begin{eqnarray}\label{A7}
{\det}_{\{NGcfsx\}}[{{\cal{S}}_{0}}^{-1}]&=&{\det}_{\{scx\}}\big[\{\gamma^{\alpha}(i\partial_{\alpha}+\breve{\mu}\delta_{\alpha
0})-m\}\{\gamma^{\alpha}(i\partial_{\alpha}-
\breve{\mu}\delta_{\alpha
0})-m\}\big],\nonumber\\
{\det}_{\{NGcfsx\}}[{{\cal{S}}_{+1}}^{-1}]&=&{\det}_{\{scx\}}\big[\{\gamma^{\alpha}(i\partial_{\alpha}+\tilde{e}\tilde{A}_{\alpha}+\breve{\mu}\delta_{\alpha
0})-m\}\{\gamma^{\alpha}(i\partial_{\alpha}+\tilde{e}\tilde{A}_{\alpha}-\breve{\mu}\delta_{\alpha0})-m\}\big],\nonumber\\
{\det}_{\{NGcfsx\}}[{{\cal{S}}_{\pm\frac{
1}{2}}}^{-1}]&=&{\det}_{\{scx\}}\big[|\Delta|^{2}+\{-\gamma^{\alpha}(i\partial_{\alpha}\pm\frac{1}{2}\tilde{e}\tilde{A}_{\alpha}+\bar{\mu}\delta_{\alpha
0})-m\}\nonumber\\
&&\qquad\qquad\times
\{\gamma^{\alpha}(i\partial_{\alpha}\pm\frac{1}{2}\tilde{e}\tilde{A}_{\alpha}-\bar{\mu}\delta_{\alpha
0})-m\}\big],
\end{eqnarray}
where we have skipped the superscript ``ext'' on the external
rotated gauge field $\tilde{A}_{\mu}$. Here, $\breve{\mu}\equiv
\mu-2\mu_8$, $\bar{\mu}\equiv \mu+\mu_{8}$ and $m\equiv m_0+\sigma$.
The determinants in (\ref{A7}) are now to be calculated in the
momentum space. To do this, a generalization of the method described
in \cite{ferrer2006} for arbitrary charges is necessary. This method
is originally developed by Ritus in \cite{ritus1972} in order to
determine the Green's function of charged fermions in the presence
of background magnetic field. It is then extended to charged vector
fields in \cite{elizalde2004}. Recently, it is used in
\cite{fukushima2009} to determine the electric-current
susceptibility of quark matter in the presence of external constant
magnetic field. As it is described in \cite{fukushima2009}, in the
Landau gauge for the external rotated gauge field, a projection
operator $P_{n}$ can be defined
\begin{eqnarray}\label{A8}
\begin{array}{rclccc}
P_{n}&\equiv&
\frac{1}{2}[f_{n_{+}}(x)+f_{n_{-}}(x)]+\frac{i}{2}[f_{n_{+}}(x)-f_{n_{-}}(x)]\gamma^{1}\gamma^{2},&\qquad&\mbox{for}&{q}B>0,\\
P_{n}&\equiv&
\frac{1}{2}[f_{n_{+}}(x)+f_{n_{-}}(x)]-\frac{i}{2}[f_{n_{+}}(x)-f_{n_{-}}(x)]\gamma^{1}\gamma^{2},&\qquad&\mbox{for}&{q}B<0,\\
\end{array}
\end{eqnarray}
that includes the basis functions $f_{n_{\pm}}(x)$ defined by
\begin{eqnarray}\label{A9}
\begin{array}{rclcrcl}
f_{n_{+}}(x)&=&\phi_{n}(x-\frac{p_{y}}{{q}B}),&\qquad&
n&=&0,1,2,\cdots,\\
f_{n_{-}}(x)&=&\phi_{n-1}(x-\frac{p_{y}}{{q}B}),&\qquad&
n&=&1,2,3\cdots.
\end{array}
\end{eqnarray}
Here, $\phi_{n}(x)$ are the standard Landau quantized wave functions
\cite{fukushima2009}
\begin{eqnarray}\label{A10}
\phi_{n}(x)=\sqrt{\frac{1}{2^n
n!}}\left(\frac{|{q}B|}{\pi}\right)^{1/4}\exp\left(-\frac{1}{2}|{q}B|x^2\right)H_{n}\left(\sqrt{|{q}B|}x\right),
\end{eqnarray}
with $H_{n}(x)$ the Hermite polynomial of degree $n$. Using the
projectors $P_{n}$ from (\ref{A8}), it is easy to show that
\begin{eqnarray}\label{A11}
\gamma_{\mu}\left(i\partial^{\mu}+{q}A^{\mu}\right)P_{n}e^{-i(p_0
t-p_y y-p_z
z)}=P_{n}\left(p_0\gamma^0-\mbox{sgn}({q}B)\sqrt{2|{q}B|n}\gamma^{2}-p_z\gamma^{3}\right)e^{-i(p_0
t-p_y y-p_z z)}.\nonumber\\
\end{eqnarray}
The r.h.s. of (\ref{A11}) is a free Dirac operator with a modified
momentum
\begin{eqnarray}\label{A12}
\bar{p}^{\mu}=(p_{0},0,\mbox{sgn}(qB)\sqrt{2|q B| n},p_{3}).
\end{eqnarray}
This shows also that the solution of the Dirac equation in the
presence of a constant magnetic field can be given by a combination
of the projection operators $P_{n}$ and the ordinary free Dirac
spinors $u(p,s)$ and $v(p,s)$ \cite{fukushima2009}.
\par
To compute the determinants in (\ref{A7}) in the momentum space, we
will use, for the charges $\tilde{q}\neq 0$, an appropriate momentum
basis, similar to (\ref{A12}), and for $\tilde{q}=0$, the ordinary
four-momentum $p^{\mu}$. In other words, we have
\begin{eqnarray}\label{A13}
\begin{array}{rclccrcl}
{\bar{p}}_{\tilde q\neq 0}^{\mu}&=&(p_{0},0,\frac{\tilde q}{|\tilde
q|}\sqrt{2|\tilde q\tilde{e} B| n},p_{3}),&\qquad&\mbox{for}&{\tilde{q}}&=&1,\pm\frac{1}{2},\\
{\bar{p}}_{\tilde{q}=0}^{\mu}&=&(p_{0},p_{1},p_{2},p_{3}),&\qquad&\mbox{for}&\tilde{q}&=&0,
\end{array}
\end{eqnarray}
where $\frac{\tilde{q}}{|\tilde{q}|}$ replaces
$\mbox{sgn}(\tilde{q}B)$ in (\ref{A12}). This leads to the
well-known quasiparticle dispersion relations in the presence of a
constant magnetic field aligned in the third direction
\cite{shovkovy2007},
\begin{eqnarray}\label{A14}
\begin{array}{rclccrcl}
E_{\tilde{q}}&=&\sqrt{2
|\tilde{q}\tilde{e}B|n+p_{3}^{2}+m^{2}}&\qquad&\mbox{for}&{\tilde{q}}&=&1,\pm\frac{1}{2},\\
E_{0}&=&\sqrt{p_{1}^{2}+p_{2}^{2}+p_{3}^{2}+m^{2}}&\qquad&\mbox{for}&{\tilde{q}}&=&0.\\
\end{array}
\end{eqnarray}
Using the momenta (\ref{A13}) and transforming (\ref{A4}) and
(\ref{A5}) into the Fourier space, the one-loop effective action
reads
\begin{eqnarray}\label{A15}
\tilde{\Gamma}^{(1)}_{\mbox{\tiny{eff}}}(\bar{p})=\sum\limits_{\kappa\in\{r,g,b\}}\tilde{\Gamma}^{(1)/\kappa}_{\mbox{\tiny{eff}}}(\bar{p}),
\end{eqnarray}
with
\begin{eqnarray}\label{A16}
\tilde{\Gamma}^{(1)/\mbox{\tiny{b}}}_{\mbox{\tiny{eff}}}(\bar{p})&=&-i\sum\limits_{\tilde{q}\in\{0,1\}}\ln{\det}_{x}[\{({E}_{\tilde{q}}
+\breve{\mu})^2-p_{0}^2\}\{({E}_{\tilde{q}}-\breve{\mu})^2-p_{0}^2\}],
\nonumber\\
\sum\limits_{\kappa\in\{r,g\}}\tilde{\Gamma}^{(1)/\kappa}_{\mbox{\tiny{eff}}}(\bar{p})&=&-2i\sum\limits_{\tilde{q}
\in\{+\frac{1}{2},-\frac{1}{2}\}}
\ln{\det}_{x}[({{E}^{(+)}_{\tilde{q}}}^{2}-{p}_{0}^2)(
{{E}^{(-)}_{\tilde{q}}}^{2}-{p}_{0}^2)].
\end{eqnarray}
Here, $E_{\tilde{q}}$ for $\tilde{q}\in\{0,1,\pm \frac{1}{2}\}$ are
defined in (\ref{A14}), and $
{E}_{\tilde{q}}^{(\pm)}\equiv\sqrt{{(E_{\tilde{q}}\pm\bar{\mu})}^2+{|\Delta|}^2}$,
for $\tilde{q}\in\{+\frac{1}{2}, -\frac{1}{2}\}$.  The factor 2 in
the last equation of (\ref{A16}) reflects the degeneracy in the
quark charges for $u_{r/g}$ as well as $d_{r/g}$ (see Table
\ref{table1}). Note that a trace over Landau levels, $n$, is
implemented in the expression on the r.h.s. of (\ref{A16}). This is
because $E_{\tilde{q}\neq 0}$ from (\ref{A14}) depends explicitly on
$n$. This trace will be performed in the next step, where the
one-loop effective action will be explicitly determined in the
momentum space. Performing the remaining determinant in the
coordinate space leads, for a constant background magnetic field, to
a space-time volume ${\cal{V}}$. At this stage, we will introduce
the effective thermodynamic (mean field) potential
$\Omega_{\mbox{\tiny{eff}}}$, that is defined by the effective
action through the relation $\Gamma_{\mbox{\tiny{eff}}}=-{\cal{V}}
\Omega_{\mbox{\tiny{eff}}}$. To determine the one-loop contribution
to the one-loop effective potential at zero temperature
$\Omega_{\mbox{\tiny{eff}}}^{(1)}$, it is convenient to determine it
first at finite temperature, and then taking the limit $T\to 0$,
consider only the zero temperature effects \cite{shovkovy2007}. For
quarks with $\tilde{q}=0$, one replaces $p_0$ by
$p_{0}=i\omega_{\ell}$,\footnote{The effect of the chemical
potential is already considered in
$\Gamma_{\mbox{\tiny{eff}}}^{(1)}$ as well as
$\Omega_{\mbox{\tiny{eff}}}^{(1)}$.} where $\omega_{\ell}$ are the
Matsubara frequency defined by $\omega_{\ell}\equiv(2\ell+1)\pi T$,
and the $p_{0}$ integration by an infinite sum over the Matsubara
frequencies. For an arbitrary function
$f(p_0,\bar{\mathbf{p}}_{\tilde{q}})$, we get therefore
\begin{eqnarray}\label{A17}
\int\frac{d^4 p}{(2\pi)^4}f(p_{0}, \bar{\mathbf{p}}_{\tilde{q}=0})=
\frac{1}{\beta}\sum\limits_{\ell=-\infty}^{+\infty}\int
\frac{d^{3}p}{(2\pi)^{3}}~f(i\omega_{\ell},\mathbf{p}),
\end{eqnarray}
where $\beta\equiv T^{-1}$ is the inverse of the temperature $T$.
For the quarks with $\tilde{q}\neq 0$, apart from a summation over
the Matsubara frequencies $\ell$, a summation over the Landau levels
$n$ is also to be considered [see (\ref{A13})]. We get therefore
\cite{providenca2008}
\begin{eqnarray}\label{A18}
\int\frac{d^4 p}{(2\pi)^4}f(p_{0}, \bar{\mathbf{p}}_{\tilde{q}\neq
0})=
\frac{|\tilde{q}\tilde{e}B|}{\beta}\sum\limits_{\ell=-\infty}^{+\infty}\sum\limits_{n=0}^{+\infty}\alpha_{n}\int
_{-\infty}^{+\infty}
\frac{dp_{3}}{8\pi^{2}}~f(i\omega_{\ell},n,p_{3}),
\end{eqnarray}
where $\alpha_{n}=2-\delta_{n0}$ reflects the fact that Landau
levels with $n>0$ are doubly degenerate \cite{ferrer2006,
shovkovy2007}. Following the above recipe, the one-loop contribution
to the thermodynamic potential is given by
\begin{eqnarray}\label{A19}
\Omega_{\mbox{\tiny{eff}}}^{(1)}=\sum\limits_{\tilde{q}\in\{0,1, \pm
\frac{1}{2}\}}{\Omega}_{\mbox{\tiny{eff}}}^{\tilde{q}},
\end{eqnarray}
where for $\tilde{q}=0$, we have
\begin{eqnarray}\label{A20}
{\Omega}_{\mbox{\tiny{eff}}}^{\tilde{q}=0}&=&
i{\cal{V}}^{-1}\ln\det\big[\beta^{4}\{({E}_{0}+\breve{\mu})^2+\omega_{\ell}^2\}\{({{E}_{0}}-\breve{\mu})^2+\omega_{\ell}^2\}\big]\nonumber\\
&=&-\frac{1}{\beta}\sum^{+\infty}_{\ell=-\infty}\int\frac{d^{3}p}{(2\pi)^{3}}\ln\big[\beta^{4}\{({E}_{0}+\breve{\mu})^2+\omega_{\ell}^2\}
\{({{E}_{0}}-\breve{\mu})^2+\omega_{\ell}^2\}\big]\nonumber\\
&=&-\frac{2}{\beta}\int\frac{d^{3}p}{(2\pi)^{3}}\left\{\beta
E_{0}+\mbox{ln}\left(1+e^{-\beta(E_{0}+\breve{\mu})}\right)+\mbox{ln}\left(1+e^{-\beta(E_{0}-\breve{\mu})}\right)\right\},
\end{eqnarray}
and for $\tilde{q}=1$, we have
\begin{eqnarray}\label{A21}
{\Omega}_{\mbox{\tiny{eff}}}^{\tilde{q}=1}&=&
i{\cal{V}}^{-1}\ln\det\big[\beta^{4}\{({E}_{+1}+\breve{\mu})^2+\omega_{\ell}^2\}\{({{E}_{+1}}-\breve{\mu})^2+\omega_{\ell}^2\}\big]\nonumber\\
&=&-\frac{\tilde{e}B}{\beta}\sum^{+\infty}_{\ell=-\infty}\sum^{+\infty}_{n=0}\alpha_{n}\int_{-\infty}^{+\infty}\frac{dp_{3}}{8\pi^{2}}\ln
\big[\beta^{4}\{({E}_{+1}+\breve{\mu})^2+\omega_{\ell}^2\}\{({{E}_{+1}}-\breve{\mu})^2+\omega_{\ell}^2\}\big]\nonumber\\
&=&-\frac{2\tilde{e}B}{\beta}\sum^{+\infty}_{n=0}\alpha_{n}\int_{-\infty}^{+\infty}\frac{d
p_{3}}{8\pi^{2}}\left\{\beta
E_{+1}+\ln\left(1+e^{-\beta(E_{+1}+\breve{\mu})}\right)+\ln\left(1+e^{-\beta(E_{+1}-\breve{\mu})}\right)\right\}.
\end{eqnarray}
Note that $E_{+1}$ depends explicitly on $n$ that labels the Landau
levels. Finally, for $\tilde{q}=\pm\frac{1}{2}$, we arrive at
\begin{eqnarray}\label{A22}
\lefteqn{\hspace{0cm}\sum_{\tilde{q}\in\{+\frac{1}{2},-\frac{1}{2}\}}{\Omega}_{\mbox{\tiny{eff}}}^{\tilde{q}}=4i{\cal{V}}^{-1}\ln\det\big[\beta^{4}\{{{E}^{+}_{+\frac{1}{2}}}^{2}+\omega_{\ell}^2\}\{{{E}^{-}_{+\frac{1}{2}}}^{2}+\omega_{\ell}^2\}\big]
}\nonumber\\
&&\hspace{-0cm}=-\frac{4\tilde{e}B}{\beta}\sum^{+\infty}_{\ell=-\infty}\sum^{+\infty}_{n=0}\alpha_{n}\int_{-\infty}^{+\infty}\frac{dp_{3}}{16\pi^{2}}
\ln\big[\beta^{4}\{{{E}^{+}_{\frac{+1}{2}}}^{2}+\omega_{\ell}^2\}\{{{E}^{-}_{\frac{+1}{2}}}^{2}+\omega_{\ell}^{2}\}\big]\nonumber\\
&&\hspace{0cm}=-\frac{4\tilde{e}B}{\beta}\sum^{+\infty}_{n=0}\alpha_{n}\int_{-\infty}^{+\infty}\frac{dp_{3}}{16\pi^{2}}\left\{\beta\left(|{E}^{+}_{\frac{+1}{2}}|+|{E}^{-}_{\frac{+1}{2}}|\right)+
2\ln\left(1+e^{-\beta{E}^{+}_{\frac{+1}{2}}}\right)+2\ln\left(1+e^{-\beta{E}^{-}_{\frac{+1}{2}}}\right)\right\}.\nonumber\\
\end{eqnarray}
where ${E}^{+}_{\frac{+1}{2}}={E}^{+}_{\frac{-1}{2}}$ and
${E}^{-}_{\frac{+1}{2}}={E}^{-}_{\frac{-1}{2}}$ are used. Plugging
(\ref{A20})-(\ref{A22}) in (\ref{A19}) and taking the limit $T\to 0$
by making use of the relation \cite{ruester}
\begin{eqnarray}\label{A23}
\lim\limits_{T\rightarrow 0} T
\ln{[1+e^{-\frac{x}{T}}]}=-x\theta(-x),
\end{eqnarray}
with $\theta(x)$ is the Heaviside $\theta$-function, the temperature
independent part of the effective potential, including the tree
level and the one-loop corrections reads
\begin{eqnarray}\label{A24}
\Omega_{\mbox{\tiny{eff}}}&=&\Omega^{(0)}+\Omega_{\mbox{\tiny{eff}}}^{(1)}=
\frac{\sigma^{2}}{4G_{S}}+\frac{|\Delta|^{2}}{4G_{D}}+\frac{B^{2}}{2}\nonumber\\
&&-2\int^{+\infty}_{-\infty}\frac{d^{3}p}{(2\pi)^{3}}\big[E_{0}-(E_{0}-\breve{\mu})
\theta(\breve{\mu}-E_{0})-
(\breve{\mu}+E_{0})\theta\left(-\breve{\mu}-E_{0}\right)
\big]\nonumber\\
&&-\tilde{e}B\sum\limits_{n=0}^{+\infty}\alpha_{n}\int^{+\infty}_{-\infty}\frac{d
p_{3}}{4\pi^{2}}\big[E_{+1}+{E}^{+}_{\frac{+1}{2}}+{E}^{-}_{\frac{+1}{2}}+
(\breve{\mu}-E_{+1})\theta(\breve{\mu}-E_{+1})-2E^{+}_{+\frac{1}{2}}\theta(-E^{+}_{+\frac{1}{2}})
\nonumber\\
&&\hspace{3.5cm} -2E^{-}_{+\frac{1}{2}}\theta(-E^{-}_{+\frac{1}{2}})
-(\breve{\mu}+E_{+1}) \theta\left(-\breve{\mu}-E_{+1}\right)\big].
\end{eqnarray}
The above result (\ref{A24}) is comparable with the result in
\cite{ebert2005}, which is derived for a similar 2SC model in the
absence of the magnetic field $\tilde{\mathbf{B}}$. In this case the
thermodynamic potential up to one-loop order at finite $T$ is given
by
\begin{eqnarray}\label{A25}
\hat{\Omega}_{\mbox{\tiny{eff}}}=\frac{\sigma^{2}}{4G_{S}}+\frac{|\Delta|^{2}}{4G_{D}}+
\sum_{\kappa\in\{r,g,b\}}\hat{\Omega}_{\mbox{\tiny{eff}}}^{(1)/\kappa},
\end{eqnarray}
where for different colors, we have
\begin{eqnarray}\label{A26}
\hat{\Omega}_{\mbox{\tiny{eff}}}^{(1)/b}&=&-2i{\cal{V}}^{-1}\ln\det\big[(\breve{E}_{+}^{2}+\omega_{\ell}^{2})(\breve{E}_{-}^{2}+
\omega_{\ell}^{2})\big]\nonumber\\
&=&-\frac{2}{\beta}\sum^{+\infty}_{\ell=-\infty}\int\frac{d^{3}p}{(2\pi)^{3}}\ln\big[\beta^{4}({\breve{E}_{+}}^{2}+
{\omega_{\ell}}^2)({\breve{E}_{-}}^{2}+{\omega_{\ell}}^2)\big]\nonumber\\
&=&-\frac{4}{\beta}\int\frac{d^{3}p}{(2\pi)^{3}}\left\{\beta E
+\ln\left(1+e^{-\beta(E+\breve{\mu})}\right)+\ln\left(1+e^{-\beta(E-\breve{\mu})}\right)\right\},
\end{eqnarray}
and
\begin{eqnarray}\label{A27}
\sum_{\kappa\in\{r,g\}}\hat{\Omega}^{(1)/\kappa}_{\mbox{\tiny{eff}}}&=&-4i{\cal{V}}^{-1}\ln\det\big[({E_{+}^{\Delta}}^{2}
+{\omega_{\ell}}^2)({E_{-}^{\Delta}}^{2}+{\omega_{\ell}}^2)\big]\nonumber\\
&=&-\frac{4}{\beta}\sum^{+\infty}_{\ell=-\infty}\int\frac{d^{3}p}{(2\pi)^{3}}\ln[\beta^{4}({E_{+}^{\Delta}}^{2}
+{\omega_{\ell}}^2)({E_{-}^{\Delta}}^{2}+{\omega_{\ell}}^2)]\nonumber\\
&=&-\frac{4}{\beta}\int\frac{d^{3}p}{(2\pi)^{3}}\left\{\beta
E_{-}^{\Delta}+\beta E_{+}^{\Delta}+2\ln\left(1+e^{-\beta
E_{-}^{\Delta}}\right)+2\ln\left(1+e^{-\beta
E_{+}^{\Delta}}\right)\right\},
\end{eqnarray}
with $E\equiv\sqrt{\mathbf{p}^{2}+m^{2}}$, $\breve{E}_{\pm}\equiv
E\pm\breve{\mu}$, and $E_{\pm}^{\Delta}\equiv
\sqrt{\left(E\pm\bar{\mu}\right)^{2}+|\Delta|^{2}}$. Using
(\ref{A23}), the temperature independent part of (\ref{A25}) reads
\begin{eqnarray}\label{A28}
\hat{\Omega}_{\mbox{\tiny{eff}}}&=&\frac{\sigma^{2}}{4G_{S}}+\frac{|\Delta|^{2}}{4G_{D}}-
4\int\frac{d^{3}p}{(2\pi)^{3}}\big[E_{+}^{\Delta}
+E_{-}^{\Delta}+E+(\breve{\mu}-E)\theta(\breve{\mu}-E)-2E_{-}^{\Delta}\theta(-E_{-}^{\Delta})-2E_{+}^{\Delta}\theta(-E_{+}^{\Delta})\nonumber\\
&&\hspace{4.5cm}-(\breve{\mu}+E)\theta\left(-\breve{\mu}-E\right)\big].
\end{eqnarray}
\section{Analytical solutions of the $\chi$SB and CSC gap equations in the LLL approximation: A comparison of $B=0$ and $B\neq 0$ cases}
\setcounter{equation}{0} \noindent In the previous section, the
one-loop effective action of the NJL model including meson
($\sigma$) and diquark ($\Delta$) condensates in the 2SC phase at
finite $\tilde{e}B$, $\mu$ and $\mu_{8}$ is computed in the mean
field approximation. This is the purpose of this paper to have a
complete understanding on the effect of external magnetic field on
the formation of these condensates. To this purpose one has to solve
the following gap equations and color neutrality conditions
\begin{eqnarray}\label{YY-1a}
\frac{\partial\Omega_{\mbox{\tiny{eff}}}(\sigma,\Delta,\mu_{8};\mu,\tilde{e}B)}{\partial\sigma}\bigg|_{\sigma_{B},\Delta_{B},\mu_{8}}&=&0,\nonumber\\
\frac{\partial\Omega_{\mbox{\tiny{eff}}}(\sigma,\Delta,\mu_{8};\mu,\tilde{e}B)}{\partial\Delta}\bigg|_{\sigma_{B},\Delta_{B},\mu_{8}}&=&0,\nonumber\\
\frac{\partial\Omega_{\mbox{\tiny{eff}}}(\sigma,\Delta,\mu_{8};\mu,\tilde{e}B)}{\partial\mu_{8}}\bigg|_{\sigma_{B},\Delta_{B},\mu_{8}}&=&0.
\end{eqnarray}
The solutions of the first two equations build the ``local'' minima
of the theory. In Sec. IV, we will solve the above equations
numerically for any value of the rotated magnetic field
$\tilde{e}B$. Keeping $(\sigma,\Delta)\neq (0,0)$ and looking for
global minima for the system described by complete
$\Omega_{\mbox{\tiny{eff}}}(\sigma,\Delta,\mu_{8};\mu,\tilde{e}B)$
from (\ref{A24}) in the presence of the rotated field, it turns out
that in the regime $300\lesssim\mu\lesssim 500$ MeV, the system
exhibits two ``global'' minima. They are given by $(\sigma_{B}\neq
0,\Delta_{B}=0,\mu_{8}=0)$ in the regime $\mu<\mu_{c}$, and
$(\sigma_{B}=0,\Delta_{B}\neq 0,\mu_{8}\neq 0)$ in the regime
$\mu>\mu_{c}$. Here, $\mu_{c}$ is a certain critical chemical
potential, and, shall be determined numerically in Sec. IV for a
wide range of  $\tilde{e}B$ [see Fig. 9]. We will denote the regime
characterized by $(\sigma_{B}\neq 0,\Delta_{B}=0,\mu_{8}=0)$ and
$(\sigma_{B}=0,\Delta_{B}\neq 0,\mu_{8}\neq 0)$, by the $\chi$SB and
the CSC phases, respectively. In this section, we will analytically
determine the solutions of the above gap equations in the limit of
strong magnetic fields $|\tilde{q}\tilde{e}B|\gg\mu^{2}$, and in the
$\chi$SB and the CSC phases separately. We will then compare these
solutions with the corresponding solutions of the gap equations in
$B=0$ case. In the above limit, the dynamics of the system is
dominated by LLL. The goal is to determine analytically the mass
gaps of the $\chi$SB and CSC phases separately. This will be done in
Sec. IV.A and IV.B, respectively. In IV.A.1 as well as IV.B.1, we
consider the case of strong magnetic field, whereas IV.A.2 as well
as IV.B.2 are devoted to $B=0$ case.
\subsection{The chiral symmetry breaking phase}
\vspace{-0cm}
\subsubsection{Strong magnetic field}
\noindent According to the descriptions from the previous paragraph,
the $\chi$SB phase is characterized by $\sigma_{B}\neq 0$ and
$\Delta_{B}=\mu_{8}=0$. To study this phase in the LLL
approximation, we will, in particular, focus on the first gap
equation from (\ref{YY-1a})
\begin{eqnarray}\label{G1}
\frac{\partial
\Omega_{\mbox{\tiny{eff}}}^{\mbox{\tiny{LLL}}}\left(\sigma,\Delta,\mu_{8};\mu,\tilde{e}B\right)}{\partial
\sigma}\Bigg|_{\sigma_{B},\Delta_{B}=\mu_{8}=0}=0,
\end{eqnarray}
or equivalently on\footnote{In \cite{berges1998}, the same procedure
is performed to study the $\chi$SB and the CSC phases separately.}
\begin{eqnarray}\label{YY-2a}
\frac{\partial\Omega_{\mbox{\tiny{$\chi$SB}}}^{\mbox{\tiny{LLL}}}(\sigma,\Delta_{B}=\mu_{8}=0;\mu,\tilde{e}B)}{\partial\sigma}\bigg|_{\sigma_{B}}=0,
\end{eqnarray}
where $\Omega_{\mbox{\tiny{$\chi$SB}}}^{\mbox{\tiny{LLL}}}$ arises
from (\ref{A24}) with $n=0$ and $(\sigma_{B}\neq 0,
\Delta_{B}=\mu_{8}=0)$. To solve (\ref{YY-2a}) analytically, let us
consider $\Omega_{\mbox{\tiny{$\chi$SB}}}^{\mbox{\tiny{LLL}}}$ first
in the momentum space
\begin{eqnarray}\label{Y2}
\lefteqn{\hspace{-1.5cm}\Omega_{\mbox{\tiny{$\chi$SB}}}^{\mbox{\tiny{LLL}}}(\sigma,\Delta_{B}=\mu_8=0)=\frac{\sigma^{2}}{4G_{S}}+\frac{B^{2}}{2}
-2\int\frac{d^{3}p}{(2\pi)^{3}}\bigg[\sqrt{p^{2}+\sigma^{2}}+(\mu-\sqrt{p^{2}+\sigma^{2}})\theta(\mu-\sqrt{p^{2}+\sigma^{2}})\bigg]
}\nonumber\\
&&~~~-3\tilde{e}B\int^{
+\infty}_{-\infty}\frac{dp_{3}}{4\pi^{2}}\bigg[\sqrt{p_{3}^{2}+\sigma^{2}}+(\mu-\sqrt{p_{3}^{2}+\sigma^{2}})
\theta(\mu-\sqrt{p_{3}^{2}+\sigma^{2}})\bigg]\nonumber\\
&&=\frac{\sigma^{2}}{4G_{S}}+\frac{B^{2}}{2}
-\bigg[\int^{\Lambda}_{0}\frac{p^{2}dp}{\pi^{2}}\sqrt{p^{2}+\sigma^{2}}+\theta(\mu-\sigma)
\int^{\sqrt{\mu^{2}-\sigma^{2}}}_{0}\frac{p^{2}dp}{\pi^{2}}(\mu-\sqrt{p^{2}+\sigma^{2}})\bigg]\nonumber\\
&&~~~
-3\tilde{e}B\bigg[\int^{\Lambda_{B}}_{0}\frac{dp_{3}}{2\pi^{2}}\sqrt{{p_{3}}^{2}+\sigma^{2}}
+\theta(\mu-\sigma)\int^{\sqrt{\mu^{2}-\sigma^{2}}}_{0}\frac{dp_{3}}{2\pi^{2}}
(\mu-\sqrt{{p_{3}}^{2}+\sigma^{2}})\bigg].
\end{eqnarray}
Here, we have introduced the momentum cutoff $\Lambda$ for the first
integral arising from the contribution of zero charged particle. In
contrast, the momentum cutoff $\Lambda_{B}\equiv \sqrt{\tilde{e}B}$
is chosen for the first integral proportional to $\tilde{e}B$, that
arises from the contribution of the remaining three charged
particles.\footnote{The charges of the particles is defined with
respect of the rotated magnetic field. They are presented in Table
I, in units of $\tilde{e}$.} Considering furthermore the effect of
the Heaviside $\theta$-functions in the integrations limits, the
corresponding momentum cutoff to the remaining two integrals is
given by $\sqrt{\mu^{2}-\sigma^{2}}$ with the assumption that
$\sigma<\mu$ (see the $\theta(\mu-\sigma)$ before these two
integrals). Performing the integrations over $p\equiv
|{\mathbf{p}}|$ and $p_3$ in (\ref{Y2}), we arrive at
\begin{eqnarray}\label{Y3}
\Omega_{\mbox{\tiny{$\chi$SB}}}^{\mbox{\tiny{LLL}}}(\sigma,\Delta_{B}=\mu_8=0)&=&\frac{\sigma^{2}}{4
G_{S}}+\frac{B^{2}}{2}
+\frac{\sigma^{4}}{8\pi^{2}}\bigg[\ln\left({\frac{\Lambda+\sqrt{\Lambda^{2}+\sigma^{2}}}{\sigma}}\right)-
\theta(\mu-\sigma)\ln\left(\frac{\mu+\sqrt{\mu^{2}-\sigma^{2}}}{\sigma}\right)\bigg]\nonumber\\
&&-\frac{\Lambda}{8\pi^{2}}\sqrt{\Lambda^{2}+\sigma^{2}}(2\Lambda^{2}
+\sigma^{2})-\frac{\mu}{24\pi^{2}}\theta(\mu-\sigma)\sqrt{\mu^{2}-\sigma^{2}}\left(2\mu^{2}-5\sigma^{2}\right)\nonumber\\
&&-\frac{3\tilde{e}B
\sigma^{2}}{4\pi^{2}}\bigg[\ln\left(\frac{\Lambda_{B}+\sqrt{\Lambda_{B}^{2}+\sigma^{2}}}{\sigma}\right)
-\theta(\mu-\sigma)\ln\left(\frac{\mu+\sqrt{\mu^{2}-\sigma^{2}}}{\sigma}\right)\bigg]\nonumber\\
&&-\frac{3\tilde{e}B}{4\pi^{2}}\bigg[\Lambda_{B}\sqrt{\Lambda_{B}^{2}+\sigma^{2}}+\theta(\mu-\sigma)\mu\sqrt{\mu^{2}-\sigma^{2}}\bigg].
\end{eqnarray}
Minimizing the above potential according to (\ref{YY-2a}), the gap
equation reads
\begin{eqnarray}\label{Y4}
0&=&\frac{\partial
\Omega_{\mbox{\tiny{$\chi$SB}}}^{\mbox{\tiny{LLL}}}}{\partial
\sigma}\bigg|_{\sigma=\sigma_{B}}=\sigma_{B}\left\{\frac{\pi^{2}}{G_{S}}-\Lambda\sigma_{B}\sqrt{\Lambda^{2}+\sigma_{B}^{2}}
-3\tilde{e}B\ln\left(\frac{\Lambda_{B}+\sqrt{\Lambda_{B}^{2}+\sigma_{B}^2}}{\sigma_{B}}\right)
\right.\nonumber\\
&&\left.+\sigma_{B}^{2}\ln\left(\frac{\Lambda+\sqrt{\Lambda^{2}+\sigma_{B}^2}}{\sigma_{B}}\right)
+\theta(\mu-\sigma_{B})\bigg[\mu\sqrt{\mu^{2}-\sigma_{B}^{2}}+(3\tilde{e}B-\sigma_{B}^{2})\ln\left(\frac{\mu+\sqrt{\mu^{2}-\sigma_{B}^{2}}}{\sigma_{B}}\right)
\bigg]\right\}.\nonumber\\
\end{eqnarray}
To find a nontrivial solution $\sigma_{B}\neq 0$ to this equation,
we expand it in the orders of the dimensionless and small parameter
$x\equiv \frac{\sigma_{B}}{\Lambda}\ll 1$ up to order
${\cal{O}}(x^3)$, and get
\begin{eqnarray}\label{Y5-a}
2\Lambda^{2}\left(\frac{1}{g_{s}}-1\right)&=&\sigma_{B}^{2}\left(\frac{5}{2}
-\ln\left(\frac{4\Lambda^{2}}{\sigma_{B}^{2}}\right)\right)+3\tilde{e}B
\ln\left(\frac{4\Lambda_{B}^{2}}{\sigma_{B}^{2}}\right)\nonumber\\
&&+\theta(\mu-\sigma_{B})
\left\{\sigma_{B}^{2}-2\mu^{2}+(\sigma_{B}^{2}-
3\tilde{e}B)\ln\left(\frac{4\mu^{2}}{\sigma_{B}^{2}}\right)\right\},
\end{eqnarray}
where the dimensionless coupling $g_{s}\equiv \frac{
G_{S}\Lambda^2}{\pi^{2}}$ is introduced. In what follows, we
consider two different regimes of $\mu\leq \sigma_{B}$ and
$\mu>\sigma_{B}$ separately. To find real solution for the
simplified gap equation (\ref{Y5-a}), we will then distinguish
various regions for the dimensionless coupling $g_{s}$.
\par\noindent
$i$) In the first regime characterized by $\mu\leq \sigma_{B}$, the
gap equation (\ref{Y5-a}) reads
\begin{eqnarray}\label{Y6-b}
2\Lambda^{2}\left(\frac{1}{g_{s}}-1\right)=\frac{5\sigma_{B}^{2}}{2}-
\sigma_{B}^{2}\ln\left(\frac{4\Lambda^{2}}{\sigma_{B}^{2}}\right)+3\tilde{e}B
\ln\left(\frac{4\Lambda_{B}^{2}}{\sigma_{B}^{2}}\right).
\end{eqnarray}
Since for $0<g_{s}< 1$, the l.h.s.  of (\ref{Y6-b}) is positive, a
nontrivial real solution arises only by the assumption
$\sigma_{B}^{2}\ln\left(\frac{4\Lambda^{2}}{\sigma_{B}^{2}}\right)\ll
3\tilde{e}B
\ln\left(\frac{4\Lambda_{B}^{2}}{\sigma_{B}^{2}}\right)$, which is
indeed justified in the LLL approximation. Neglecting therefore the
first two terms on the r.h.s. of (\ref{Y6-b}), we arrive at
\begin{eqnarray}\label{Y7-b}
\sigma_{B}^{2}=4\tilde{e}B
e^{-\frac{2\Lambda^{2}}{3\tilde{e}B}\left(\frac{1}{g_{s}}-1\right)}.
\end{eqnarray}
Note that the assumption
$\sigma_{B}^{2}\ln\left(\frac{4\Lambda^{2}}{\sigma_{B}^{2}}\right)\ll
3\tilde{e}B \ln\left(\frac{4\Lambda_{B}^{2}}{\sigma_{B}^{2}}\right)$
does not set any limitation on the relation between two momentum
cutoffs $\Lambda$ and $\Lambda_{B}$. Depending on whether $\Lambda$
is larger or smaller than $\Lambda_{B}$, different regimes are to be
distinguished for the coupling $g_{s}$:
\begin{eqnarray}\label{Y8-b}
\left\{\begin{array}{ccccccl}
\mbox{For}&&\Lambda\leq\Lambda_{B},&&\mbox{we
get}&&0<g_{s}<\frac{\Lambda^{2}}{\Lambda^{2}-3\tilde{e}B\ln\left(\frac{\Lambda}{2\Lambda_{B}}\right)}< 1.\\
\mbox{For}&&\Lambda>\Lambda_{B},&&\mbox{we
get}&&0<g_{s}<\frac{\Lambda^{2}}{\Lambda^{2}+3\tilde{e}B\ln 2}< 1
\end{array}
\right.
\end{eqnarray}
The dynamical mass $\sigma_{B}$ from (\ref{Y7-b}) is, apart from
numerical factors, the same as the dynamical mass of the NJL model
in the presence of constant magnetic field from \cite{miransky1995}.
The additional factor $1/3$, that arises in the exponent of
(\ref{Y7-b}) corresponds to three different quark charges
$\tilde{q}=1,\pm\frac{1}{2}$ that have, in the regime of LLL
dominance ($n=0$) equal contributions to the effective potential in
the $\chi$SB phase.
\par
Let us consider again the gap equation (\ref{Y6-b}) for the case
$g_{s}>1$. In this case a nontrivial solution may exist only for
$\Lambda$ in the same order of magnitude as $\Lambda_{B}$. To find
the solution, we rewrite first the gap equation (\ref{Y6-b}) as
\begin{eqnarray}\label{Y9-b}
2\Lambda^{2}\left(\frac{1}{g_{s}}-1\right)=-\sigma_{B}^{2}\ln\left(\frac{4\Lambda^{2}}{e^{\frac{5}{2}}\sigma_{B}^{2}}\right)
+3\tilde{e}B\left(\ln\left(\frac{3\Lambda_{B}^{2}}{2\Lambda^{2}}\right)+\ln\left(\frac{8\Lambda^{2}}{3\sigma_{B}^{2}}\right)\right).
\end{eqnarray}
Expanding now the second term on the r.h.s. in the orders of
$y\equiv \frac{2\Lambda^{2}}{3\Lambda_{B}^{2}}-1\simeq 0$ up to
${\cal{O}}(y^2)$, we arrive at
\begin{eqnarray}\label{Y10-b}
2\Lambda^{2}\left(\frac{1}{g_{s}}-1\right)\approx-\sigma_{B}^{2}\ln\left(\frac{4\Lambda^{2}}
{e^{\frac{5}{2}}\sigma_{B}^{2}}\right)-3\tilde{e}B
\left(\frac{2\Lambda^{2}}{3\Lambda^{2}_{B}}-1\right)+
3\tilde{e}B\ln\left(\frac{8\Lambda^{2}}{3\sigma_{B}^{2}}\right).
\end{eqnarray}
Using the same approximation
$\sigma_{B}^{2}\ln\left(\frac{4\Lambda^{2}}{\sigma_{B}^{2}}\right)\ll
3\tilde{e}B
\ln\left(\frac{4\Lambda_{B}^{2}}{\sigma_{B}^{2}}\right)$, we can
neglect the first term on the r.h.s. of (\ref{Y10-b}), and arrive at
\begin{eqnarray}\label{Y11-b}
\frac{2\Lambda^{2}}{g_{s}}\approx 3\tilde{e}B
\ln\left(\frac{8\Lambda^{2}}{3\sigma_{B}^{2}}\right)+3\tilde{e}B,
\end{eqnarray}
whose solution reads
\begin{eqnarray}\label{Y12-b}
\sigma_{B}^{2}={\cal{C}}
\Lambda^{2}e^{-\frac{2\Lambda^{2}}{3\tilde{e}Bg_{s}}},\qquad\mbox{with}\qquad
{\cal{C}}=\frac{8}{3}e\approx 7.25.
\end{eqnarray}
$ii$) In the regime characterized by $\sigma_{B}<\mu$, the gap
equation is given by
\begin{eqnarray}\label{Y13-b}
\Lambda^{2}\left(\frac{1}{g_s}-1\right)=\frac{7\sigma_{B}^{2}}{4}-\mu^{2}-\frac{3\tilde{e}B}{2}
\ln\frac{\mu^{2}}{\Lambda_{B}^{2}}+\frac{\sigma_{B}^{2}}{2}\ln\frac{\mu^{2}}{\Lambda^{2}}.
\end{eqnarray}
It arises by expanding (\ref{Y4}) in the orders of
$x=\frac{\sigma_{B}}{\Lambda}$ up to order ${\cal{O}}(x^{3})$. As it
turns out a real solution may be found by expanding (\ref{Y13-b}) in
the orders $w\equiv \frac{\mu^{2}}{\Lambda^{2}}-1\simeq 0$ and
$z\equiv \frac{\mu^{2}}{\Lambda_{B}^{2}}-1\simeq 0$ up to
${\cal{O}}(w^2)$ as well as ${\cal{O}}(z^{2})$. The mass gap can be
computed directly from the resulting equation and reads
\begin{eqnarray}\label{Y15-b}
\sigma_{B}^{2}\simeq\Lambda^{2}\left(\frac{1}{g_s}-1\right)-\frac{3\tilde{e}B}{2}+\frac{5\mu^{2}}{2}.
\end{eqnarray}
Note that a real solution for $\sigma_{B}$ in this regimes arises
only when $\tilde{e}B$ and $g_{s}$ satisfy the following conditions:
\begin{eqnarray}\label{Y16-b}
\begin{array}{ccrcl}
\mbox{For}&&0<g_s< 1:&&\left\{
\begin{array}{lclcl}
\mu^{2}<\Lambda_{B}\leq \frac{5\mu^{2}}{3},&&\mbox{and}&&
\frac{2\Lambda^{2}}{3\tilde{e}B+2\Lambda^{2}-3\mu^{2}}<g_{s}< 1,\\
\frac{5\mu^{2}}{3}<\Lambda_{B},&&\mbox{and}&&
\frac{2\Lambda^{2}}{3\tilde{e}B+2\Lambda^{2}-3\mu^{2}}<g_{s}<\frac{2\Lambda^{2}}{3\tilde{e}B+2\Lambda^{2}-5\mu^{2}},\\
\end{array}
\right. \\
\mbox{For}&&g_{s}>1:&&\left\{
\begin{array}{lclcl}
\mu^{2}<\Lambda_{B}<\frac{5\mu^{2}}{3},&&\mbox{and}&&
1<g_{s}<\frac{2\Lambda^{2}}{3\tilde{e}B+2\Lambda^{2}-5\mu^{2}}.\\
\end{array}
\right.\\
\end{array}
\end{eqnarray}
The above conditions arise without specifying any relation between
$\Lambda$ and $\Lambda_{B}$.
\subsubsection{Zero magnetic field}
\noindent To clarify the effect of strong magnetic fields on the
mass gap, we will present in what follows the analytical results of
the gap equation corresponding to the effective potential
(\ref{A28}) at zero magnetic field.\footnote{In \cite{ebert2005} the
full gap equations of the 2SC model including the mesons is solved
numerically for $B=0$. As it turns out the system exhibits, as in
$B\neq 0$ case, a phase transition from the $\chi$SB to the CSC
phase. Here, the $\chi$SB phase is characterized by $(\sigma_{0}\neq
0, \Delta_{0}=0)$ and the CSC by $(\sigma_{0}=0, \Delta_{0}\neq
0)$.} Setting, as in the previous section, in the $\chi$SB phase,
$\Delta_{0}=\mu_{8}=0$ in the corresponding effective potential
(\ref{A28}), the resulting potential in the momentum space reads
\begin{eqnarray}\label{Y17-b}
\hat{\Omega}_{\mbox{\tiny{$\chi$SB}}}(\sigma,
\Delta_{0}=\mu_8=0)&=&\frac{\sigma^{2}}{4G_{S}}-6\int^{\Lambda}_{0}\frac{p^{2}d
p}{\pi^{2}}\left(\sqrt{{{p}}^{2}+\sigma^2}+(\mu-\sqrt{{{p}}^{2}+\sigma^2})\theta(\mu-\sqrt{{{p}}^{2}+\sigma^2})\right).\nonumber\\
\end{eqnarray}
After performing the $p$ integration, we arrive at
\begin{eqnarray}\label{Y18-b}
\lefteqn{\hspace{-1.5cm}\hat{\Omega}_{\mbox{\tiny{$\chi$SB}}}(\sigma,
\Delta_{0}=\mu_8=0)=\frac{\sigma^{2}}{4
G_{S}}+\frac{3m^{4}}{4\pi^{2}}\bigg[\ln\left(\frac{\Lambda+\sqrt{\Lambda^{2}+\sigma^{2}}}{\sigma}\right)-
\theta(\mu-\sigma)\ln\left(\frac{\mu+\sqrt{\mu^{2}-\sigma^{2}}}{\sigma}\right)\bigg]}\nonumber\\
&&-\frac{3\Lambda}{4\pi^{2}}(2\Lambda^{2}+\sigma^{2})\sqrt{\Lambda^{2}+\sigma^{2}}+
\frac{\mu}{4\pi^{2}}(5\sigma^{2}-2\mu^{2})\theta(\mu-\sigma)\sqrt{\mu^{2}-\sigma^{2}}.
\end{eqnarray}
The corresponding gap equation reads then
\begin{eqnarray}\label{Y19-b}
0&=&\frac{\partial \hat{\Omega}_{\mbox{\tiny{$\chi$SB}}}}{\partial
\sigma}\bigg|_{\sigma=\sigma_{0}}=\frac{\pi^{2}}{6G_{S}}-\Lambda\sqrt{\Lambda^{2}+\sigma_{0}^{2}}+\sigma_{0}^{2}
\ln\left(\frac{\Lambda+\sqrt{\Lambda^{2}+\sigma_{0}^{2}}}{\sigma_{0}}\right)\nonumber\\
&&\qquad\qquad
+\theta(\mu-\sigma_{0})\bigg[\mu\sqrt{\mu^{2}-\sigma_{0}^{2}}-\sigma_{0}^{2}\ln\left({\frac{\mu+\sqrt{\mu^{2}-\sigma_{0}^{2}}}{\sigma_{0}}}\right)\bigg].
\end{eqnarray}
Defining, similar to what we did in the $B\neq 0$ case, a
dimensionless parameter $\hat{x}\equiv \frac{\sigma_0}{\Lambda}\ll
1$, and expanding the gap equation (\ref{Y19-b}) in the orders of
$\hat{x}$ up to ${\cal{O}}(\hat{x}^{3})$, we arrive at
\begin{eqnarray}\label{Y20-b}
\hspace{0cm}0=\frac{\partial
\hat{\Omega}_{\mbox{\tiny{$\chi$SB}}}}{\partial
\sigma}\bigg|_{\sigma=\sigma_{0}}=2\Lambda^{2}\left(\frac{1}{\hat{g}_{s}}-1\right)-
\sigma_0^{2}\left(1-\ln\frac{4\Lambda^{2}}{\sigma_0^{2}}\right)+
\theta(\mu-\sigma_0)\left(-\sigma_0^{2}+2\mu^{2}-\sigma_0^{2}
\ln\frac{4\mu^{2}}{\sigma_0^{2}}\right),\nonumber\\
\end{eqnarray}
where we have introduced the dimensionless coupling
$\hat{g}_{s}\equiv \frac{6G_{S}\Lambda^{2}}{\pi^{2}}=6g_{s}$. To
find a real solution for the mass gap $\sigma_{0}$, we have to
distinguish two different regimes of the chemical potential $\mu$.
They are characterized by $\mu\leq \sigma_{0}$ and $\sigma_{0}<\mu$.
\par\noindent
$i$) In the first regime characterized by $\mu\leq \sigma_{0}$, a
real solution $\sigma_{0}<\Lambda$ arises only for
$1<\hat{g}_{s}<\hat{g}_{1}$ with $\hat{g}_{1}\equiv \frac{1}{1-\ln
2}\simeq 3.26$. It reads
\begin{eqnarray}\label{Y21-b}
\sigma_{0}^{2}=4\Lambda^{2}\exp\left(W_{-1}\left(\frac{1}
{2\hat{g}_{s}}-\frac{1}{2}\right)\right).
\end{eqnarray}
It corresponds to one of the two real branches of the Lambert $W(x)$
function, $W_{0}(x)$ and $W_{-1}(x)$, which is known to be the
function satisfying (see \cite{knuth} for more details on the
Lambert $W$-function)\footnote{According to the explanations in
\cite{knuth}: If $x$ is real, then for $-1/e\leq x<0$, there are two
possible real values of $W(x)$. One denotes the branch satisfying
$-1\leq W(x)$ by $W_0(x)$, and the branch satisfying $W(x)\leq -1$
by $W_{-1}(x)$.}.
\begin{eqnarray}\label{Y22-b}
W(x)e^{W(x)}=x.
\end{eqnarray}\par\noindent
$ii$) In the second regime characterized by $\sigma_{0}<\mu$, the
gap equation is
\begin{eqnarray}\label{Y23-b}
\Lambda^{2}\left(\frac{1}{\hat{g}_{s}}-1\right)=\sigma_{0}^{2}-\mu^{2}+\frac{\sigma_{0}^2}{2}
\ln\left(\frac{\mu^{2}}{\Lambda^{2}}\right).
\end{eqnarray}
It arises from (\ref{Y19-b}) by an expansion in the orders of
$\hat{x}=\frac{\sigma_{0}}{\Lambda}$ up to ${\cal{O}}(\hat{x}^{3})$.
Introducing a small parameter
$\hat{w}\equiv\frac{\mu^{2}}{\Lambda^{2}}-1$ and expanding
(\ref{Y23-b}) in the orders of $\hat{w}$ up to
${\cal{O}}(\hat{w}^{2})$, yields the mass gap
\begin{eqnarray}\label{Y24-b}
\sigma_{0}^{2}=2\Lambda^{2}\left(\frac{1}{\hat{g}_{s}}-1\right)+2\mu^{2}.
\end{eqnarray}
Note that a real solution for $\sigma_{0}$ arises only for
$\hat{g}_{s}>1$, satisfying
\begin{eqnarray}\label{Y25-b}
\frac{2\Lambda^{2}}{2\Lambda^{2}-\mu^{2}}<\hat{g}_{s}<\frac{\Lambda^{2}}{\Lambda^{2}-\mu^{2}}.
\end{eqnarray}
In Sec. V, we will perform a numerical analysis to study the
$\chi$SB phase for any arbitrary magnetic field. We will show that,
similar to the $B\neq 0$ case, the second regime characterized by
$\sigma_{0}<\mu$ belongs to the color symmetry breaking phase and is
indeed irrelevant for the present $\chi$SB phase. Comparing
therefore only the relevant part of the solutions, i.e. (\ref{Y7-b})
for $B\neq 0$ with (\ref{Y21-b}) for $B=0$, we note that, in
contrast to $B=0$ case, in the presence of strong magnetic fields,
the formation of chiral symmetry breaking bound state $\sigma_B$ is
possible even for small dimensionless $\chi$SB coupling $0<g_{s}<1$.
This is in fact one of the consequences of the phenomenon of
magnetic catalysis \cite{miransky1995}.
\subsection{The color superconducting phase}
\subsubsection{Strong magnetic field}
\noindent According to our explanation in the paragraph below
(\ref{YY-1a}), the CSC phase is characterized by $(\sigma_{B}=0,
\Delta\neq 0, \mu_{8}\neq 0)$. The corresponding effective potential
arises from (\ref{A24}) by setting $n=0$, and $\sigma_{B}=0$. In the
momentum space, it is given by
\begin{eqnarray}\label{Y26-b}
\Omega_{\mbox{\tiny{CSC}}}^{\mbox{\tiny{LLL}}}(\sigma_{B}=0,
\Delta,\mu_{8})&=&\frac{\Delta^{2}}{4
G_{D}}+\frac{B^{2}}{2}-2\int^{\infty}_{0}\frac{d^{3}p}{(2\pi)^{3}}[p+(\breve{\mu}-p)\theta(\breve{\mu}-p)-(\breve{\mu}+p)
\theta\left(-\breve{\mu}-p\right)]\nonumber\\
&&-\tilde{e}B\int^{\infty}_{0}\frac{d
p_{3}}{2\pi^{2}}\bigg[p_{3}+(\breve{\mu}-p_{3})\theta(\breve{\mu}-p_{3})-(\breve{\mu}+p_{3})
\theta\left(-\breve{\mu}-p_{3}\right)\nonumber\\
&&+\sqrt{(p_{3}+\bar{\mu})^{2}+\Delta^{2}}+\sqrt{(p_{3}-\bar{\mu})^{2}
+\Delta^{2}}\bigg].
\end{eqnarray}
Performing the $p_{3}$ and $p$ integrations by introducing the
momentum cutoffs $\Lambda$ and $\Lambda_{B}$ for the
$p=|{\mathbf{p}}|$ as well as $p_{3}$ integrations,\footnote{See our
description below (\ref{Y2}) for the choice of the momentum
cutoffs.} we arrive at
\begin{eqnarray}\label{Y27-b}
\Omega_{\mbox{\tiny{CSC}}}^{\mbox{\tiny{LLL}}}(\sigma_{B}=0,
\Delta,\mu_{8})&=&\frac{\Delta^{2}}{4 G_{D}}+\frac{B^{2}}{2}
-\frac{3\Lambda^{4}+(3\tilde{e}B+\breve{\mu}^{2})\breve{\mu}^{2}}{12\pi^{2}}\nonumber\\
&&-\frac{\tilde{e}B}{4\pi^{2}}\bigg[\Lambda^{2}_{B}+(\Lambda_{B}-\bar{\mu})\sqrt{\Delta^{2}+(\Lambda_{B}-\bar{\mu})^{2}}
+(\Lambda_{B}+\bar{\mu})\sqrt{\Delta^{2}+
(\Lambda_{B}+\bar{\mu})^{2}}\nonumber\\
&&+\Delta^{2}\ln\left(\frac{\sqrt{\Delta^{2}+(\Lambda_{B}+\bar{\mu})^{2}}+(\Lambda_{B}+\bar{\mu})}
{\sqrt{\Delta^{2}+(\Lambda_{B}-\bar{\mu})^{2}}-(\Lambda_{B}-\bar{\mu})}\right)\bigg].
\end{eqnarray}
In the CSC phase, we have a set of two coupled equations: the color
neutrality condition,  $\frac{\partial
\Omega_{\mbox{\tiny{CSC}}}^{\mbox{\tiny{LLL}}}}{\partial
\mu_{8}}=0$, and the gap equation $\frac{\partial
\Omega_{\mbox{\tiny{CSC}}}^{\mbox{\tiny{LLL}}}}{\partial \Delta}=0$.
The goal is to solve these two equations to determine $\mu_8$ and
$\Delta$ as a function of the external magnetic field $B$ and the
chemical potential $\mu$. Let us first consider the color neutrality
condition
\begin{eqnarray}\label{Y28-b}
0=\frac{\partial
\Omega_{\mbox{\tiny{CSC}}}^{\mbox{\tiny{LLL}}}}{\partial
\mu_{8}}\bigg|_{\Delta_{B}}&=&
4\breve{\mu}^{3}+3\tilde{e}B\left(2\breve{\mu}+\sqrt{\Delta_B^{2}
+(\Lambda_{B}-\bar{\mu})^{2}}-\sqrt{\Delta_B^{2}
+(\Lambda_{B}+\bar{\mu})^{2}}\right),
\end{eqnarray}
where $\bar{\mu}=\mu+\mu_8$ and $\breve{\mu}=\mu-2\mu_{8}$. Defining
three dimensionless (small) parameters, $x\equiv
\frac{\mu}{\Lambda}$, $y\equiv \frac{\Delta_B}{\Lambda}$ as well as
$z\equiv \frac{\mu_8}{\Lambda}$, and expanding (\ref{Y28-b}) in the
orders of $x,y$ and $z$ up to ${\cal{O}}(x^{4}), {\cal{O}}(y^{3})$
and ${\cal{O}}(z^2)$, $\mu_{8}$ can be determined from the resulting
equation and is given by
\begin{eqnarray}\label{Y30-b}
\mu_{8}=\frac{2\mu^3}{3(3\tilde{e}B+4\mu^{2})}.
\end{eqnarray}
As for the gap equation corresponding to $\Delta_{B}$, we have
\begin{eqnarray}\label{Y31-b}
0=\frac{\Omega_{\mbox{\tiny{CSC}}}^{\mbox{\tiny{LLL}}}}{\partial
\Delta}\bigg|_{\Delta_{B}}&=&\frac{\pi^{2}}{G_{D}}-\tilde{e}B
\ln\left(\frac{\sqrt{\Delta_B^{2}+(\Lambda+\bar{\mu})^{2}}+(\Lambda_B+\bar{\mu})}{\sqrt{\Delta_B^{2}
+(\Lambda_B-\bar{\mu})^{2}}-(\Lambda_B-\bar{\mu})}\right).
\end{eqnarray}
After expanding (\ref{Y31-b}) in the orders of
$y\equiv\frac{\Delta_{B}}{\Lambda}$ and $z\equiv
\frac{\mu_{8}}{\Lambda}$ up to ${\cal{O}}(y^{3})$ and
${\cal{O}}(z^{2})$, and replacing $\mu_{8}$ from (\ref{Y30-b}) in
the resulting equation, we arrive at
\begin{eqnarray}\label{Y32-b}
\frac{\Lambda^{2}}{g_{d}}+\frac{4\mu^{4}}{9\tilde{e}B}-\tilde{e}B\ln
\left(\frac{4(\tilde{e}B-\mu^{2})}{\Delta_{B}^{2}}\right)=0,
\end{eqnarray}
where the dimensionless diquark coupling in the CSC phase
$g_{d}\equiv\frac{G_{D}\Lambda^{2}}{\pi^{2}}$ is introduced. The
diquark mass gap $\Delta_{B}$ can then be determined directly from
(\ref{Y32-b}) and reads
\begin{eqnarray}\label{Y33-b}
\Delta_{B}^{2}=
4(\Lambda_{B}^{2}-\mu^{2})\exp\left(-\frac{\Lambda^{2}}{\tilde{e}B}\frac{1}{g_{d}}\right).
\end{eqnarray}
This result is comparable with the results by \cite{ferrer2006} for
the three-flavor CFL model. In particular, in both models the
exponents are proportional to $(\tilde{e}Bg_{d})^{-1}$. The
dependence of $\Delta_{B}$ on the magnetic field demonstrates the
effect of magnetic catalysis \cite{miransky1995}, that states that
even for small value of the dimensionless diquark coupling $g_{d}$,
the presence of a strong magnetic field leads to color symmetry
breaking and the formation of diquark mass $\Delta_{B}$.
\subsubsection{Zero magnetic field}
\noindent We consider, as next, the effective potential of the 2SC
model in the absence of magnetic field from (\ref{A28}) in the CSC
phase by setting $(\sigma_{0}=0, \Delta_{0}\neq 0, \mu_{8}\neq 0)$.
In the momentum space, the resulting potential is then given by
\begin{eqnarray}\label{Y34-b}
\lefteqn{\hat{\Omega}_{\mbox{\tiny{CSC}}}(\sigma_{0}=0,
\Delta,\mu_{8})=\frac{\Delta^{2}}{4 G_{D}}
-2\int^{\Lambda}_{0}\frac{p^{2} d
p}{\pi^{2}}\bigg[\sqrt{\Delta^{2}+(p+\bar{\mu})^{2}}+
\sqrt{\Delta^{2}+(p-\bar{\mu})^{2}}+p}
\nonumber\\
&&\hspace{3cm}+(\breve{\mu}-p)\theta(\breve{\mu}-p)-(\breve{\mu}+p)\theta\left(-\breve{\mu}-p\right)\bigg]\nonumber\\
&=&\frac{\Delta^{2}}{4
G_{D}}-\frac{\breve{\mu}^{4}}{6\pi^{2}}-\frac{\Lambda^{4}}{2\pi^{2}}
-\frac{1}{12\pi^{2}}[3\Lambda(\Delta^{2}+2\Lambda^{2})+\bar{\mu}(13\Delta^{2}-2\Lambda^{2})
-2\bar{\mu}^{2}(\Lambda+\bar{\mu})]\sqrt{\Delta^{2}+(\Lambda-\bar{\mu})^{2}}\nonumber\\
&&+[3\Lambda(\Delta^{2}+2\Lambda^{2})-\bar{\mu}(13\Delta^{2}-2\Lambda^{2})-
2\bar{\mu}^{2}(\Lambda-\bar{\mu})]\sqrt{\Delta^{2}+(\Lambda+\bar{\mu})^{2}}\nonumber\\
&&-3\Delta^{2}(\Delta^{2}-4\bar{\mu}^{2})\ln\left(\frac{[\Lambda+\bar{\mu}+\sqrt{\Delta^{2}+(\Lambda+\bar{\mu})^{2}}]
[\Lambda-\bar{\mu}+\sqrt{\Delta^{2}+(\Lambda-\bar{\mu})^{2}}]}{\Delta^{2}}\right).
\end{eqnarray}
In this case, the color neutrality condition reads
\begin{eqnarray}\label{Y35-b}
0&=&\frac{\partial\hat{\Omega}_{\mbox{\tiny{CSC}}}}{\partial
\mu_{8}}\bigg|_{\Delta_{0},\hat{\mu}_8}=
2\breve{\mu}^{3}-3\Delta_{0}^{2}\bar{\mu}\ln\left(\frac{[\Lambda+\bar{\mu}+\sqrt{\Delta_{0}^{2}+(\Lambda+\bar{\mu})^{2}}][\Lambda-\bar{\mu}+\sqrt{\Delta_{0}^{2}
+(\Lambda-\bar{\mu})^{2}}]}{\Delta_{0}^{2}}\right)\nonumber\\
&&-[2\Delta_{0}^{2}-\Lambda^{2}-\bar{\mu}(\Lambda+\bar{\mu})]\sqrt{\Delta_{0}^{2}+(\Lambda-\bar{\mu})^{2}}+[2\Delta_{0}^{2}-\Lambda^{2}+\bar{\mu}(\Lambda-\bar{\mu})]
\sqrt{\Delta_{0}^{2}+(\Lambda+\bar{\mu})^{2}}.\nonumber\\
\end{eqnarray}
After expanding in the orders of $\hat{x}\equiv\frac{\mu}{\Lambda}$,
$\hat{y}\equiv \frac{\Delta_{0}}{\Lambda}$, and
$\hat{z}\equiv\frac{\hat{\mu_{8}}}{\Lambda}$ up to order
${\cal{O}}\left(\hat{x}^{4}\right)$, ${\cal{O}}(\hat{y}^{3})$, as
well as ${\cal{O}}(\hat{z}^{2})$, we arrive at
\begin{eqnarray}\label{Y36-b}
\hat{\mu}_{8}=\frac{\Delta_{0}^{2}}{3\mu}+\frac{\Delta_{0}^{2}}{6\mu}\ln\left(\frac{\Delta_{0}^{2}}{4\Lambda^{2}}\right),
\end{eqnarray}
where $\Delta_{0}$ satisfies the gap equation
\begin{eqnarray}\label{Y37-b}
0&=&\frac{\hat{\Omega}_{\mbox{\tiny{CSC}}}}{\partial
\Delta}\bigg|_{\Delta_{0},\hat{\mu}_{8}}=-\frac{\pi^{2}}{2G_{D}}
+(\Lambda+3\bar{\mu})\sqrt{\Delta_{0}^{2}+(\Lambda-\bar{\mu})^{2}}+(\Lambda-3\bar{\mu})\sqrt{\Delta_{0}^{2}
+(\Lambda+\bar{\mu})^{2}}\nonumber\\
&&-(\Delta_{0}^{2}-2\bar{\mu}^{2})\ln\left(\frac{[\Lambda+\bar{\mu}+\sqrt{\Delta_{0}^{2}+(\Lambda+\bar{\mu})^{2}}]
[\Lambda-\bar{\mu}+\sqrt{\Delta_{0}^{2}+(\Lambda-\bar{\mu})^{2}}]}{\Delta_{0}^{2}}\right).
\end{eqnarray}
Using the same method as above and expanding (\ref{Y37-b}) in the
orders of $\hat{y}$ and $\hat{z}$ up to order
${\cal{O}}(\hat{y}^{3})$, as well as ${\cal{O}}(\hat{z}^{2})$, we
get
\begin{eqnarray}\label{Y38-b}
0&=&\frac{\hat{\Omega}_{\mbox{\tiny{CSC}}}}{\partial
\Delta}\bigg|_{\Delta_{0},\hat{\mu}_{8}}=\Lambda^{2}\left(1-\frac{1}{\hat{g}_{d}}\right)-3\mu^{2}-\mu^{2}
\ln\left(\frac{\Delta_{0}^{2}}{4(\Lambda^{2}-\mu^{2})}\right),
\end{eqnarray}
where $\hat{g}_{d}\equiv \frac{4G_{D}\Lambda^{2}}{\pi^{2}}=4g_{d}$.
Solving (\ref{Y38-b}), the diquark mass for vanishing magnetic
fields is then given by
\begin{eqnarray}\label{Y39-b}
\Delta^{2}_{0}&=&
{\cal{C}}_{2}\left(\Lambda^{2}-\mu^{2}\right)\exp\left(-\frac{\Lambda^{2}}{\mu^{2}}\left(\frac{1}{\hat{g}_{d}}-1\right)\right),\qquad\mbox{with}\qquad
{\cal{C}}_{2}=4e^{-3}\simeq 0.2.
\end{eqnarray}
The qualitative behavior of $\Delta_{0}$ as a function of $\mu$
coincides with the results from \cite{ebert2005, ebert2010}. The
 color chemical potential $\hat{\mu}_{8}$ arises by replacing (\ref{Y39-b}) in (\ref{Y36-b}).
These results are to be compared with (\ref{Y33-b}) [for the diquark
mass gap $\Delta_{B}$] as well as (\ref{Y30-b}) [for the color
chemical potential $\mu_{8}$] in the presence of strong magnetic
field.
\section{Numerical results for arbitrary magnetic field}
\par\noindent
\setcounter{equation}{0}In the previous section, we have presented
analytical solutions for the order parameters $\sigma$ and $\Delta$
corresponding to $\chi$SB and CSC phases, as well as for the color
chemical potential $\mu_{8}$ in the presence of strong magnetic
fields in the LLL approximation. We have then compared our results
with the mass gaps arising from the thermodynamic potential of the
2SC model in the absence of magnetic field in order to emphasize the
effect of strong magnetic fields on the formation of bound states
$\sigma$ and $\Delta$ in the superconducting 2SC model. In this
section, we will study numerically the effect of any arbitrary
magnetic field on quark matter without restricting ourselves to LLL
approximation. In particular, we are interested on the dependence of
the mass gaps on the external magnetic field $\tilde{e}B$ and the
chemical potential $\mu$. To do this, we set, as in the previous
section, $m_{0}=0$ and choose $G_{D}<G_{S}$. Comparing our numerical
results with the analytical results from Sec. IV, we will determine
numerically the range of the magnetic field strength for which the
LLL approximation is reliable. At the end of this section, we will
study the phase diagram of the model in a $\mu_{c}-\tilde{e}B$
plane, and determine the type of various phase transition between
the $\chi$SB and the CSC phases for a wide range of $\tilde{e}B$.
\par
Let us start with the one-loop effective potential (\ref{A24})
arising from a mean field approximation in the presence of an
arbitrary magnetic field. To perform the momentum integrations
numerically, we have to fix the free parameters of the model, the
momentum cutoff $\Lambda$ and the couplings $G_{S}$ and $G_{D}$. Our
specific choice of the parameters is \cite{huang2002}
\begin{eqnarray}\label{B1}
\Lambda=0.6533~\mbox{GeV},~~G_{S}=5.0163~\mbox{GeV}^{-2},\qquad\mbox{and}\qquad
G_{D}=\frac{3}{4}G_{S}.
\end{eqnarray}
For vanishing magnetic field $\tilde{e}B=0$, they yield the $\chi$SB
gap $\sigma_{0}\simeq 323.8$ MeV at $\mu=250$ MeV, and the 2SC gap
of $\Delta_{0}\simeq 126$ MeV at $\mu=460$ MeV.\footnote{Although
our free parameters $\Lambda, G_{D}$, and $G_{S}$ coincide with the
parameters used in \cite{huang2002}, the  numerical value of
$\sigma_{0}$ is different from what is reported in \cite{huang2002}.
The reason for this difference is apparently in the choice of the
cutoff function. Whereas we use smooth cutoff function (\ref{B2}),
in \cite{huang2002} a sharp momentum cutoff is used to perform the
momentum integrations numerically.} Smooth cutoff functions (form
factor)
\begin{eqnarray}\label{B2}
f_{\Lambda}=\frac{1}{1+\exp\left(\frac{|{\mathbf{p}}|-\Lambda}{A}\right)},\qquad\mbox{and}\qquad
f_{\Lambda,B}^{n}=\frac{1}{1+\exp\left(\frac{\sqrt{p_{3}^{2}+2|\tilde{q}\tilde{e}B|n}-\Lambda}{A}\right)},
\end{eqnarray}
are then introduced to perform numerically the momentum $p$
integrations corresponding to zero charged particles and charged
particles, respectively.\footnote{In (\ref{A24}), the integrals
proportional to $\tilde{e}B$  and including a summation over Landau
levels $n$ arises from charged quarks with charges $\tilde{q}=\pm
\frac{1}{2},+1$.} In (\ref{B3}), $A$ is a free parameter and is
chosen to be $A=0.05\Lambda$. Similar smooth cutoff function (form
factor) is also used in \cite{warringa2007}. Here, as in
\cite{warringa2007}, the free parameter $A$ determines the sharpness
of the cutoff scheme. In what follows, we will first study the
behavior of mass gaps as well as magnetizations in the $\chi$SB and
CSC regimes as functions of $\tilde{e}B$ and for fixed chemical
potentials.
\begin{figure}[hbt]
\includegraphics[width=8cm,height=6cm]{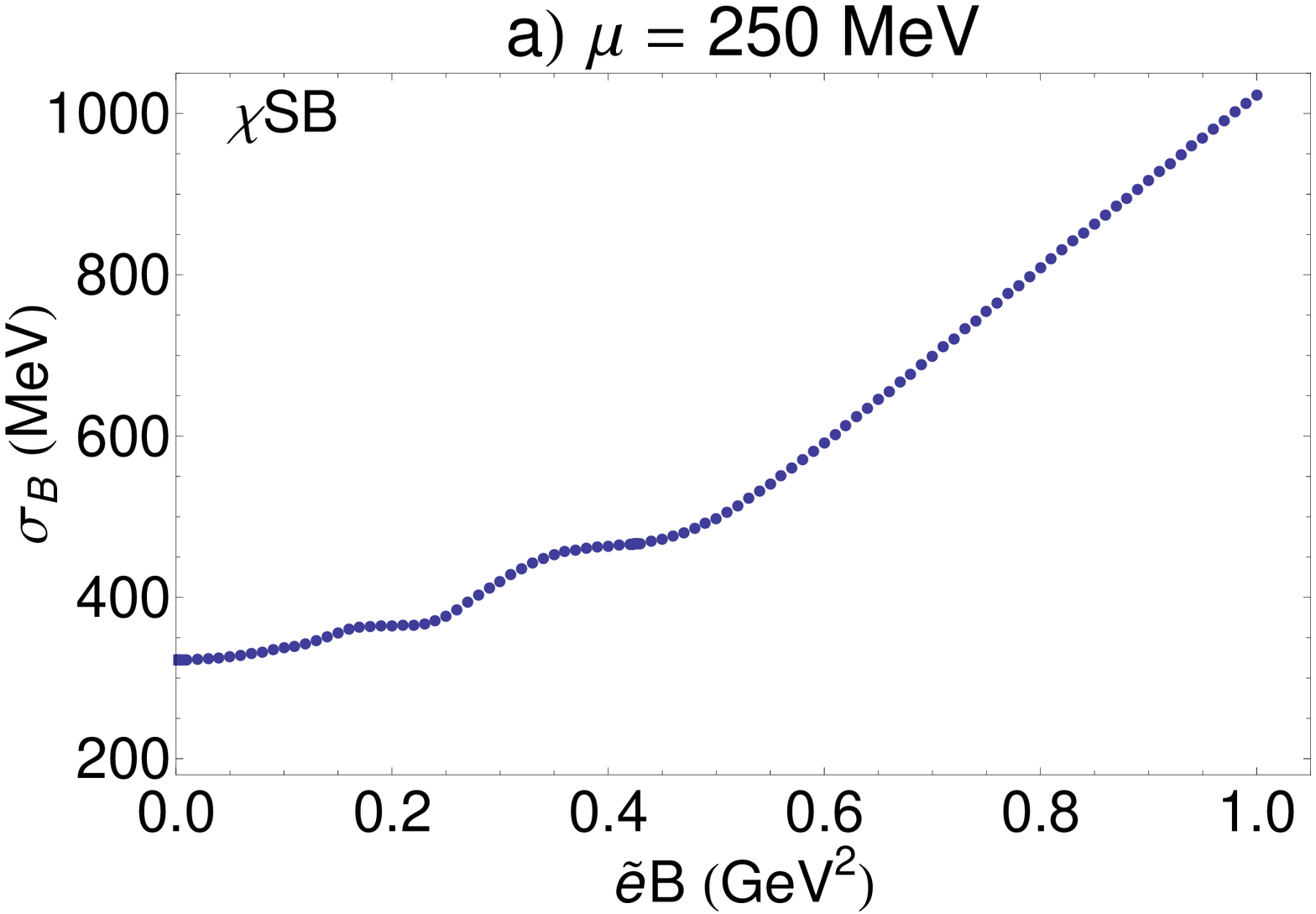}
\hspace{0.3cm}
\includegraphics[width=8cm,height=6cm]{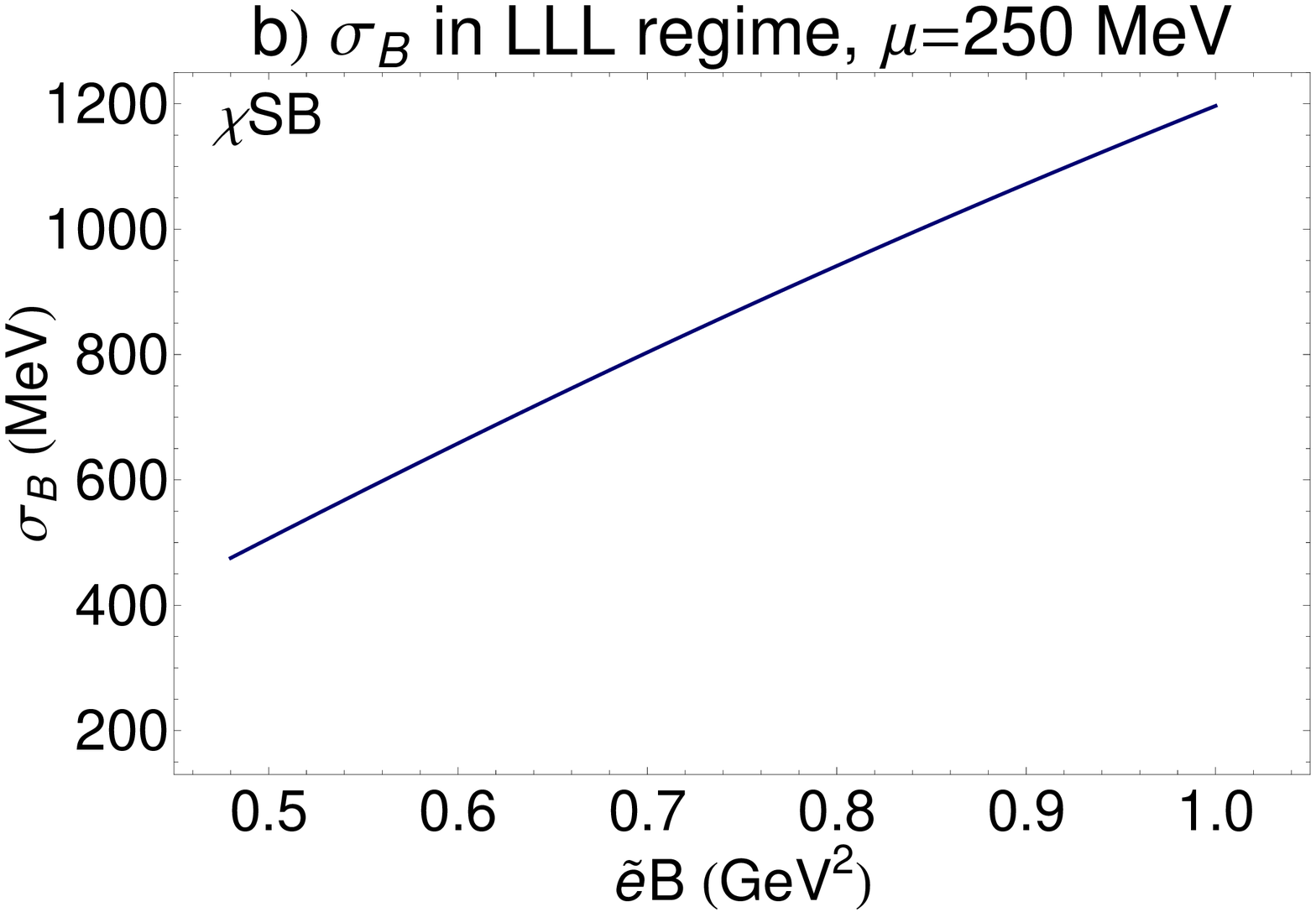}
\caption{a) The dependence of $\sigma_{B}$ on $\tilde{e}B$ in the
$\chi$SB phase for $\mu=250$ MeV;  b) The analytical result of
$\sigma_{B}$ in the regime of LLL dominance from (\ref{Y7-b}) is
plotted for $\tilde{e}B\in\{0.45,1\}~\mbox{GeV}^2$ and $\mu=250$
MeV. }
\end{figure}
\par\noindent
In Fig. 1a, the $\chi$SB mass gap $\sigma$ is plotted as a function
of $\tilde{e}B$ and for fixed chemical potential $\mu=250$ MeV.
Small oscillations for small value of $\tilde{e}B$ arise from the
well-known van Alfven--de Haas (vAdH) effect \cite{alfven}. They
occur when the Landau levels pass the quark Fermi surface. They are
also observed in \cite{inagaki2003} for the $\chi$SB mass gaps. Note
that the oscillations are sharper, the smaller the value of the free
parameter $A$ in (\ref{B3}) is chosen [see also \cite{ruggieri2009}
for a discussion on the effect of free parameters in smooth cutoff
functions (form factors)].\footnote{We have also checked our results
for $A=0.001\Lambda$ (quasi-sharp cutoff), where instead of small
oscillations, small discontinuities appear in the regime
$\tilde{e}B\lesssim 0.4$ GeV$^{2}$.} As for $\tilde{e}B\gtrsim 0.45$
GeV$^2$, where the dependence of $\sigma_{B}$ on $\tilde{e}B$ is
almost linear, we enter in the regime of LLL dominance. The
qualitative behavior of $\sigma_{B}$ as a function of $\tilde{e}B$
for strong magnetic fields can be checked by comparing our numerical
result from Fig. 1a with the analytical result for $\sigma_{B}$ from
(\ref{Y7-b}).\footnote{For our specific choice of $G_{S}$ and
$\Lambda$ from (\ref{B1}), the dimensionless coupling $0<g_{s}<1$.
On the other hand, since no mixed phase is assumed here,
$\sigma_{B}$ from (\ref{Y7-b}) in the regime $\mu<\sigma_{B}$ is the
only relevant mass gap that can be compared with $\sigma_{B}$
arising from our numerical calculation.} The latter is plotted in
Fig. 1b for the same interval of the magnetic field, i.e.
$\tilde{e}B\in\{0.45,1\}~\mbox{GeV}^2$. Similarly, in Fig. 2a, the
CSC mass gap $\Delta_{B}$ is plotted as a function of $\tilde{e}B$
for $\mu=460$ MeV. Same small vAdH oscillations appear for small
$\tilde{e}B\lesssim 0.47$ GeV$^{2}$. They are also observed in
\cite{warringa2007} and \cite{shovkovy2007} for the diquark in the
CFL superconducting phase. Small oscillations in Fig. 2a, end up in
a linear regime, that starts, as in the previous case, at
$\tilde{e}B\gtrsim 0.47$ GeV$^{2}$. The qualitative behavior of
$\Delta_{B}$ in this regime can be compared with the analytical
result (\ref{Y33-b}), that arises in the LLL approximation (Fig.
2b).
\begin{figure}[hbt]
\includegraphics[width=8cm,height=6cm]{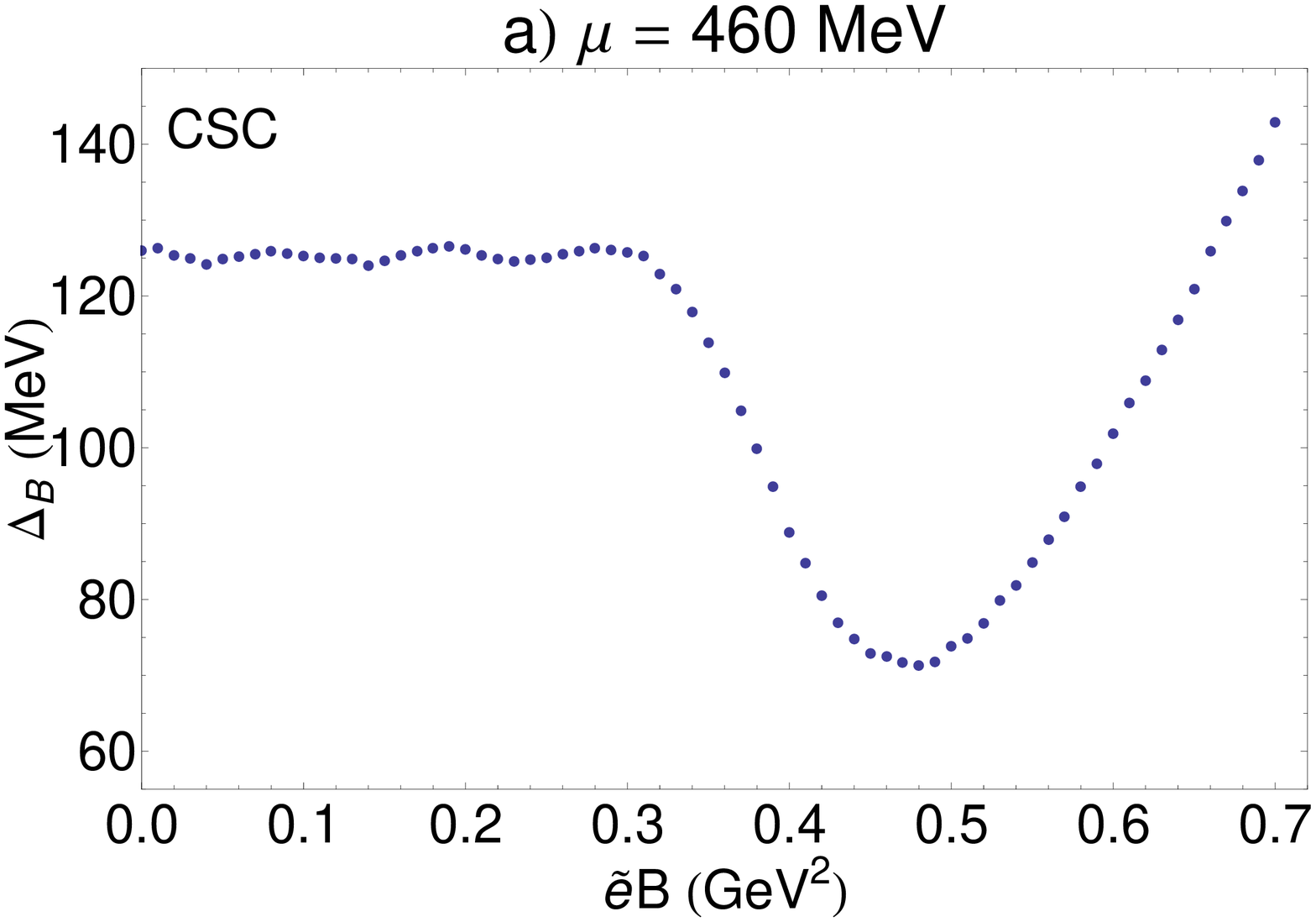}
\hspace{0.3cm}
\includegraphics[width=8cm,height=6cm]{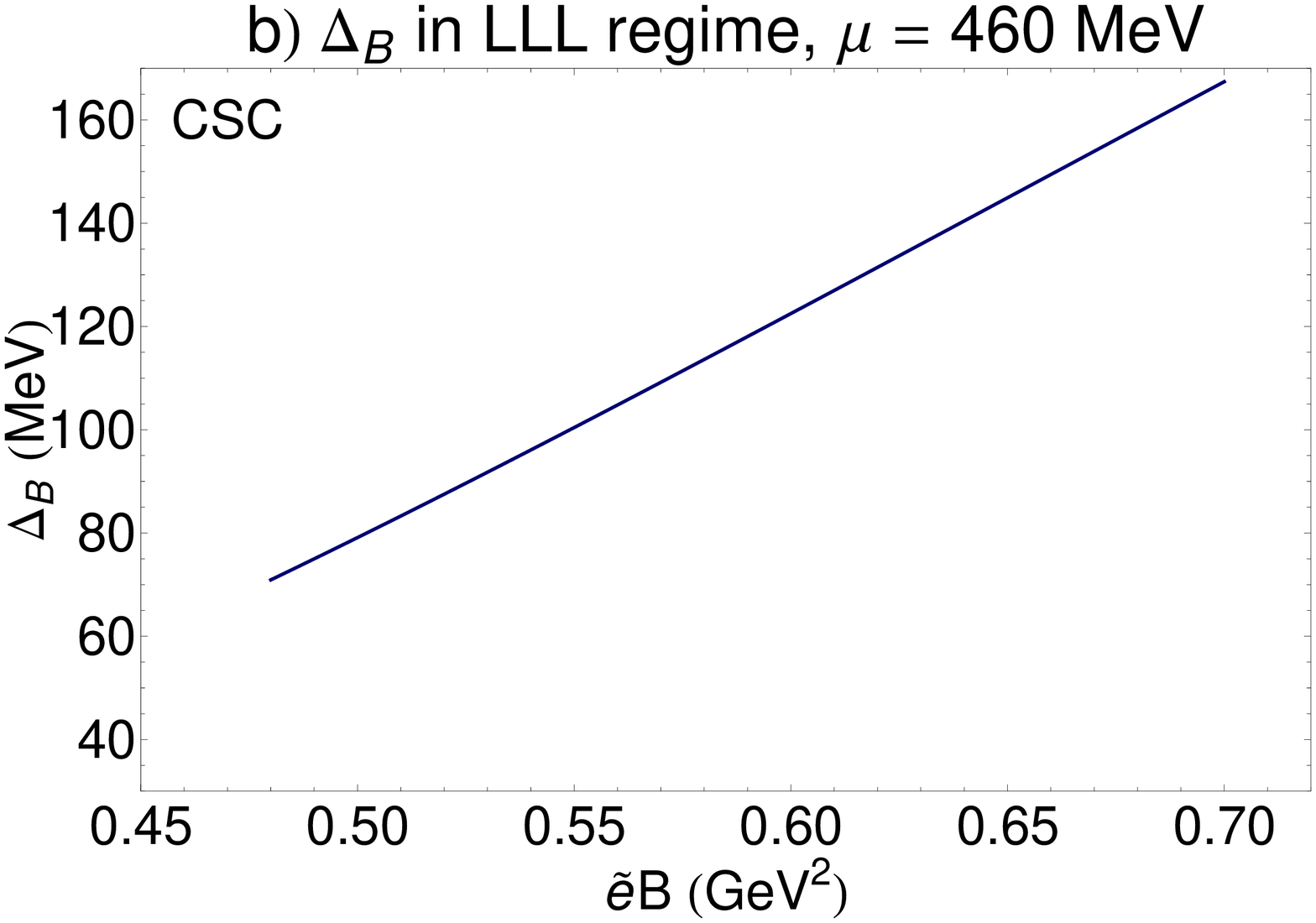}
\caption{a) The dependence of $\Delta_{B}$ on $\tilde{e}B$ in the
CSC phase for $\mu=460$ MeV. b) The analytical result of
$\Delta_{B}$ in the regime of LLL dominance from (\ref{Y33-b}) is
plotted for $\tilde{e}B\in\{0.47,1\}~\mbox{GeV}^2$.}
\end{figure}
\par\noindent
In Fig. 3, the dependence of the color chemical potential $\mu_{8}$
on $\tilde{e}B$ is plotted for $\mu=460$ MeV in the CSC phase. The
vAdH oscillations in Fig. 3 are similar to the oscillations of
$\mu_{8}$ in the regime of small magnetic fields that are observed
in \cite{warringa2007} in the superconducting CFL model.
\begin{figure}[htb]
\includegraphics[width=8cm,height=6cm]{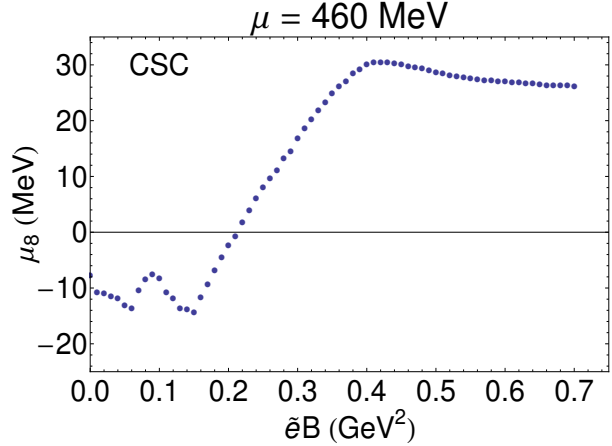}
\caption{The dependence of $\mu_8$ on $\tilde{e}B$ in the CSC phase
for $\mu=460$ MeV. The vAdH oscillations are similar to the
oscillations that are observed in \cite{warringa2007} in the
superconducting CFL model.}
\end{figure}
\par
In summary, comparing the above numerical results with our
analytical results from Sec. IV.A and IV.B for non-zero magnetic
field, it turns out that there exists a threshold magnetic field
$(\tilde{e}B)_{t}\simeq 0.45-0.50$ GeV$^{2}$, where the qualitative
behavior of our numerical results coincides with the qualitative
behavior of the analytical results for $\chi$SB and CSC mass gaps
$\sigma_{B}$ and $\Delta_{B}$.\footnote{Note that the similarity in
the numerical and analytical results for
$\tilde{e}B>(\tilde{e}B)_{t}$ is only qualitative. This is because
of various approximations that are carried out to determine the
analytical results [see Sec. IV for more details].} This regime, for
which the LLL approximation seems to be reliable, will be denoted
from now on by ``the linear regime''.\footnote{Note that in Figs. 2
and 3, the threshold magnetic field satisfies the requirement of LLL
approximation $(\tilde{e}B)_{t}\gg \mu^{2}$.}
\par
Using the above data, the magnetization of the 2SC superconducting
medium can be studied as a function of $\tilde{e}B$ and for fixed
chemical potential $\mu$. Fig. 4 shows the dependence of the product
of the magnetization ${\mathbf{M}}=M{\mathbf{e}}_{3}$ with $M\equiv
-\frac{\partial\Omega_{\mbox{\tiny{eff}}}^{(1)}}{\partial B}$ and
the rotated magnetic field $\tilde{\mathbf{B}}=B{\mathbf{e}}_{3}$ as
a function of $\tilde{e}B$ for two different chemical potential
$\mu=250$ MeV in the $\chi$SB regime and $\mu=460$ MeV in the CSC
regime (Figs. 4a and 4b). Here $\Omega_{\mbox{\tiny{eff}}}^{(1)}$ is
the one-loop part of the effective potential (\ref{A24}). For
simplicity, we will use the definition ${\mathbf{M}}\cdot
\tilde{\mathbf{B}}\equiv
-\tilde{e}B\frac{\partial\Omega_{\mbox{\tiny{eff}}}^{(1)}}{\partial
\tilde{e}B}$. Equivalently, one can define the magnetization by
introducing the Gibbs free energy density ${\cal{G}}$ in the
presence of a constant magnetic field $B$
\begin{eqnarray}\label{B3}
{\cal{G}}(\sigma,\Delta;B,\mu)=\frac{B^{2}}{2}+\Omega_{\mbox{\tiny{eff}}}^{(1)}(\sigma,\Delta;B,\mu)-H
B,
\end{eqnarray}
where $H$ is the external magnetic field \cite{shovkovy2007}.
Whereas in vacuum $H=B$, in a medium with finite magnetization
density, the external magnetic field $H$ is different from the
induced magnetic field $B$. Minimizing ${\cal{G}}$ with respect to
$B$ and evaluating the result at the minimum of the potential, we
get the well-known relation
${\mathbf{M}}={\mathbf{B}}-{\mathbf{H}}$, where ${\mathbf{M}}$ is
the magnetization. Note that the minimum of the potential in the
$\chi$SB phase is given by $(\sigma_{B}\neq 0, \Delta_B=\mu_8=0)$
and in the CSC phase by $(\sigma_{B}=0, \Delta_{B}\neq 0,
\mu_{8}\neq 0)$. The magnetization of the superconducting CFL phase
is studied as a function of $eB/\mu^{2}$ for $\mu=500$ MeV in
\cite{shovkovy2007}, where the same vAdH oscillations as appears in
Fig. 4 are observed.
\begin{figure}[hbt]
\includegraphics[width=8cm,height=6cm]{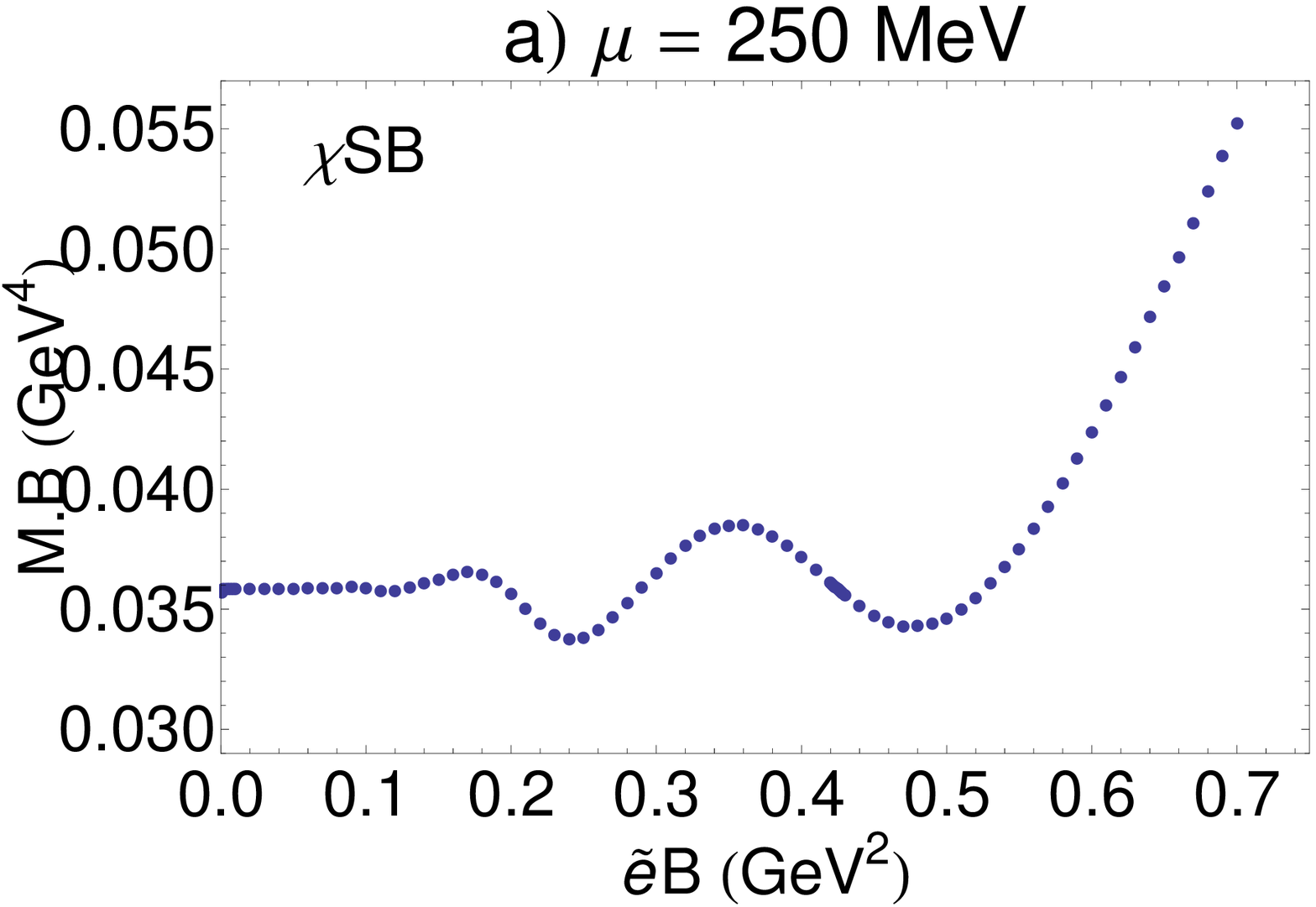}
\hspace{0.3cm}
\includegraphics[width=8cm,height=6cm]{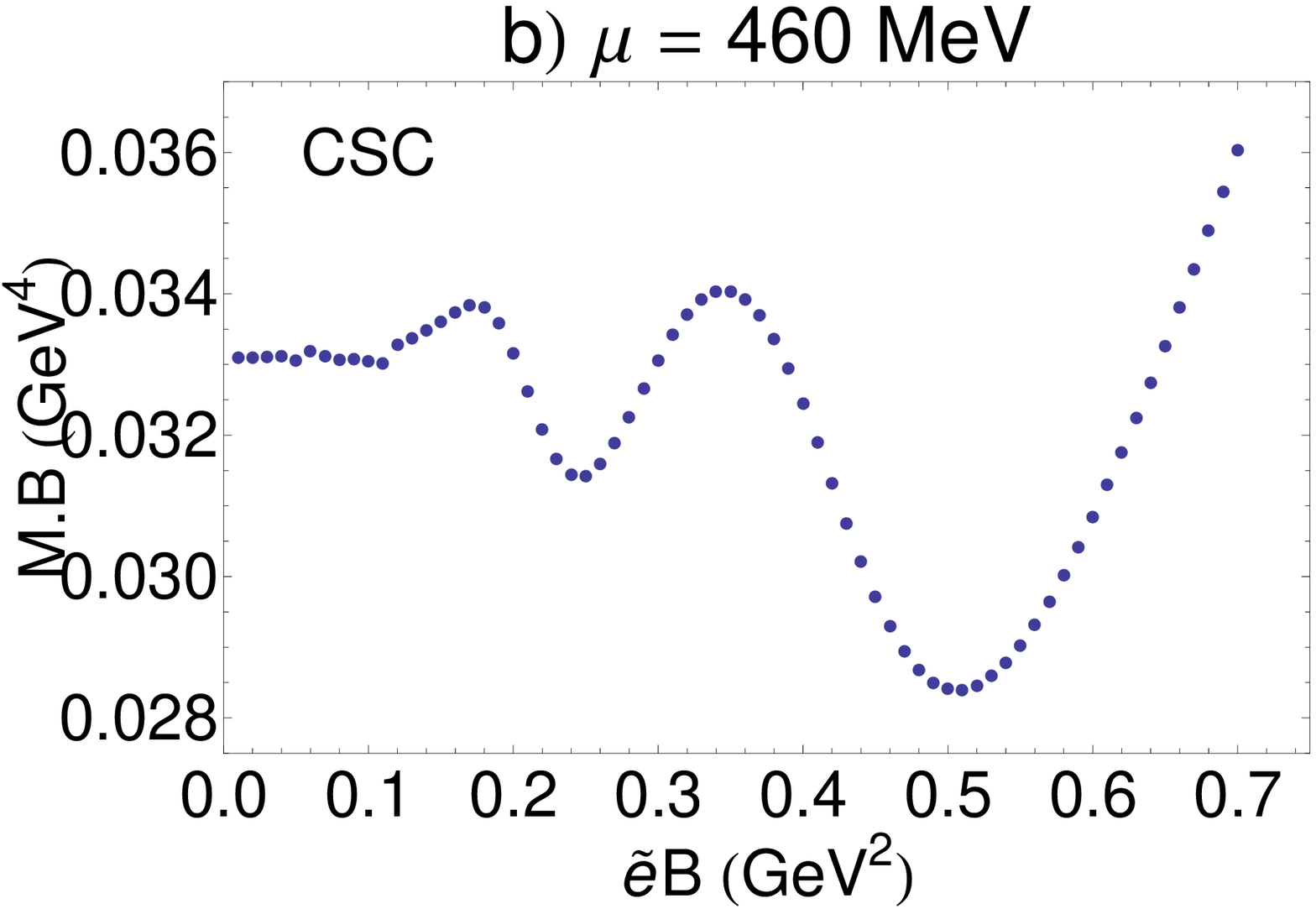}
\caption{The dependence of the product
${\mathbf{M}}\cdot\tilde{\mathbf{B}}$ on the magnetic field of fixed
chemical potential a) $\mu=250$ MeV in the $\chi$SB phase and b)
$\mu=460$ MeV in the CSC phase. The linear regime in both phases
starts at ${\tilde{e}}B\simeq 0.45-0.50$ GeV$^{2}$.}
\end{figure}
\par
In what follows, we will first study the $\mu$-dependence of
$\sigma_{B}$ and $\Delta_{B}$. We then present the phase diagram
$\mu_{c}-\tilde{e}B$ of the 2SC quark matter at zero temperature.
Let us start with the case of zero magnetic field. In Fig. 5, the
$\mu$-dependence of $\sigma_{0}$ in the $\chi$SB phase, as well as
$\Delta_{0}$ and $\mu_{8}$ in the CSC phase are plotted for zero
magnetic field. For our specific choice of free parameters $\Lambda,
G_{S}$ and $G_{D}$, $\sigma_{0}=323.8$ MeV for $\mu\leq \mu_{c}$.
Here, the critical chemical potential $\mu_{c}=325$ MeV and the
value of $\Delta_{0}$ for $\mu\simeq \mu_{c}$ is $\Delta_{0}=78.0$
MeV. Our results coincides qualitatively with the numerical results
presented in \cite{ebert2005} (see also \cite{ebert2010} for a
recent investigation of Cooper-pairing in NJL-type
models).\footnote{In \cite{ebert2005}, the quark mass $m_0\neq 0$
and therefore a mixed phase appears for $\mu\geq 340$ MeV.}
\begin{figure}[hbt]
\includegraphics[width=8cm,height=6cm]{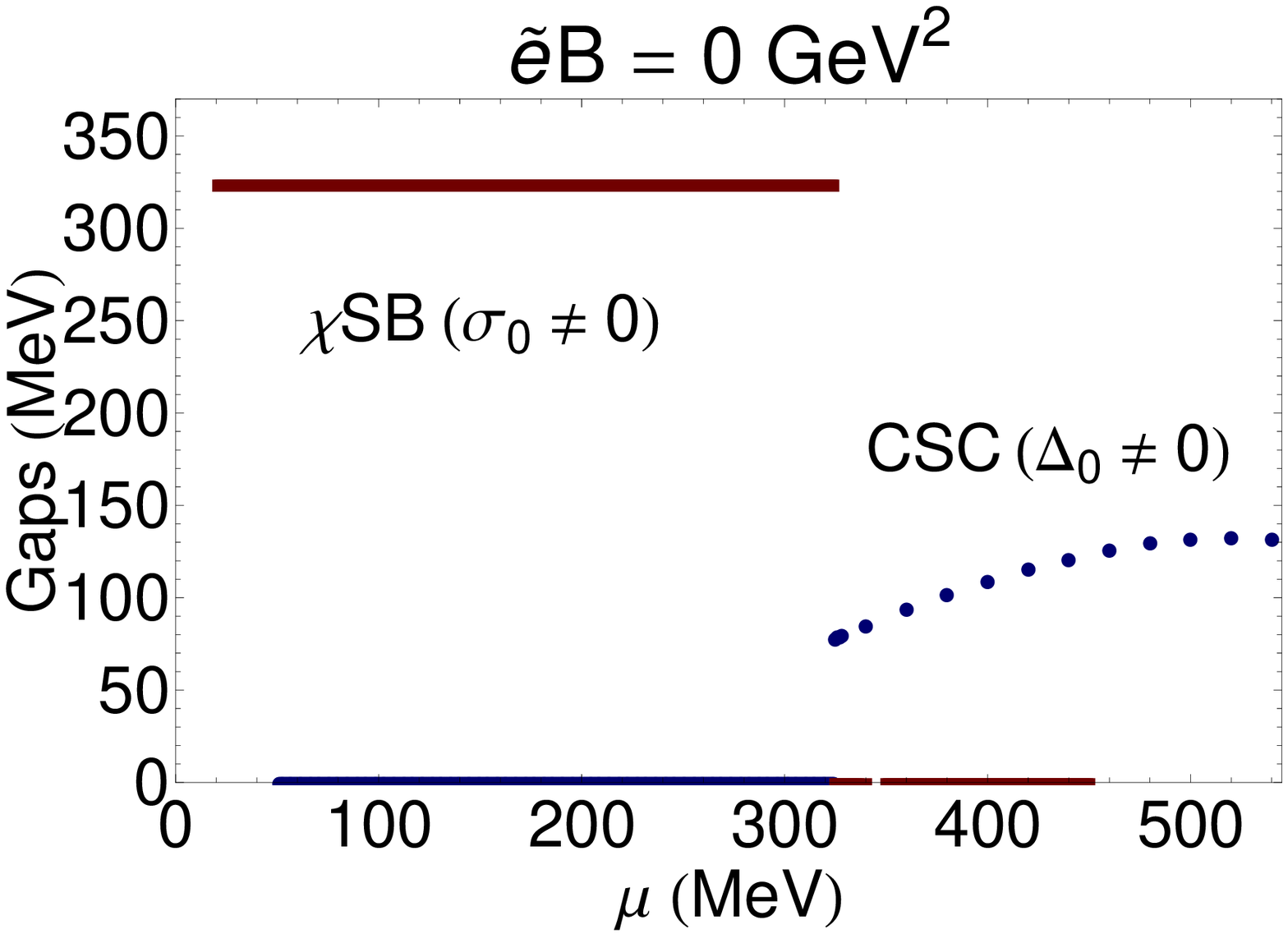}
\hspace{0.3cm}
\includegraphics[width=8cm,height=6cm]{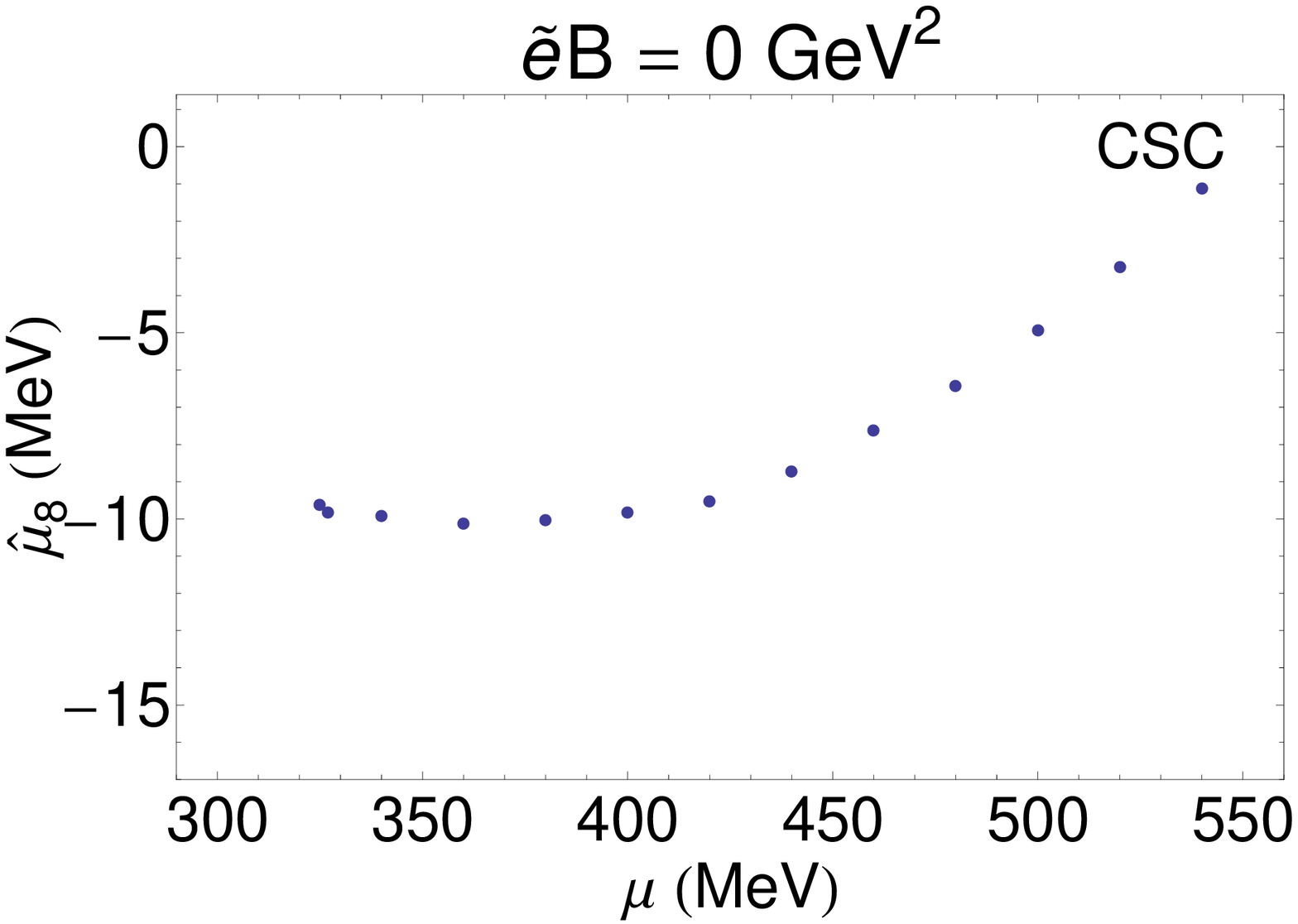}
\caption{The $\mu$-dependence of $\sigma_{0}$ in the $\chi$SB phase,
and $\Delta_{0}$ in the CSC phase for $\tilde{e}B=0$ (left panel).
The $\mu$-dependence of $\mu_{8}$ for $\tilde{e}B=0$ (right panel).}
\end{figure}
\par\noindent
We can compare the $\mu$-dependence of $\Delta_{0}^{2}$ arising from
our numerical calculation with the relation (\ref{Y39-b}) arising
from our analytical results for vanishing magnetic field. To do this
we have fitted our numerical data with a function
\begin{eqnarray}\label{B4}
\Delta_{0}^{2}(\mu)=(a-b
\mu^{2})\exp\left(-\frac{c}{\mu^{2}}\right),
\end{eqnarray}
similar to (\ref{Y39-b}). Here, $a,b$ and $c$ are free parameters.
The numerical values of these parameters arising from our fit are in
good agreement with the expected analytical values arising from
(\ref{Y39-b}) (see Fig. 6 and Table II). This can be quantified by
defining
\begin{eqnarray}\label{eta}
\eta\equiv\bigg|\frac{\mbox{Analytical value $-$ Numerical
value}}{\mbox{(Analytical value $+$ Numerical value)/2}}\bigg|,
\end{eqnarray}
as a measure for the variation of the numerical value with respect
to the average of analytical and numerical values. In Table II,
$\eta_{a}, \eta_{b}$ and $\eta_{c}$ are less than $50$\%. Note that
the difference between the analytical and numerical values of $a,b$
and $c$ lies on the approximations that are made to determine
analytically $\Delta_{0}^{2}$ in (\ref{Y39-b}).
\begin{figure}[hbt]
\includegraphics[width=8cm,height=6cm]{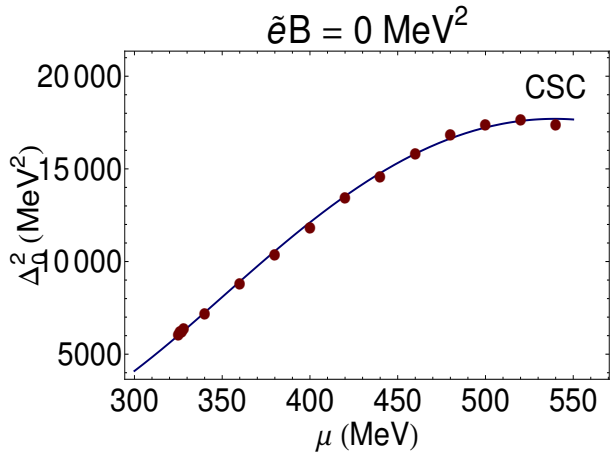}
\caption{The dots are the numerical values of $\Delta_{0}^{2}$. The
solid line is the corresponding fit of $\Delta_{0}^{2}(\mu)$ from
(\ref{B4}). The fit parameters $a,b$ and $c$ are listed in Table II.
The regression parameter $R^{2}$, as a measure of reliability of the
numerical fit, is in this case $R^{2}=0.999852$.  }
\end{figure}
\begin{table}
\begin{tabular}{| c || c| c |c|| c| c |c||c|c|c|}
\hline &\multicolumn{3}{|c||}{\textbf{Analytical
parameters}}&\multicolumn{3}{c||}{\textbf{Numerical fit parameters}}
&\multicolumn{3}{c|}{\textbf{$\eta$ in \%}}\\
\hline $\tilde{e}B$ (GeV$^{2}$) & $a$ (MeV$^2$)&$ b$ &$c$ (MeV$^2$)& $a$ (MeV$^2$)& $b$ &$c$ (MeV$^2$)&$\eta_{a}$&$
\eta_{b}$&$\eta_{c}$\\
\hline\hline
$0$ & $8.49\times10^{4}$ & $0.19$ & $2.29\times10^{5}$ & $7.84\times10^{4}$ & $0.13$ & $2.52\times10^{5}$&$8$&$38$&$10$ \\
\hline
\end{tabular}
\caption{Numerical fit data for $\Delta_{0}^{2}(\mu)$ from
(\ref{B4}). The numerical values of the parameters arising from our
fit are in good agreement with the expected analytical values
arising from (\ref{Y39-b}) [see $\eta_{a}, \eta_{b}$ and $\eta_{c}$
with $\eta$ defined in (\ref{eta})].}
\end{table}
\par
Let us now concentrate on the case of non-vanishing magnetic field.
In Table III, we have summarized our numerical results for critical
chemical potential $\mu_{c}$, the mass gap $\sigma_{B}$ for $\mu\leq
\mu_{c}$ and the 2SC gap $\Delta_{B}$ at $\mu\simeq \mu_{c}$. The
critical chemical potential $\mu_{c}$ and the $\chi$SB mass gaps
$\sigma_{B}(\mu\leq \mu_{c})$ increase by increasing the external
magnetic field. In the vicinity of the phase transition from
$\chi$SB to CSC phase, the CSC mass gap $\Delta_{B}(\mu\simeq
\mu_{c})$ also increases by increasing the magnetic field.
\begin{table}[hbt]
\begin{tabular}{|c|c|c|c|c|}\hline
$\tilde{e}B$ in GeV$^{2}$&$\mu_{c}$ in
MeV&$\sigma_{B}(\mu\leq\mu_{c})$ in MeV&$\Delta_{B}(\mu\simeq
\mu_{c})$ in
MeV\\
\hline\hline
 $0.002$&$316.0$&$312.0$&$70.0$\\ \hline
 $0.005$&$287.0$&$311.0$&$61.0$\\ \hline
 $0.01$&$324.2$&$322.4$&$78.0$\\ \hline
 $0.04$&$322.3$&$326.4$&$77.6$\\ \hline
 $0.05$&$321.9$&$327.8$&$77.9$\\ \hline
 $0.10$&$311.3$&$338.4$&$76.9$\\ \hline
 $0.20$&$296.0$&$366.0$&$50.8$\\ \hline
 $0.30$&$315.0$&$420.7$&$73.2$\\ \hline
 $0.40$&$329.0$&$464.8$&$80.4$\\ \hline
 $0.46$&$329.6$&$477.0$&$82.6$\\ \hline
 $0.48$&$334.0$&$487.0$&$85.7$\\ \hline
 $0.50$&$340.8$&$499.0$&$88.8$\\ \hline
 $0.60$&$392.4$&$593.0$&$116.9$\\ \hline
 $0.70$&$450.7$&$700.0$&$147.7$\\ \hline
\end{tabular}
\caption{Numerical results for critical chemical potential
$\mu_{c}$, the mass gap $\sigma_{B}$ for $\mu\leq \mu_{c}$ and the
2SC gap $\Delta_{B}$ at $\mu\simeq \mu_{c}$. }
\end{table}
\begin{figure}[t]
\includegraphics[width=5cm,height=4cm]{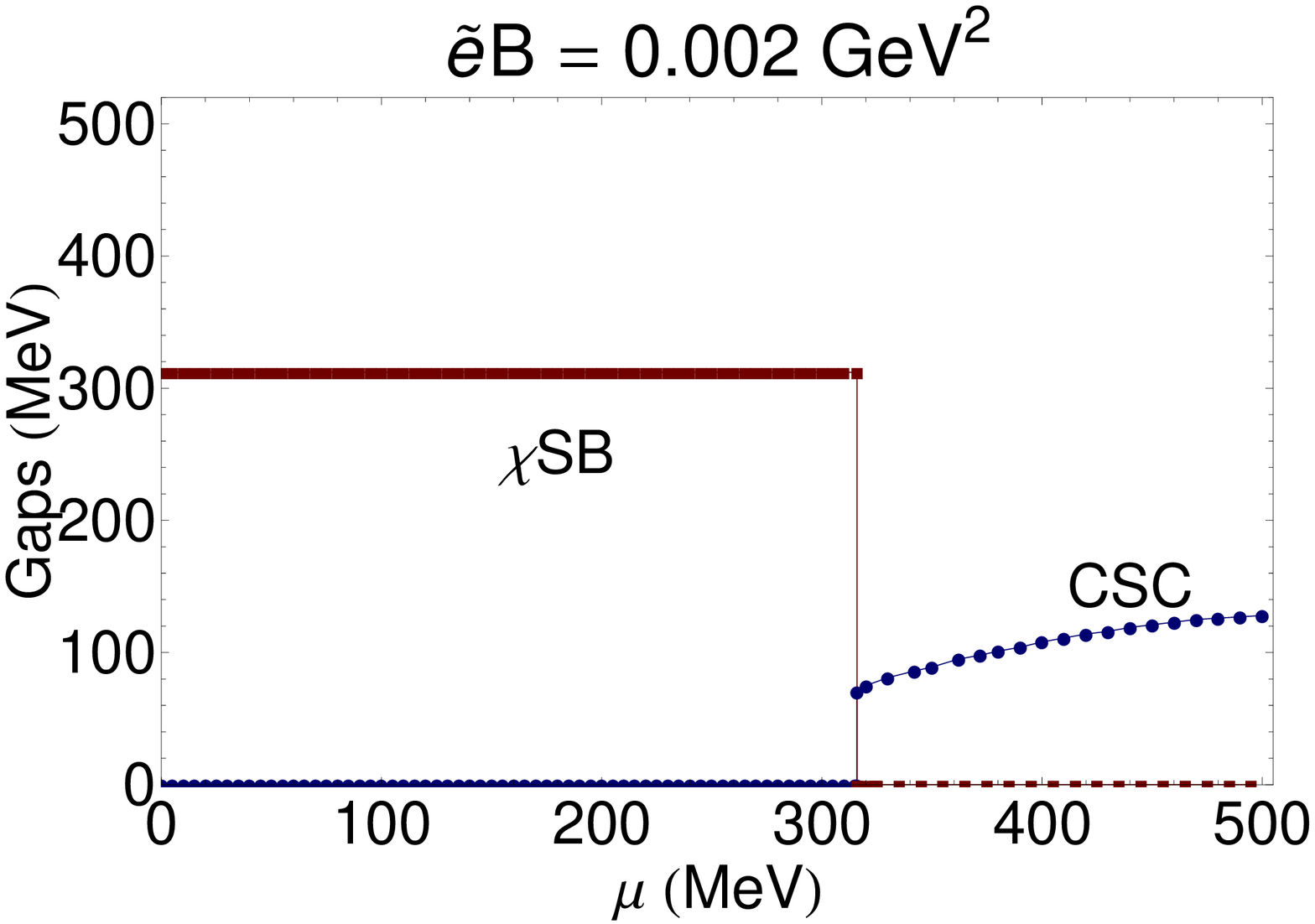}
\hspace{0.2cm}
\includegraphics[width=5cm,height=4cm]{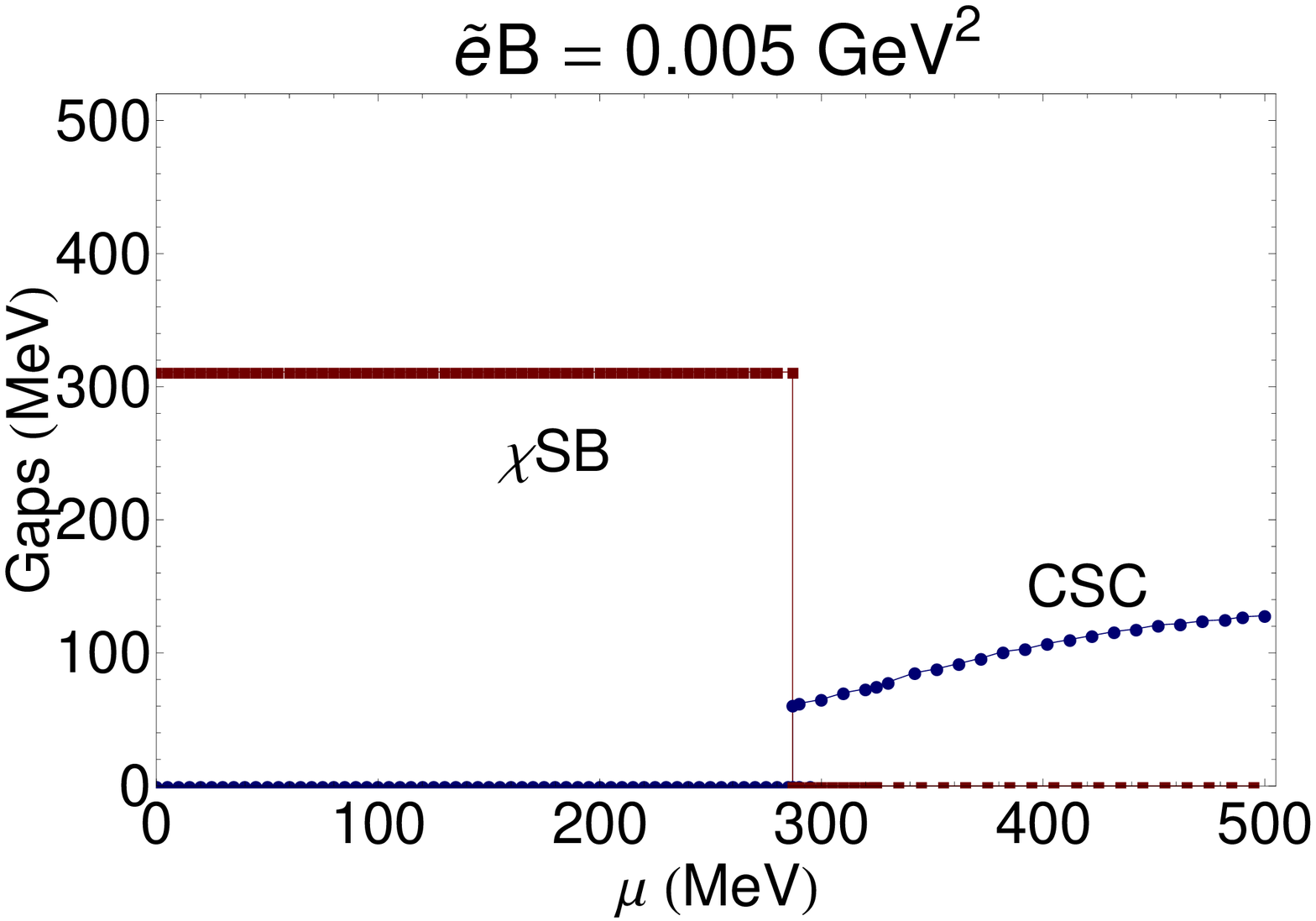}
\hspace{0.2cm}
\includegraphics[width=5cm,height=4cm]{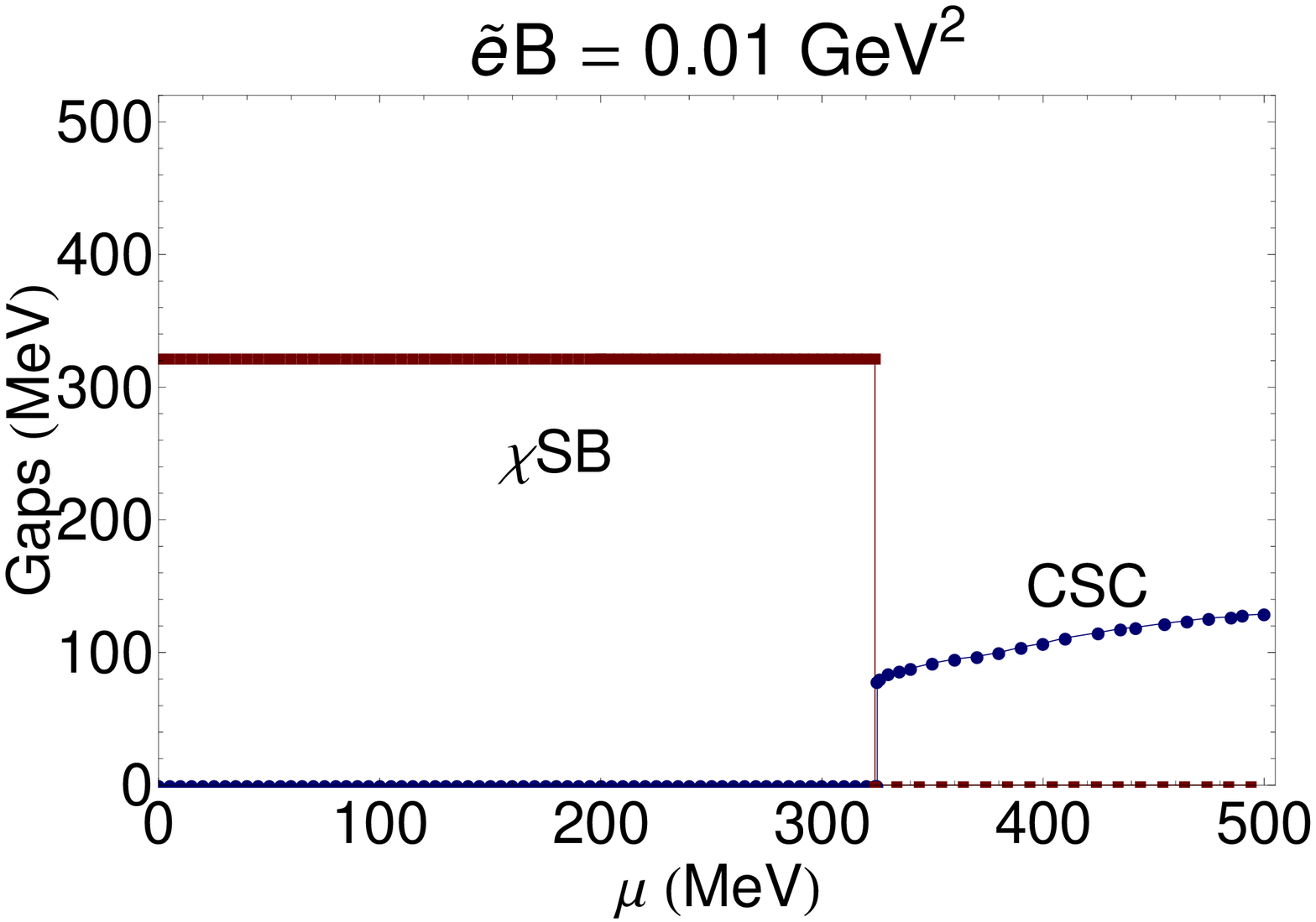}
\par\vspace{0.5cm}
\includegraphics[width=5cm,height=4cm]{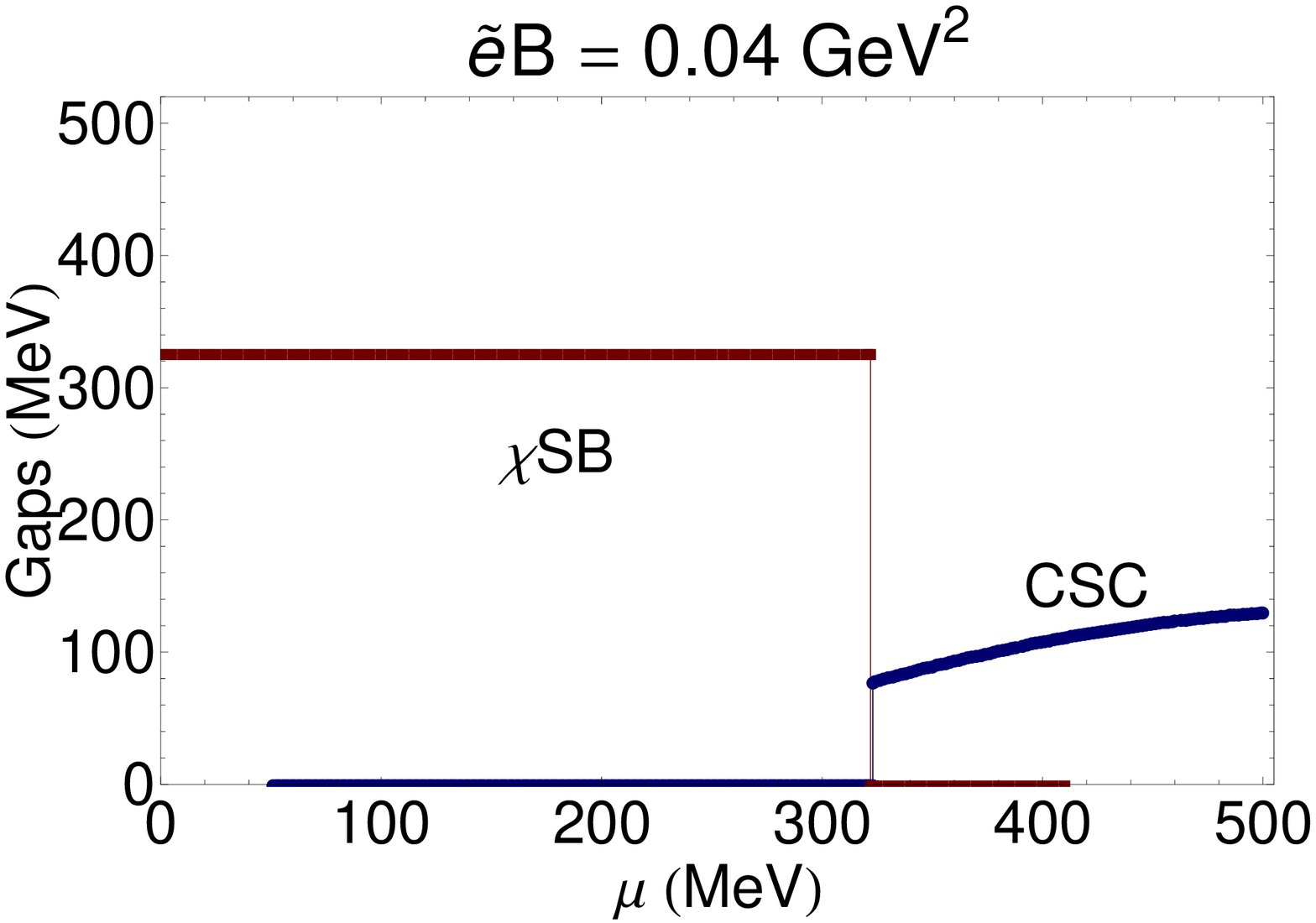}
\hspace{0.2cm}
\includegraphics[width=5cm,height=4cm]{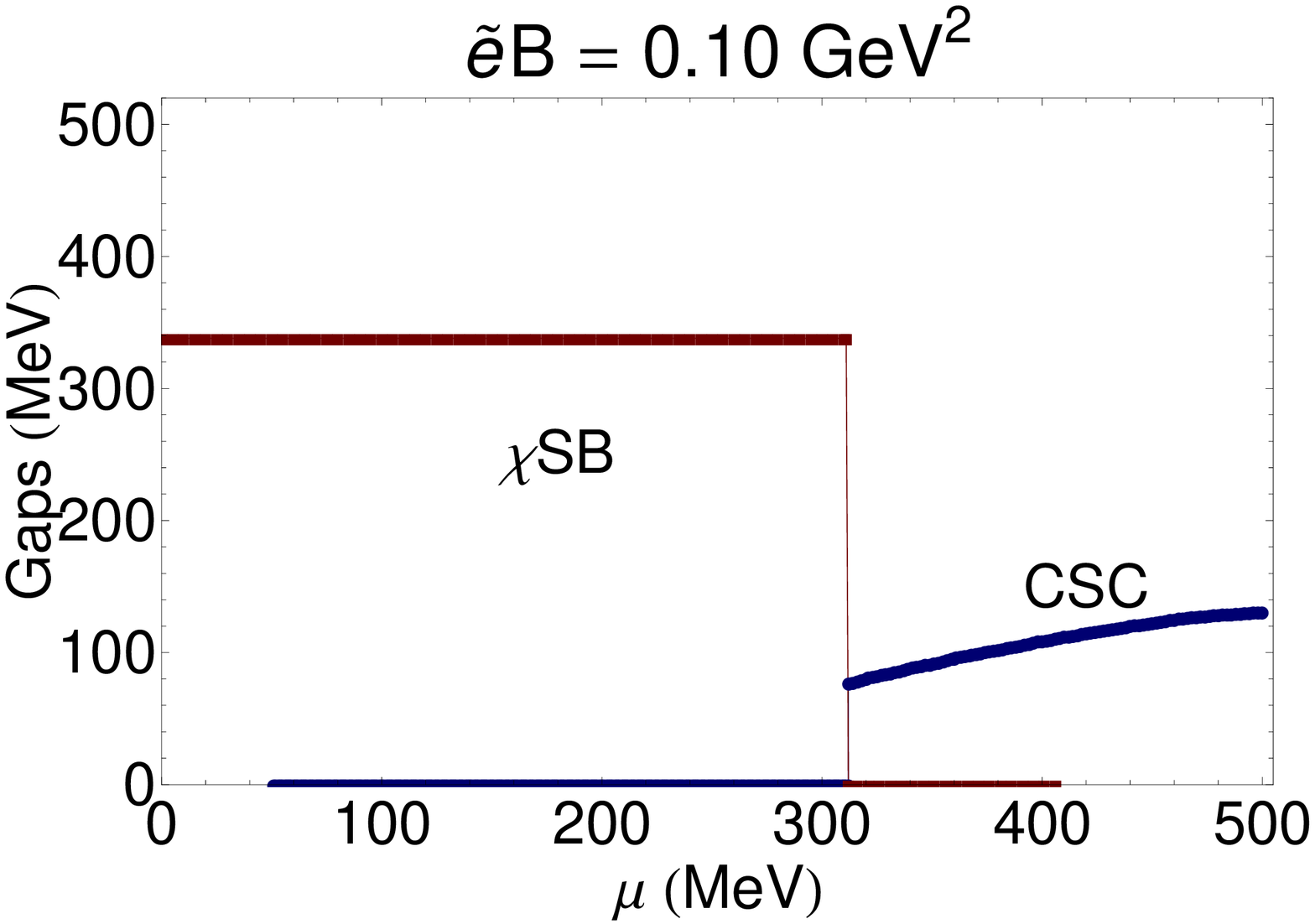}
\hspace{0.2cm}
\includegraphics[width=5cm,height=4cm]{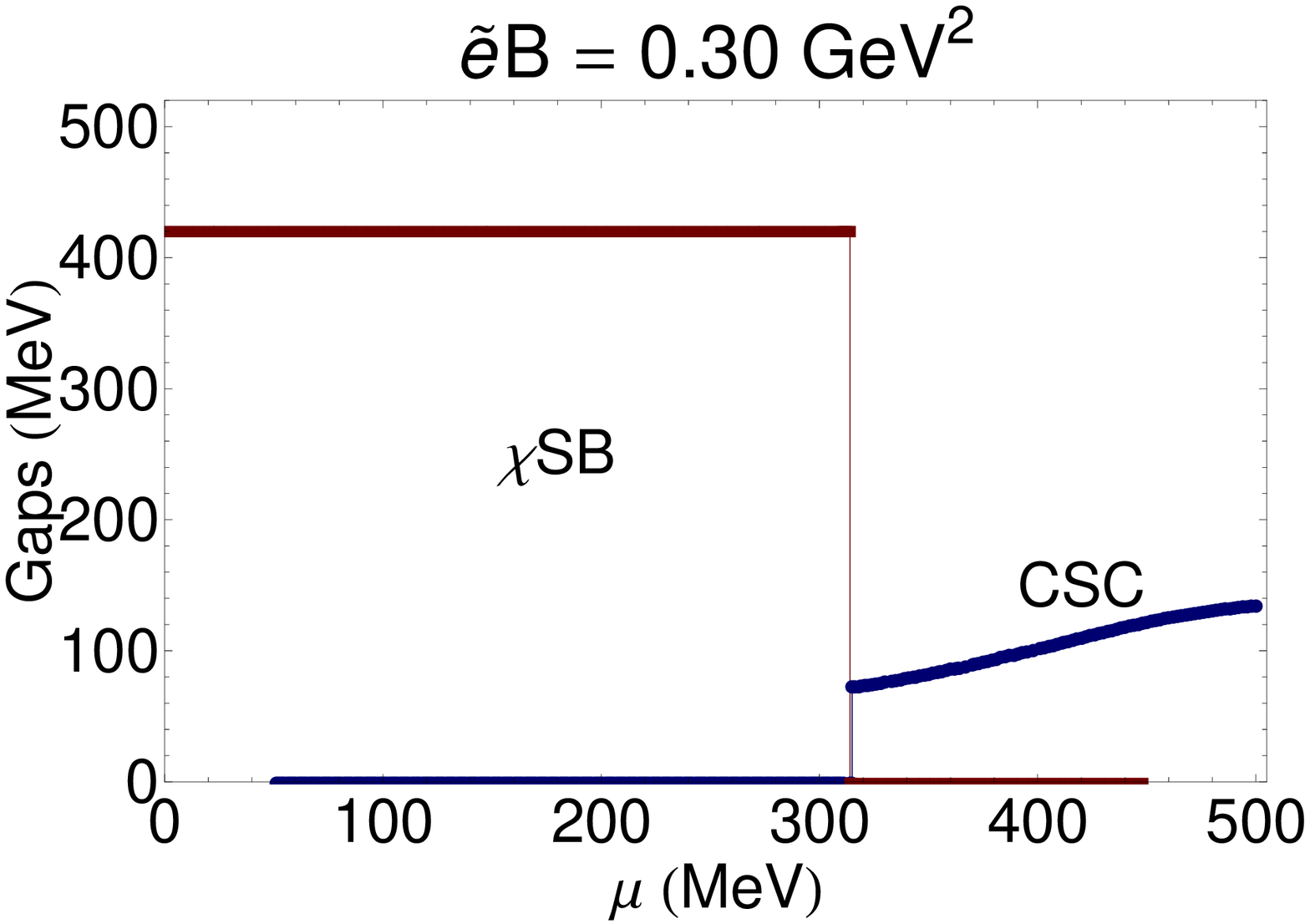}
\par\vspace{0.5cm}
\includegraphics[width=5cm,height=4cm]{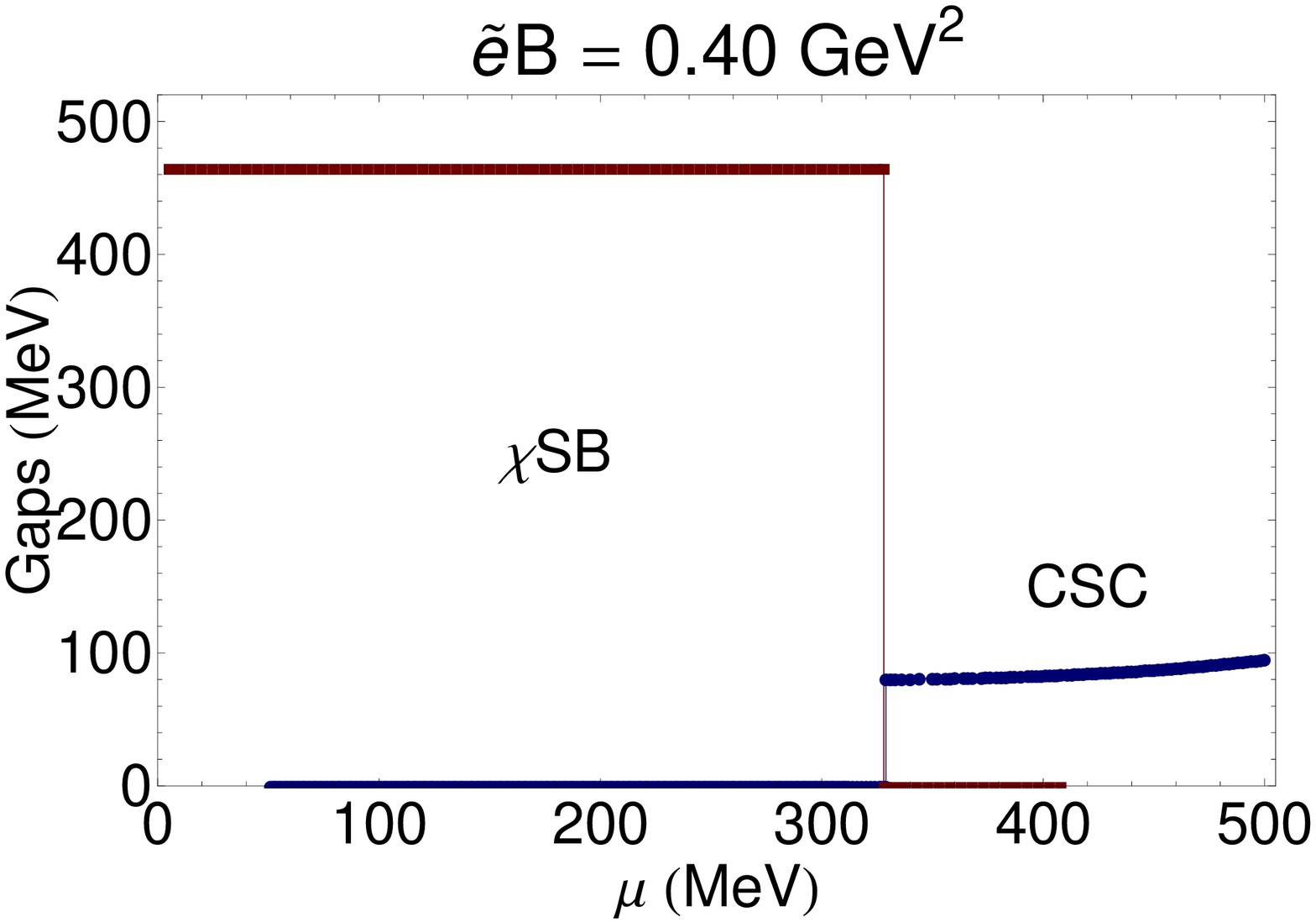}
\hspace{0.2cm}
\includegraphics[width=5cm,height=4cm]{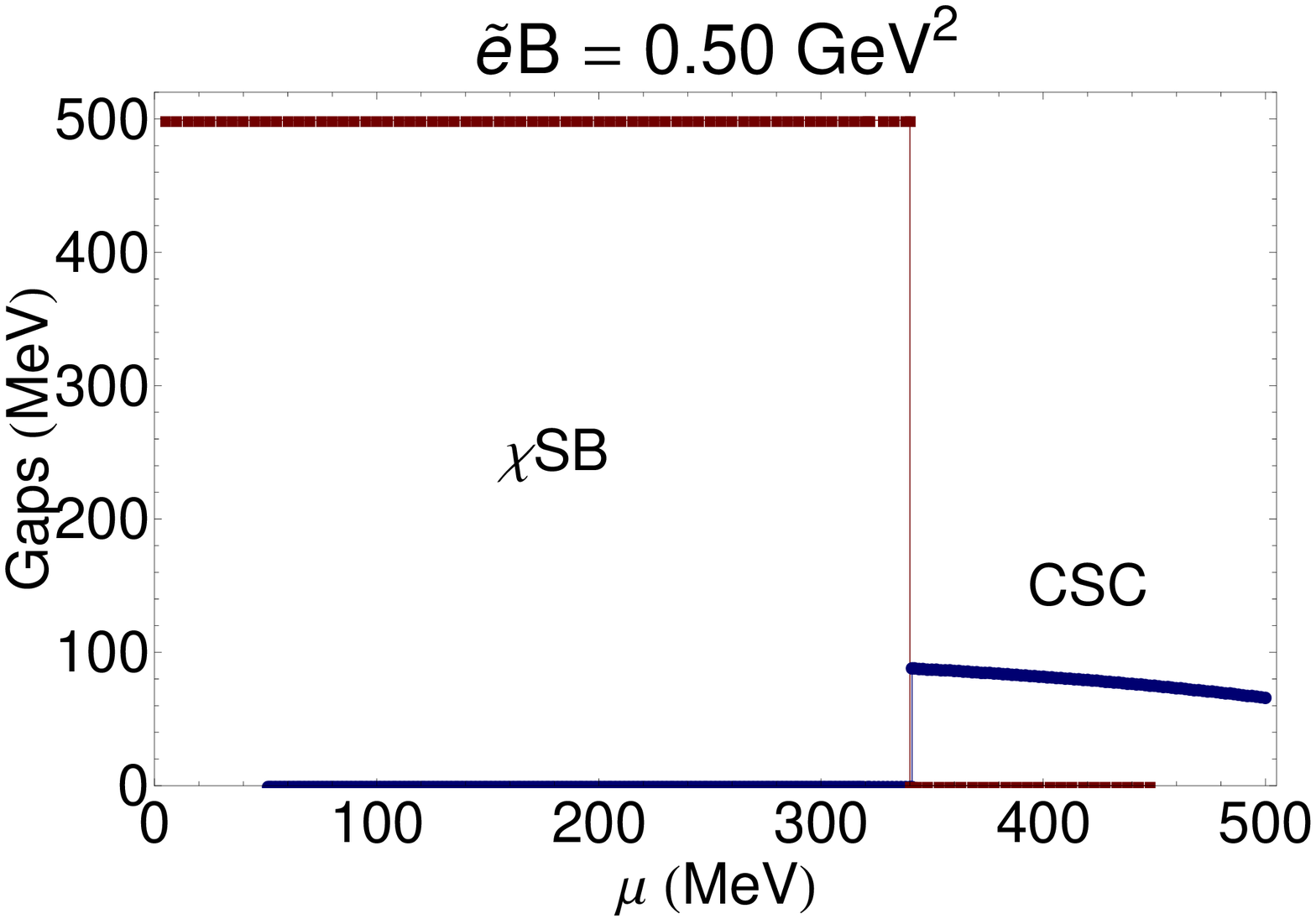}
\hspace{0.2cm}
\includegraphics[width=5cm,height=4cm]{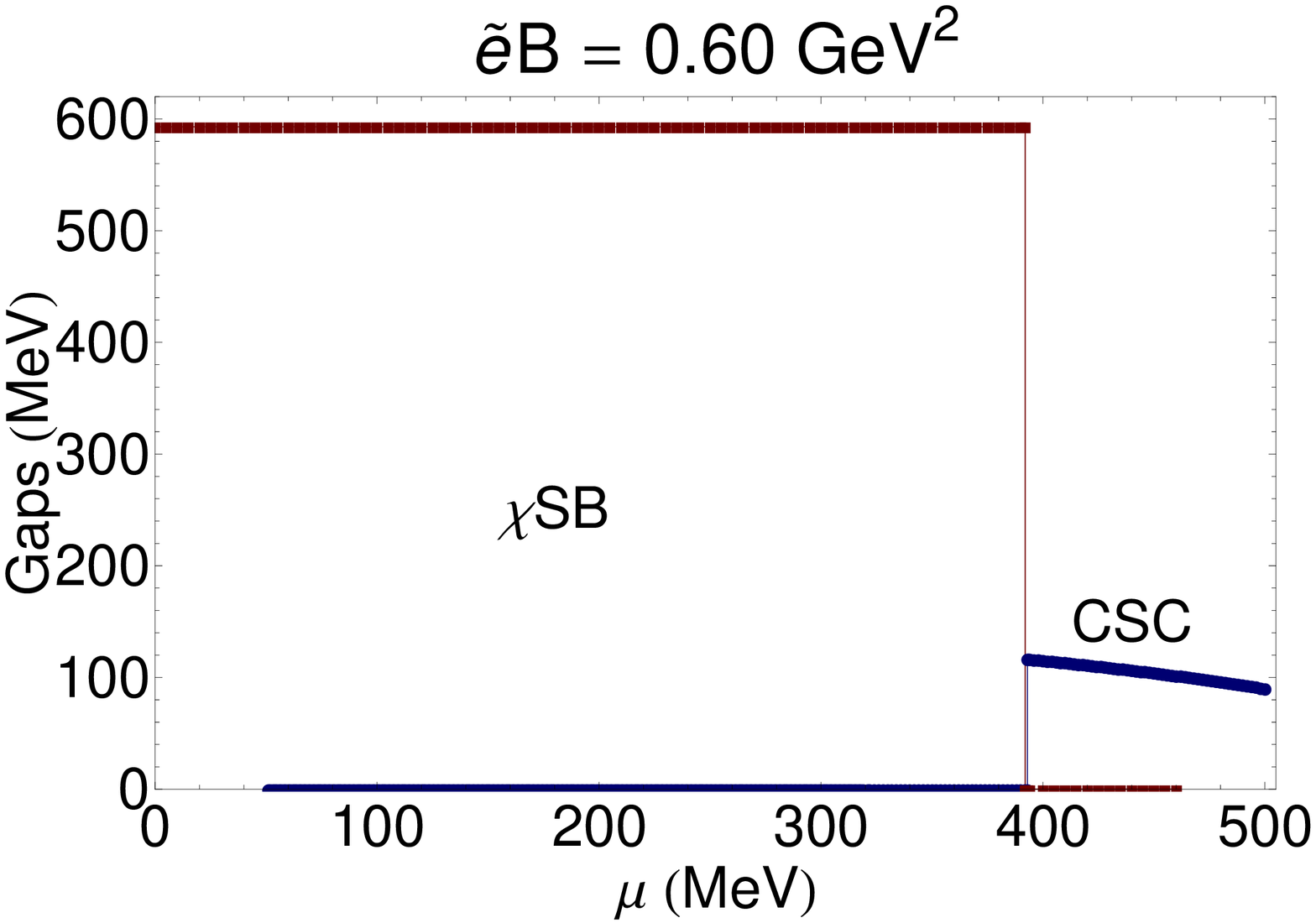}
 \caption{The $\mu$-dependence of $\sigma_{B}$ in the
$\chi$SB phase (red lines), and $\Delta_{B}$ in the CSC phase (blue
lines) for different values of $\tilde{e}B$. The $\chi$SB mass gap
$\sigma_{B}(\mu,\tilde{e}B)$ is constant in $\mu\leq \mu_{c}$ and
increases for increasing $\tilde{e}B$. The critical chemical
potential $\mu_{c}$ increases for increasing $\tilde{e}B$. For our
specific choice of parameters ($m_{0}=0,G_{D}<G_{S}$) no mixed phase
appears. The CSC mass gap $\Delta_{B}$ exists therefore only at
$\mu>\mu_{c}$. The slopes of the curves appearing at $\mu>\mu_{c}$
are decreasing for increasing $\tilde{e}B$. The first order nature
of the phase transition between $\chi$SB and CSC phases is visible.
The dependence of the $\chi$SB and CSC gaps for vanishing magnetic
field is also considered here to have a comparison with the
$\mu$-dependence of the gaps for non-vanishing $\tilde{e}B$.}
\end{figure}
\par\noindent
The $\mu$-dependence of $\sigma_{B}$ and $\Delta_{B}$ are presented
also in Fig. 7. There is a first order phase transition from the
$\chi$SB to the CSC phase [see also Fig. 9 for more detail on the
phase structure in $\mu_{c}-\tilde{e}B$ plane]. Because of our
specific choice $m_{0}=0$ and $G_{D}<G_{S}$, no mixed broken phase
appears at $\mu>\mu_{c}$ \cite{huang2002}, and the $\chi$SB mass gap
$\sigma_{B}(\mu)$ is constant for $\mu\leq \mu_{c}$. For small value
of $\tilde{e}B$, the CSC mass gap $\Delta_{B}(\mu,\tilde{e}B)$, is
increasing with $\mu$. The magnetic field enhances the chiral
symmetry breaking. This is known as the phenomenon of magnetic
catalysis \cite{miransky1995}, which is also observed in
\cite{inagaki2003}. In the linear regime, i.e. for
$\tilde{e}B\gtrsim 0.45$ GeV$^{2}$, $\Delta_{B}$ is decreasing with
$\mu$. To study the linear regime in detail, we have fitted our
numerical data for $\Delta_{B}$ as a function of $\mu$ and fixed
$\tilde{e}B$, with a function similar to (\ref{Y33-b})
\begin{eqnarray}\label{B5}
\Delta_{B}^{2}=(a-b \mu^{2}),
\end{eqnarray}
where $a$ and $b$ are free parameters, that depend on $\tilde{e}B$.
In Table IV, we have compared the expected analytical results for
the parameters $a$ and $b$, with the corresponding results from
fitting our numerical data with (\ref{B5}) for different
$\tilde{e}B$. For $\tilde{e}B\gtrsim 0.44$ GeV$^{2}$, $\eta_{a}$ and
$\eta_{b}$ are less than $50$\%.
\begin{table}
\begin{tabular}{| c || c| c || c| c ||c|c|}
\hline &\multicolumn{2}{|c||}{\textbf{Analytical
parameters}}&\multicolumn{2}{c||}{\textbf{Numerical fit parameters}}
&\multicolumn{2}{|c|}{\textbf{$\eta$ in \%}}\\
\hline
$\tilde{e}B$ (GeV$^{2})$ & $a$ (MeV$^2$)& $b$ & $a$ (MeV$^2$)& $b$&$\eta_{a}$&$\eta_{b}$ \\
\hline\hline
$0.04$ & $5.27\times10^{-24}$ & $1.32\times10^{-28}$ & $-1.06\times10^{3}$ & $-0.077$&$200$&$200$ \\
\hline
$0.10$ & $1.62\times10^{-6}$ & $1.62\times10^{-11}$ & $-8.37\times10^{2}$ & $-0.077$&$200$&$200$ \\
\hline
$0.30$ & $1.91\times10^{2}$ & $6.37\times10^{-4}$ & $-4.53\times10^{3}$ & $-0.094$&$218$&$201$\\
\hline
$0.40$ & $2.27\times10^{3}$ & $0.006$ & $+4.17\times10^{3}$ & $-0.018$&$59$&$400$\\
\hline
$0.44$ & $4.53\times10^{3}$ & $0.010$ & $+7.38\times10^{3}$ & $+0.008$&$48$&$22$ \\
\hline
$0.46$ & $6.14\times10^{3}$ & $0.013$ & $+8.47\times10^{3}$ & $+0.015$&$32$&$14$ \\
\hline
$0.50$ & $1.05\times10^{4}$ & $0.021$ & $+1.09\times10^{4}$ & $+0.025$&$4$&$17$ \\
\hline
$0.60$ & $3.03\times10^{4}$ & $0.050$ & $+2.26\times10^{4}$ & $+0.058$&$29$&$15$ \\
\hline
\end{tabular}
\caption{Numerical fit data for $\Delta_{B}^{2}$ as a function of
$\mu$ from (\ref{B5}). In the linear regime, i.e. for
$\tilde{e}B\gtrsim 0.45$ GeV$^{2}$, the numerical values of the
parameters arising from our fit are in good agreement with the
expected analytical values of the parameters from (\ref{Y33-b}) [see
$\eta_{a}$ and $\eta_{b}$ with $\eta$ defined in (\ref{eta})].}
\end{table}
\par\noindent
\begin{figure}[hbt]
\includegraphics[width=5cm,height=4cm]{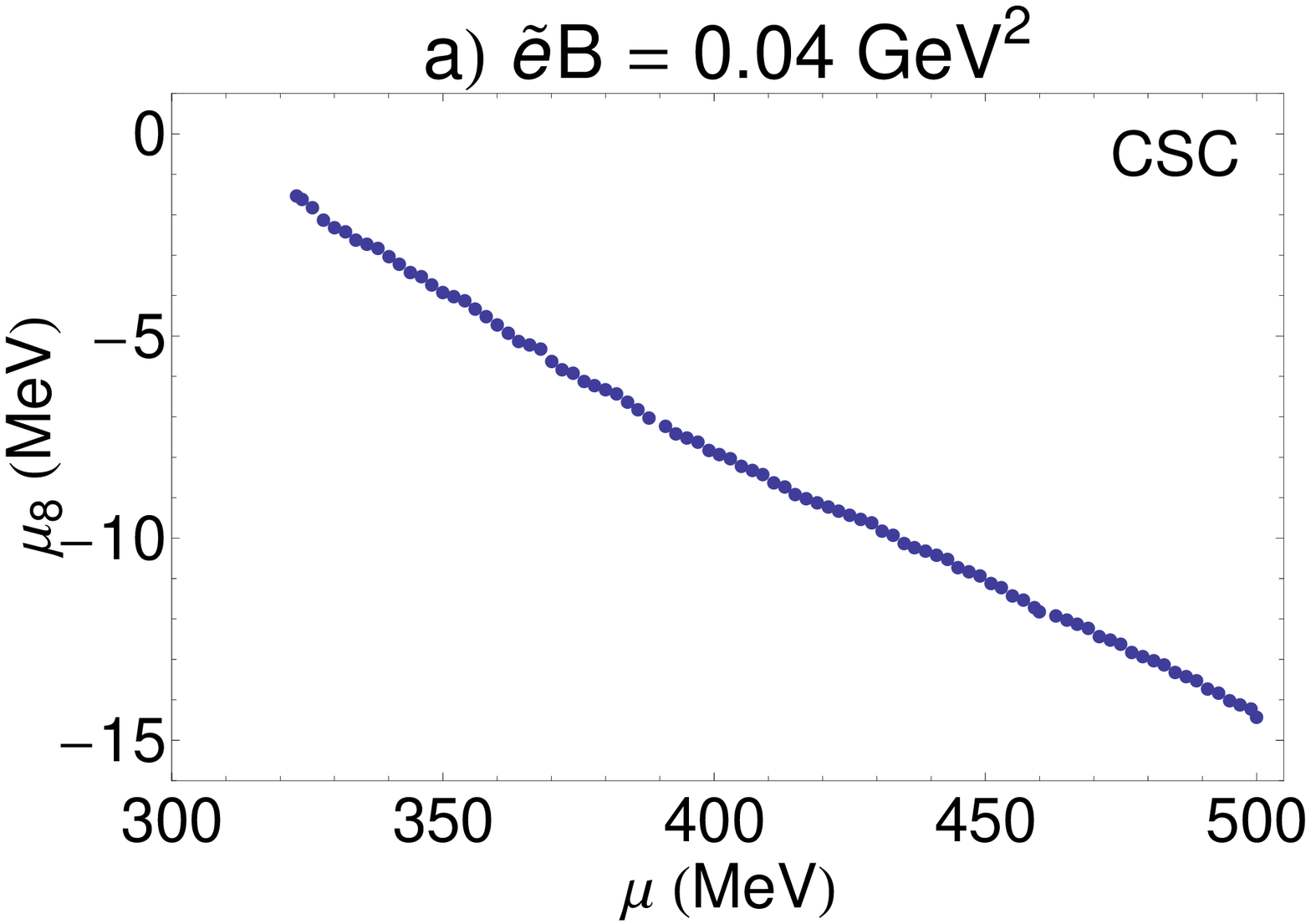}
\hspace{0.2cm}
\includegraphics[width=5cm,height=4cm]{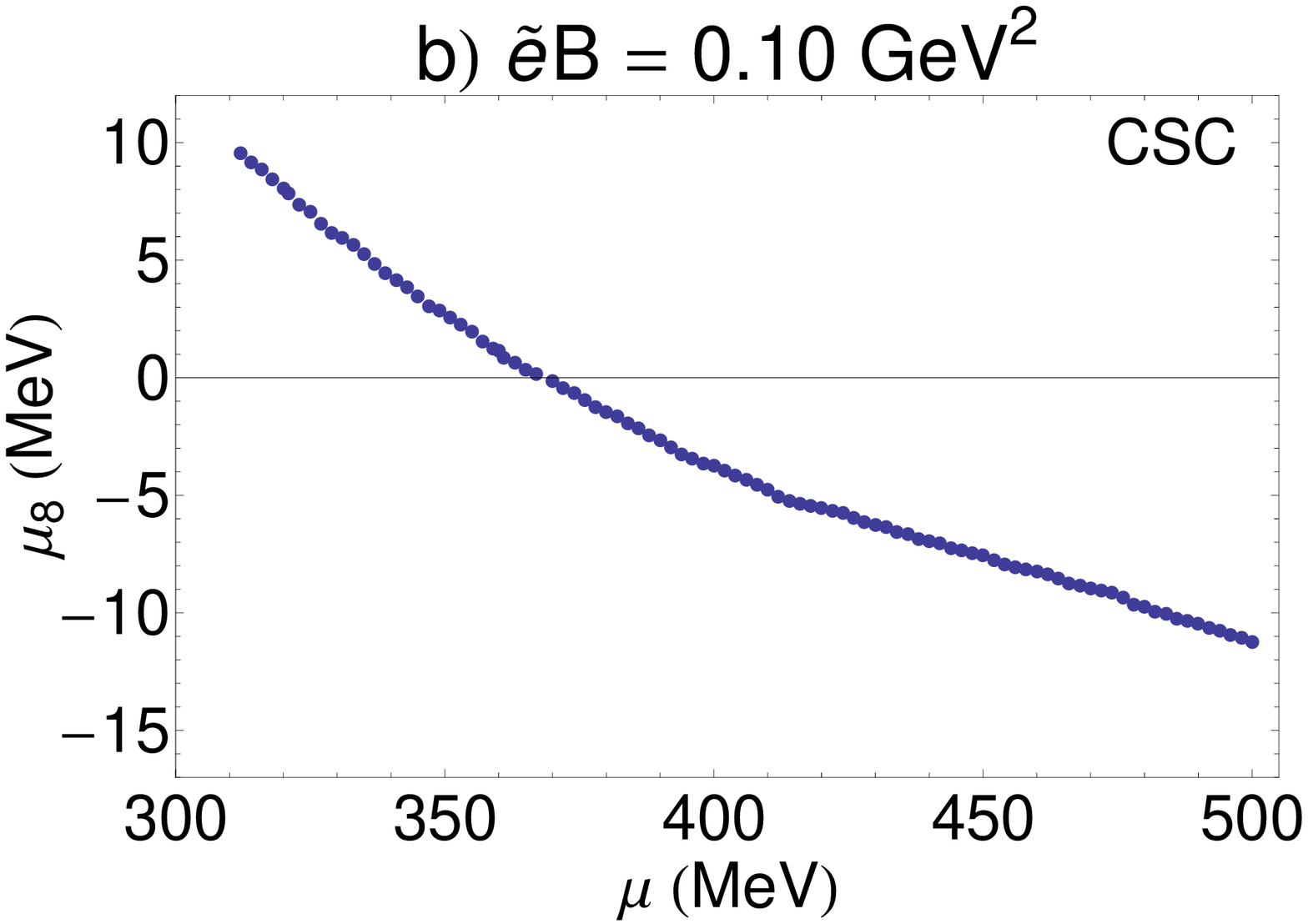}
\hspace{0.2cm}
\includegraphics[width=5cm,height=4cm]{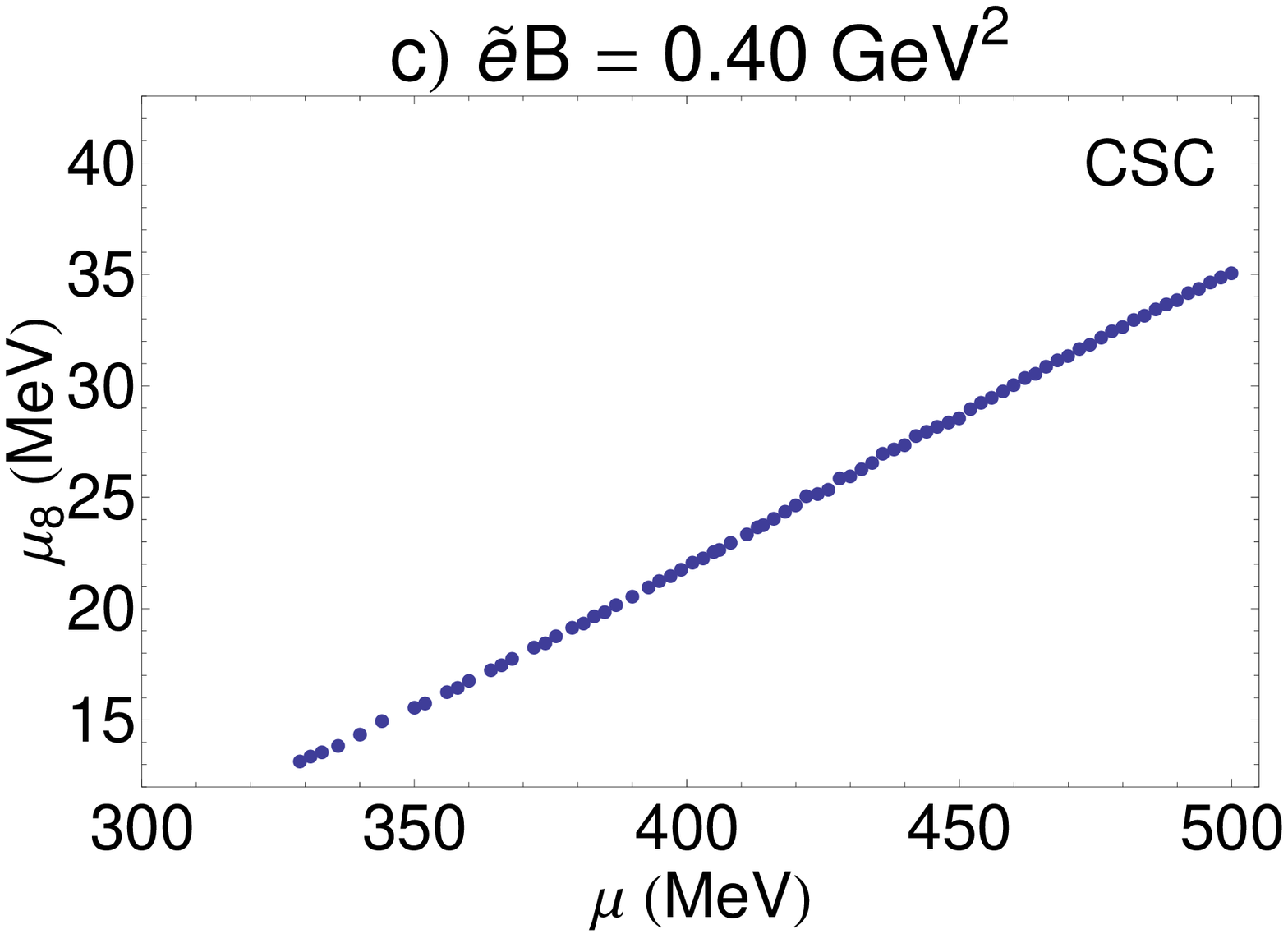}
\hspace{0.2cm}
\par\vspace{0.5cm}
\includegraphics[width=5cm,height=4cm]{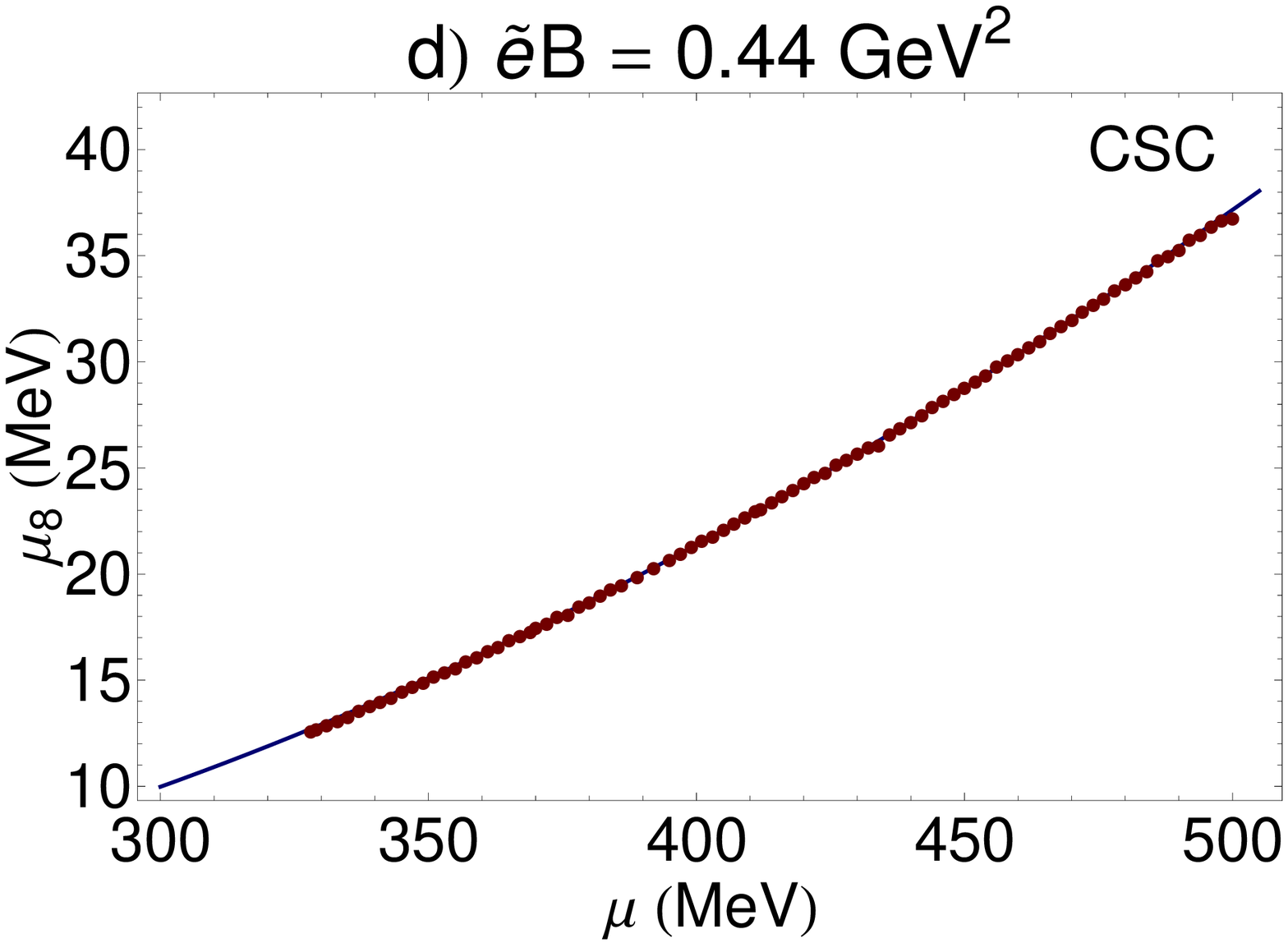}
\hspace{0.2cm}
\includegraphics[width=5cm,height=4cm]{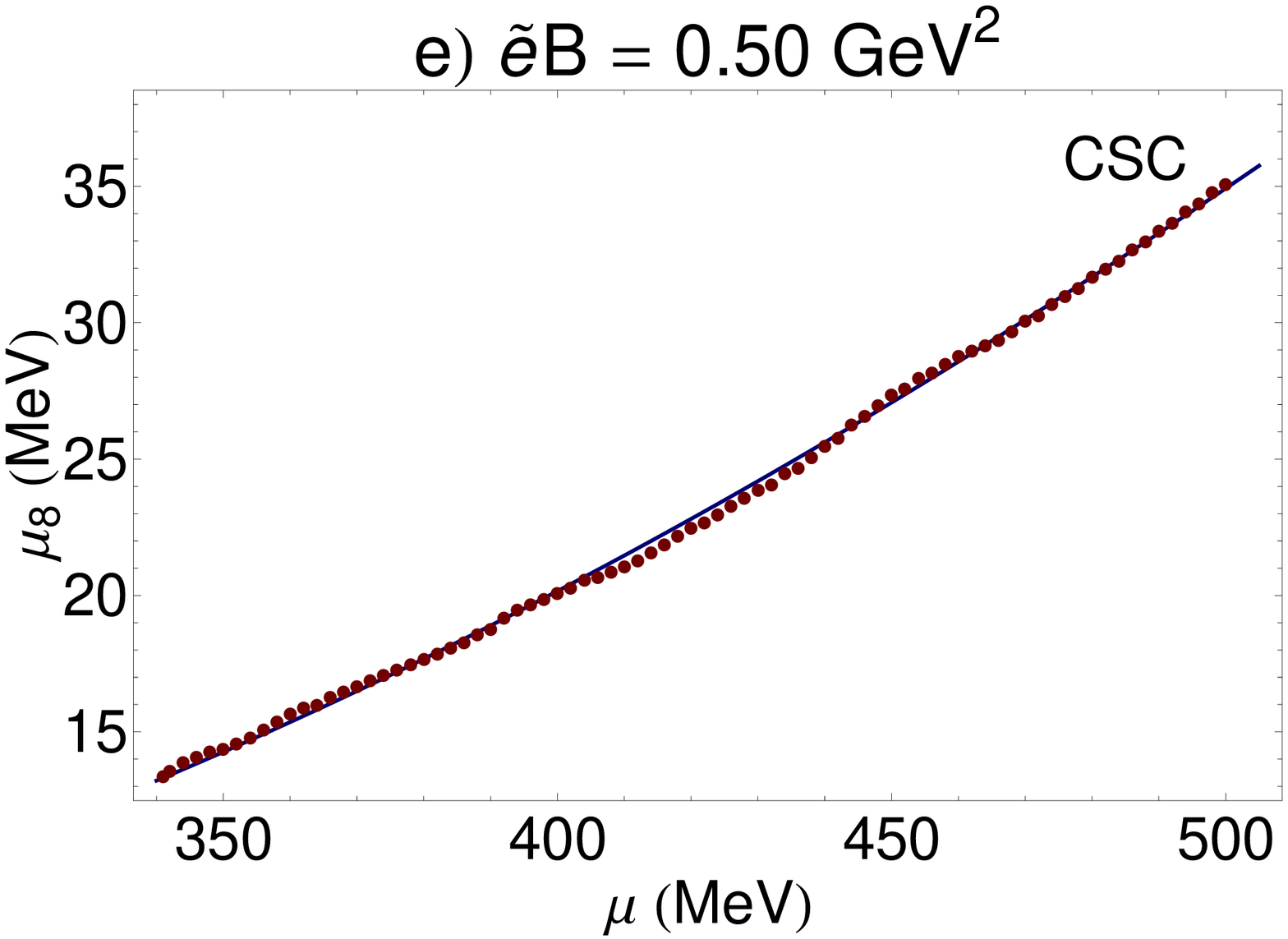}
\hspace{0.2cm}
\includegraphics[width=5cm,height=4cm]{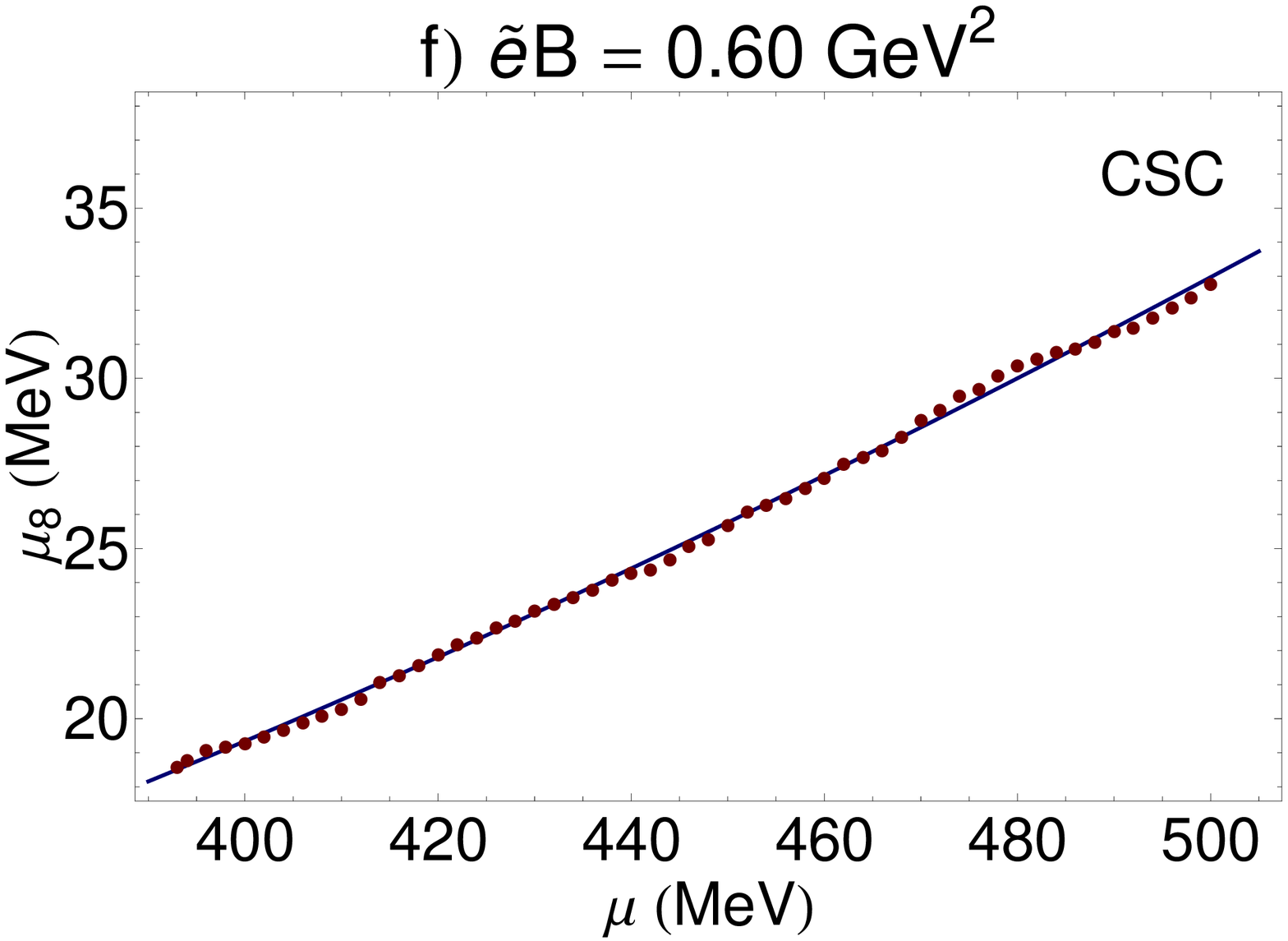}
\caption{The $\mu$-dependence of the color chemical potential
$\mu_{8}$ as a function. The numerical data for $\mu_{8}$ are fitted
in Figs $8d-8f$ by (\ref{B6}). In the linear regime, the fitted
curves [solid (blue) lines in $8d-8f$] are in good agreement with
our numerical data. For $\tilde{e}B=0.44,0.5$ GeV$^{2}$ and $0.6$
GeV$^{2}$, the regression parameter $R^{2}$, as a measure of
reliability of numerical fits are
$R^{2}=0.999991,0.999924,0.999952$, respectively. }
\end{figure}
\par
In Fig. 8, the $\mu$-dependence of the  color chemical potential
$\mu_{8}$ for different $\tilde{e}B$ is demonstrated. As in the
previous cases, we expect that, in the linear regime
$\tilde{e}B\gtrsim 0.45$ GeV$^{2}$, the $\mu$ dependence of
$\mu_{8}$ is given by a function similar to (\ref{Y30-b}), that
arises analytically in the LLL approximation. We define therefore a
function
\begin{eqnarray}\label{B6}
\mu_{8}=\frac{\mu^{3}}{a+b\mu^{2}},
\end{eqnarray}
with arbitrary, $\tilde{e}B$-dependent parameters $a$ and $b$. In
Table V, we have compared the data that arise numerically by fitting
the numerical values of $\mu_{8}$ with (\ref{B6}) for different
$\tilde{e}B$. As in the previous case, the difference between the
numerical fit data and the expected analytical values of $a$ and $b$
arising from (\ref{Y30-b}) minimizes in the linear regime for
$\tilde{e}B\gtrsim 0.40$ GeV$^{2}$ ($\eta_{a}$ and $\eta_{b}$ in
Table V are less than $50$\%.).
\begin{table}
\begin{tabular}{| c | c |c || c |c ||c|c|}
          \hline
&\multicolumn{2}{|c||}{\textbf{Analytical
parameters}}&\multicolumn{2}{c||}{\textbf{Numerical fit parameters}}
&\multicolumn{2}{|c|}{\textbf{$\eta$ in \%}}\\
\hline $\tilde{e}B$ (GeV$^{2}$) & $a$ (MeV$^2$)& $b$ & $a$ (MeV$^2$)& $b$&$\eta_{a}$&$\eta_{b}$\\
\hline\hline
$0.04$ & $1.80\times10^{5}$ & $6$ & $-1.01\times10^{7}$ & $+7.53$&$207$&$23$  \\
\hline
$0.10$ & $4.50\times10^{5}$ & $6$ & $-4.93\times10^{7}$ & $+161.14$&$204$&$186$ \\
\hline
$0.30$ & $1.35\times10^{6}$ & $6 $& $+3.04\times10^{6}$ & $+13.04$&$77$&$74$ \\
\hline
$0.40$ & $1.80\times10^{6}$ &  $6 $ & $+1.93\times10^{6}$ &$+6.27$&$7$&$4$ \\
\hline
$0.44$ & $1.98\times10^{6}$ &  $6 $ & $+2.34\times10^{6}$ & $+4.11$&$17$&$37$ \\
\hline
$0.46$ & $2.07\times10^{6}$ &  $6 $ & $+2.43\times10^{6}$ & $+3.89$&$16$&$43$ \\
\hline
$0.50$ & $2.25\times10^{6}$ &  $6 $ & $+2.46\times10^{6}$ & $+4.49$&$9$&$29$ \\
\hline
$0.60$ & $2.70\times10^{6}$ &  $6 $ & $+2.45\times10^{6}$ & $+5.35$&$10$&$11$ \\
\hline
\end{tabular}
\caption{Numerical fit data for $\mu_{8}$ as a function of $\mu$
from (\ref{B6}). In the linear regime, i.e. for $\tilde{e}B\gtrsim
0.40$ GeV$^{2}$, the numerical values of the parameters arising from
our fit are in good agreement with the expected analytical values of
the parameters from (\ref{Y30-b}) [see $\eta_{a}$ and $\eta_{b}$
with $\eta$ defined in (\ref{eta})].}
\end{table}
\par
Finally, we will present the phase structure of the model in a
$\mu_c-\tilde{e}B$ plane in Fig. 9. In particular, we are interested
on the effect of the  color chemical potential $\mu_{8}$ on the
phase structure of the model. In Fig. 9a (Fig. 9b) the phase
structure for $\mu_{8}=0$ ($\mu_{8}\neq 0$) is plotted. Because of
our specific choice of parameters, we expect $\chi$SB and CSC phase
without mixing. A normal phase can also exist, where the mass gaps
$\sigma_{B}$ and $\Delta_{B}$ corresponding to $\chi$SB and CSC
phases vanish identically.\footnote{As it is known from [5], in the
regime of large chemical potential, $\mu\gtrsim 500$ MeV, the 2SC
phase goes over into the three-flavor CFL phase. In the present
two-flavor model, we only assume that a normal phase may exist, and,
if so a phase transition will occur from the color superconducting
2SC phase into this normal phase (see Fig. 9). Hence, the present
results concerning the transition from CSC to the normal phase is
only of theoretical nature. To include the CFL phase, we have to
extend the model to three-flavor superconductivity including up,
down and strange quarks. This is indeed beyond the scope of the
present paper and is planned for future publications.} To check
this, we consider the gap equations and the color neutrality
condition (\ref{YY-1a}). We have looked for the global minima of the
system in two different regimes: $\mu_{c}\simeq 350-450$ MeV and
$\mu_{c}\simeq 750-800$ MeV. As it turns out, in the first regime
corresponding to $\mu_{c}\simeq 350-450$ MeV, the minima of
$\Omega_{\mbox{\tiny{eff}}}$ from (\ref{A24}) are given by
$(\sigma_{B}\neq 0,\Delta_{B}=0,\mu_{8}=0)$ for $\mu<\mu_{c}$ as
well as $(\sigma_{B}=0,\Delta_{B}\neq 0,\mu_{8}\neq 0)$ for
$\mu>\mu_{c}$. In the second regime corresponding to $\mu_{c}\simeq
750-800$ MeV, however, the global minima are
$(\sigma_{B}=0,\Delta_{B}\neq 0,\mu_{8}\neq 0)$ for $\mu<\mu_{c}$,
as well as $(\sigma_{B}=0,\Delta_{B}=0,\mu_{8}=0)$ for
$\mu>\mu_{c}$. We conclude therefore that a phase transition from
$\chi$SB to CSC phase occurs in the first regime at $\mu_{c}\simeq
350-450$ MeV, and a phase transition from CSC to the normal phase
occurs in the second regime $\mu_{c}\simeq 750-800$ MeV. In the
following, we denote the value of $\Omega_{\mbox{\tiny{eff}}}$ at
the global minima by $\Omega_{\mbox{\tiny{$\chi$SB}}}\equiv
\Omega_{\mbox{\tiny{eff}}}(\sigma_{B},0,0;\mu,\tilde{e}B)$,
$\Omega_{\mbox{\tiny{CSC}}}\equiv
\Omega_{\mbox{\tiny{eff}}}(0,\Delta_{B},\mu_8;\mu,\tilde{e}B)$, and
$\Omega_{\mbox{\tiny{Normal}}}\equiv
\Omega_{\mbox{\tiny{eff}}}(0,0,0;\mu,\tilde{e}B)$ corresponding to
the $\chi$SB, CSC and the normal phase, respectively. For different
values of $(\mu,\tilde{e}B)$, the $\chi$SB phase is defined by
$\Omega_{\mbox{\tiny{$\chi$SB}}}\leq \Omega_{\mbox{\tiny{CSC}}}$ and
the CSC phase by $\Omega_{\mbox{\tiny{CSC}}}\leq
\Omega_{\mbox{\tiny{$\chi$SB}}}$. Moreover, the exact value of
$\mu_{c}$ for the first order phase transition from $\chi$SB to the
CSC phase [the lower (red) solid line in Fig. 9a and 9b] and from
the CSC phase to the $\chi$SB phase [(green) solid line in Fig. 9b]
are then defined by
$\Omega_{\mbox{\tiny{$\chi$SB}}}=\Omega_{\mbox{\tiny{CSC}}}$ and
$\Omega_{\mbox{\tiny{CSC}}}=\Omega_{\mbox{\tiny{Normal}}}$,
respectively \cite{sato1998}. As for the second order phase
transition between the CSC and the normal phase, an analysis similar
to \cite{berges1998} is performed.
\begin{figure}[hbt]
\includegraphics[width=8cm,height=6cm]{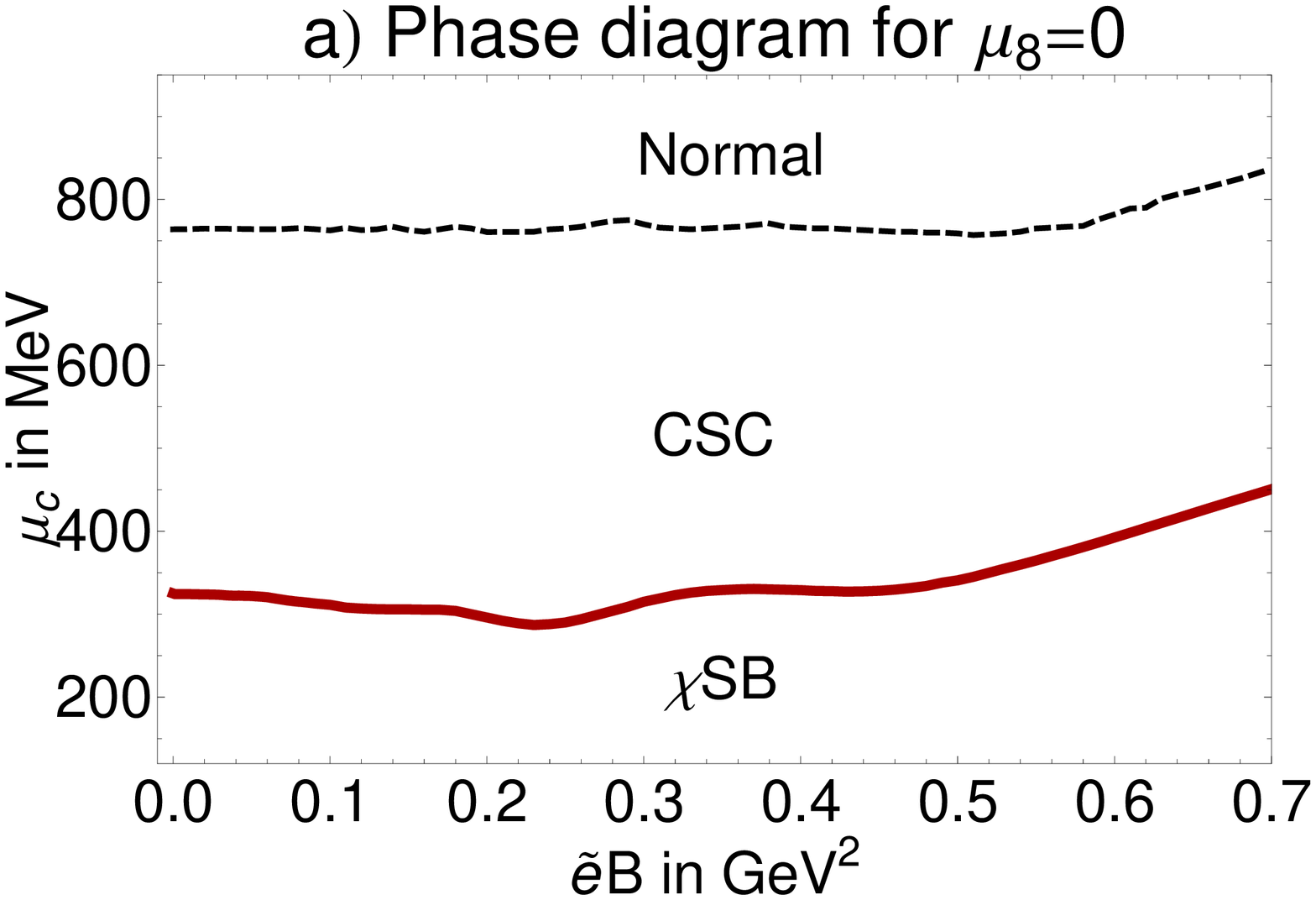}
\includegraphics[width=8cm,height=6cm]{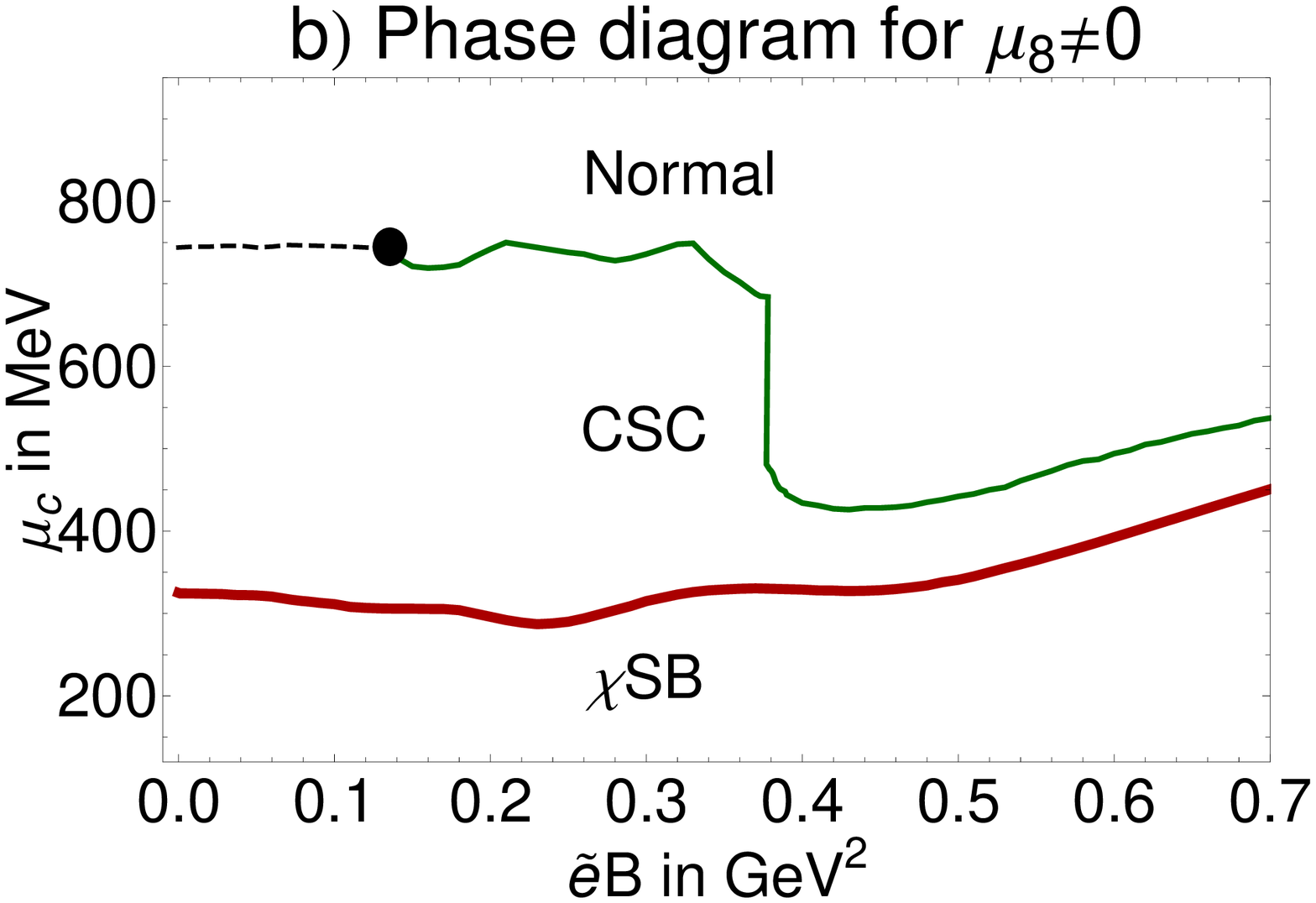}
\hspace{0.3cm}
\caption{The phase diagram of 2SC model is plotted in a
$\mu_{c}-\tilde{e}B$ plane for a) $\mu_{8}=0$ and b) $\mu_{8}\neq
0$. The solid (red) lines in a) and b)  indicate first order phase
transitions between the $\chi$SB and the CSC phase. The dashed
(black) lines in a) and b) are second order critical lines between
the CSC and the normal phase. As it is shown in  b), for
$\mu_{8}\neq 0$ and at $\mu_{c}\simeq 755$ MeV and $\tilde{e}B\simeq
0.13$ GeV$^{2}$, the second order phase transition goes over into a
first order phase transition between the CSC and the normal phase
[solid (green) line]. At $\tilde{e}B\simeq 0.4$ GeV$^{2}$, $\mu_{c}$
suddenly decreases and increases once again by increasing the
external magnetic fields in the CSC regime.}
\end{figure}
In Fig. 9a, the phase diagram of the 2SC model in a
$\mu_{c}-\tilde{e}B$ plane is plotted for vanishing $\mu_{8}$. A
first order phase transition occurs between the $\chi$SB and the CSC
phase in the regime $\mu_{c}\simeq 350-450$ MeV (solid red line).
This confirms the results by \cite{mandal2009}, where a first order
phase transition is observed for fixed value of $\tilde{e}B=0.05$
GeV$^{2}$, and various $G_{D}/G_{S}$. The transition from the CSC
into the normal phase is of second order and occurs in the regime
$\mu_{c}\simeq 750-800$ MeV (dashed black line). According to the
phase diagram for nonvanishing $\mu_{8}$ in Fig. 9b, however,
whereas the transition from the $\chi$SB to CSC phase is still of
first order (solid red line), the second order phase transition for
small values of $\tilde{e}B$ (dashed black line) goes over into a
first order phase transition at $\mu_{c}\simeq 755$ MeV and
$\tilde{e}B\simeq 0.13$ GeV$^{2}$ (solid green line). Moreover, at
$\tilde{e}B\simeq 0.4$ GeV$^{2}$, $\mu_{c}$ suddenly decreases and
increases once again by increasing the external magnetic fields in
the CSC regime. The CSC regime is nevertheless suppressed in the
linear regime $\tilde{e}B\gtrsim 0.45$ GeV$^{2}$ by the external
magnetic field (see Fig. 9b).
\section{Concluding remarks}
\par\noindent In this paper, we have studied the effect of constant magnetic
fields on the formation of bound states in the chiral as well as the
color symmetry breaking phase. In the first part of the paper, after
introducing a two-flavor NJL type model including meson and diquark
condensates, we have computed the one-loop effective action and the
thermodynamic potential of the theory at zero temperature and finite
density. Neglecting the quark mass $m_{0}$ and choosing the
diquark-to-chiral coupling ratio $G_{D}/G_{S}<1$ \cite{huang2002},
we can consider the $\chi$SB and CSC phases separately. The $\chi$SB
and CSC mass gaps $\sigma_{B}$ and $\Delta_{B}$ as well as the color
chemical potential $\mu_{8}$ are determined analytically in the
limit of very strong magnetic fields. In this limit, the dynamics of
the system is dominated by the lowest Landau level and therefore the
effect of all higher Landau levels are negligible. According to
\cite{miransky1995}, in this limit, as a result of dimensional
reduction from $D$ to $D-2$ dimensions, the formation of bound
states and consequently a dynamical symmetry breaking will be
possible even for weak interactions between two fermions. This is
the phenomenon of magnetic catalysis, discussed widely in the
literature in the past few years \cite{cosmology, condensed,
particle}. Denoting the dimensionless coupling constants in the
$\chi$SB and CSC phases by $g_{s}\sim G_{S}\Lambda^{2}$ and
$g_{d}\sim G_{D}\Lambda^{2}$,  we have determined the mass gaps for
different regimes of $g_{s}$ and $g_{d}$. Here, $\Lambda$ is certain
momentum cutoff. In \cite{ebert2005}, the $\chi$SB and CSC mass gaps
of a similar 2SC model was determined numerically for vanishing
magnetic field. Introducing a large momentum cutoff $\Lambda$ and
performing appropriate approximations, we have determined
analytically the mass gaps $\sigma_{0}$ and $\Delta_{0}$ as well as
$\mu_{8}$ corresponding to $\chi$SB and CSC phases for zero magnetic
field too.
\par
In the second part of the paper, a detailed numerical analysis is
performed to explore the effect of any arbitrary magnetic field on
the mass gaps $\sigma$ and $\Delta$ and the color chemical potential
$\mu_{8}$. First, the dependence of $\sigma$ and $\Delta$ on various
$\tilde{e}B\in [0,1]$ GeV$^{2}$ is plotted for fixed $\mu=250$ MeV
and $\mu=460$ MeV in the $\chi$SB and CSC phases, respectively. For
small values of $\tilde{e}B$, we observe small van Alfven-de Haas
oscillations, that appear, according to \cite{warringa2007,
shovkovy2007} also in the CFL phase for $\mu=500$ MeV. Same
oscillations appears also in the magnetization $M$ for the same
fixed value of chemical potentials. At $\tilde{e}B\simeq 0.4-0.5$
GeV$^{2}$, the oscillations end up in a linear regime, where we
believe that the dynamics of the system is described exclusively by
the LLL. This can be checked by comparing qualitatively the
numerical dependence of $\sigma_{B}$ and $\Delta_{B}$ for
$\tilde{e}B\geq 0.45$ GeV$^{2}$ and fixed $\mu$. The
$\mu$-dependence of $\sigma$ and $\Delta$ are then considered for
various $\tilde{e}B$. Our numerical results for vanishing magnetic
field coincide with the numerical results presented in
\cite{ebert2005}. The $\mu$-dependence of $\sigma_{B}$ and
$\Delta_{B}$ as well as $\mu_{8}$ are then considered for various
finite $\tilde{e}B$. The numerical results in the linear regime,
i.e. for $\tilde{e}B\gtrsim 0.45$ GeV$^{2}$ are comparable with our
before mentioned analytical results in the limit of large
$\tilde{e}B$. This is shown using appropriate numerical fits. The
phase structure of the $\chi$SB and CSC phases in a
$\mu_{c}-\tilde{e}B$ plane is also presented. We are in particular
interested on the effect of the color chemical potential $\mu_{8}$
on the phase diagram of the model. For both $\mu_{8}=0$ as well as
$\mu_{8}\neq 0$, the transition from the $\chi$SB phase into the CSC
phase is of first order, and occurs in the regime $\mu_{c}\simeq
350-450$ MeV and for $\tilde{e}B\in [0,0.7]$ GeV$^{2}$ (see Fig.
9a). This confirms the result in \cite{mandal2009}, where a first
order phase transition is observed between the $\chi$SB and CSC
phase for fixed $\tilde{e}B=0.05$ GeV$^{2}$ and various
$G_{D}/G_{S}$ ratios. Assuming that the CSC phase goes over into a
normal phase at $\mu>500$ MeV,\footnote{See footnote 23.} it turns
out that whereas for $\mu_{8}=0$, this transition is of second
order, for nonvanishing $\mu_{8}$, a second order phase transition
occurs first for small $\tilde{e}B$. It goes then over into a first
order phase transition at $\mu_{c}\simeq 755$ MeV and
$\tilde{e}B\simeq 0.13$ GeV$^{2}$. At $\tilde{e}B\simeq 0.4$
GeV$^{2}$, $\mu_{c}$ suddenly decreases and increases once again by
increasing the external magnetic fields. The CSC phase is
nevertheless suppressed in the linear regime $\tilde{e}B\gtrsim
0.45$ GeV$^{2}$ by the external magnetic field (see Fig. 9b).
\par
At the end, let us just emphasize that the study of color
superconductivity in the presence of constant magnetic fields has
not only astrophysical consequences in forming the structure of
compact star cores, it may be also relevant for future heavy ion
collision experiments. Recently in \cite{blaschke2010}, the
accessibility of the color superconducting 2SC phase in the heavy
ion collisions is investigated. It is stated that for high enough
collision energies the 2SC may be accessible in future collision
experiments. On the other hand, there are various evidences of the
creation of very strong magnetic fields in non-central heavy ion
collisions \cite{skokov, kharzeev51-STAR}. It would be interesting
to perform similar analysis as in \cite{blaschke2010} considering
the presence of constant magnetic fields. To do this, the effect of
finite temperature on the phase diagram of the 2SC superconducting
phase in the presence of constant magnetic fields has also to be
considered. This will be reported in future publications
\cite{sadooghi2010}.
\section{Acknowledgments}
\par\noindent  Sh.~F. thanks F. Farahpour for useful discussions on
numerical results, and H. Hadipour for discussions on high
$T_{c}$-superconductivity, N.~S. thanks M. Bahmanabadi and S.~Rahvar
for discussions on the numerical fit data. Both authors thank F.
Ardalan, H. Arfaei, A.~E. Mosaffa for discussions and E.~J.~Ferrer
for email correspondence.


\end{document}